\documentclass[11pt,preprint]{aastex}

\usepackage{ulem}
\usepackage{graphicx}

\shorttitle{The $L$-$\sigma$ Relation of Local H{\sc ii} Galaxies}
\shortauthors{Bordalo \& Telles}
\begin{document}
\title{The $L$-$\sigma$ Relation of Local H{\sc ii} Galaxies}
\author{V. Bordalo\altaffilmark{1} and E. Telles\altaffilmark{1}}
\affil{Observat\'orio Nacional, Rua Gal. Jos\'e Cristino, 77, CEP
20912-400, S\~ao Crist\'ov\~ao, Rio de Janeiro, Brasil;
vschmidt@on.br, etelles@on.br}

\begin{abstract}
We present for the first time a new data set of emission
line widths for 118 star-forming regions in H{\sc ii} galaxies
(H{\sc ii}Gs). This homogeneous set is used to investigate the
$L$-$\sigma$ relation in conjunction with optical spectrophotometric
observations.
We were able to classify their nebular emission line profiles due to
our high resolution spectra. Peculiarities in the line profiles such
as sharp lines, wings, asymmetries, and in some cases more than one
component in emission were verified. From a new independent
homogeneous set of spectrophotometric data we derived physical
condition parameters and performed the statistical principal
component analysis. We have investigated the potential role of
metallicity (O/H), H$\beta$ equivalent width
($W_{\scriptsize\mbox{H}\beta}$) and ionization ratio [OIII]/[OII]
to account for the observational scatter of $L$-$\sigma$ relation.
Our results indicate that the $L$-$\sigma$ relation for H{\sc ii}Gs
is more sensitive to the evolution of the current starburst event
(short-term evolution) and dated by $W_{\scriptsize\mbox{H}\beta}$
or even the [OIII]/[OII] ratio. The long-term evolution measured by
O/H also plays a potential role in determining the
luminosity of the current burst for a given velocity dispersion and
age as previously suggested. Additionally, galaxies showing Gaussian
line profiles present more tight correlations indicating that they
are best targets for the application of the parametric relations as
an extragalactic cosmological distance indicator. Best fits for a
restricted homogeneous sample of 45 H{\sc ii}Gs provide us a set of
new extragalactic distance indicators with an RMS scatter compatible
with observational errors of $\delta\log
L_{\scriptsize\mbox{H}\alpha}$ = 0.2 dex or 0.5 mag. Improvements
may still come from future optimized observational programs to
reduce the observational uncertainties on the predicted
luminosities of H{\sc ii}Gs in order to achieve the precision
required for the application of these relations as tests of
cosmological models.

\end{abstract}

\keywords{galaxies: HII --  starburst, galaxy: dynamics -- kinematics
dynamics}

\section{Introduction}\label{sec1}

Giant H{\sc ii} regions (GH{\sc ii}Rs) and H{\sc ii} galaxies (H{\sc
ii}Gs) have been intensively studied for almost forty years not only
because they are natural laboratories to test astrophysical models
\citep[][and references therein]{sta03} but also because they
present tight scaling relations which can be useful as extragalactic
distance estimators
\citep{mel00,mel03,sie05,pli09}.

The correlations between nebular diameter, luminosity ($L$) and
velocity dispersion ($\sigma$) of the ionized gas in GH{\sc ii}Rs
were found by \cite{mel77,mel78,mel79} and further investigated by
\cite{ter81}. Several other works from independent groups have
confirmed the existence of these relations for GH{\sc ii}Rs in the
Local Group's magellanic irregular galaxies and in some nearby
spirals but there is no agreement about the calibration coefficients
(slope and zero point), mainly due to different sample selection,
observational data quality and linear fit algorithms
\citep{hip86,mel87,ars88,fue00,bos02,roz06}. For H{\sc ii}Gs these
studies are more scarce. The first $L$-$\sigma$ calibration was
obtained by \cite*{mel88} (hereafter MTM) with an RMS scatter of
$\delta \log L_{\scriptsize\mbox{H}\beta} \sim 0.30$, which is a
landmark for follow up achievements. In fact, part of this observed
scatter has been proposed to be associated with a second parameter,
namely oxygen abundance (MTM) or core radius \citep{tel93}.

Despite the fact that the physical origin of the observed supersonic
line widths has been a topic of intense debate in the literature,
with no consensus \citep{ter81,ten93,chu94,sca99,mel99,ten06,bor09},
the $L$-$\sigma$ relation remains potentially as a powerful
alternative empirical extragalactic distance estimator to the
classical Tully-Fisher for spirals and
$D_{\scriptsize\mbox{n}}$-$\sigma$ relations for ellipticals. This
is still more exciting since Tully-Fisher and
$D_{\scriptsize\mbox{n}}$-$\sigma$ relations can only be applied up
to redshift $\sim$ 1, where the relation is less affected by natural
galaxy evolution with the look-back time \citep{fer11,fer10}. On the
other hand, light curves of Supernovae Type Ia (SNIa), which are
today the most used technique to obtain such cosmological distances
encounters lack of target objects at redshifts above
$\sim1.2$ \citep{rie07}.

There are, therefore, two possible roles for the $L$-$\sigma$
relation: (1) to obtain distances to nearby galaxies, mainly in the
Local Group, where peculiar velocities are significant compared to
cosmological recession velocities and distances to nearby galaxy
clusters, where H{\sc ii}Gs can be found in their neighborhood; (2)
to obtain distances to intermediate and high redshift galaxies
(cosmological distances) to probe the dark energy equation of state
parameter through Hubble diagram analysis. Several issues, however,
should be further investigated for this latter goal with precision
required in the era of ``concordance cosmology'': (i) {\it
the origin of $\sigma$} - the physical mechanism which produces the
observed supersonic motions in the ISM (e.g. gravity, turbulence,
feedback, stochastic effects of the ISM, etc.); (ii) {\it  the
validity of the relation to high redshifts} - the identification of
bona-fide H{\sc ii}Gs at great distances due to the conspicuous
emission lines should allow the validity test, once a homogeneous
set of kinematic and spectrophotometric data is gathered; (iii) {\it
systematic effects} - evolutionary and other effects may affect the
relation and may be parameterized (e.g. age of the starburst,
metallicity, etc.); (iv) {\it observational errors} -
spectrophotometric calibration, distance, line width, errors may
still be suppressed so that the calibration can be competitive with
other distance indicators of cosmological interest, and so that the
intrinsic scatter in the relation can be assessed; (v) {\it
zero-point calibration} - further improvement of the zero-point for
H{\sc ii}Gs may be achieved by a revision of the distances for
GH{\sc ii}Rs in view of modern observations, and at the same time,
test the hypothesis that both classes follow the same relation as
proposed by MTM.

Several recent works have investigated the kinematics of
star-forming galaxies at high redshifts $z \sim 2 - 3$
\citep{law09,erb04,pet01}. These authors have shown that most of the
galaxies found exhibit high local velocity dispersions $\sim 60 -
100$ km s$^{-1}$, suggesting that even for those galaxies with clear
velocity gradients, rotation about a preferred kinematic axis may
not be the dominant means of physical support. This is also been
confirmed in local H{\sc ii}Gs \citep{moi10,bor09,mar07,mai99}.
Despite the fact that most of these distant galaxies have suffered a
more violent, and probably continuous star-formation, which led them
to become the normal galaxies found in the local Universe, many of
their juvenile physical properties are very similar to those found
in local H{\sc ii}Gs. These provide us, therefore, empirical support
to speculate about the ambitious goal for using the $L$-$\sigma$
relation to determine distances of high redshift galaxies.

In this paper, we present for the first time a large homogeneous
data set of emission line width measurements of over 100
local H{\sc ii}Gs ($z$ $<$ 0.1), doubling the sample of MTM with
velocity widths obtained from high resolution spectra. These were
combined with a complete set of spectrophotometric data obtained
mostly from \cite*{keh04} (hereafter KTC) to produce a new
calibration for the $L$-$\sigma$ relation. We investigated the
potential role of different systematic effects over the $L$-$\sigma$
relation, such as age, metallicity, aperture and non-Gaussianity of
the emission line profiles. We argue that the detailed study of
these effects is crucial to identify a homogeneous sample for which
the relation is valid, and bring light on its physical
interpretation.

The paper is organized as follows. Section 2 describes the data,
observations and reductions. The results are presented in \S 3. In
\S 4 we present our data analysis. We present a discussion about our
results in \S 5 and summarize the conclusions in \S 6. The
Hubble Constant adopted throughout this work is $H_{0}$ = 71 km
s$^{-1}$ Mpc$^{-1}$.

\section{Data Sample, Observations and Reductions}\label{sec2}

\subsection{The Sample}\label{sec2_1}

We have selected most objects from the Spectrophotometric Catalogue
of H{\sc ii} Galaxies (SCHG) \citep[hereafter T91]{ter91}. It
contains many galaxies from Curtis Schmidt-Thin Prism Survey of
Tololo \citep[Tol]{smi76} and University of Michigan Survey
\citep[UM]{mac77}. The SCHG also contains a few galaxies from the
Fairall, Markarian and Zwicky lists \citep[F80, MRK, Zw,
respectively]{fai80,mar67,zwi66}. We have also selected H{\sc ii}Gs
from smaller surveys such as those produced by \cite{kun81} (POX),
\cite{maz91} (CTS) and \cite{sur98} (SC98). Cambridge UK Schmidt
galaxies (Cam) have been selected from \cite{cam86}. Note that some
of the H{\sc ii}Gs had their names cataloged in more than
one of these lists. Additionally, we have selected some classical
starburst galaxies | spiral galaxies with H{\sc ii} nuclear regions
or simply nuclear starburst galaxies | from Montreal Blue Galaxy
Survey (MBG) \citep{coz93}. However, the objects NGC 6970, IC 5154,
ESO 533-G 014 and MCG -01-57-017 were only presented in a private
list from Roger Coziol. Nuclear starburst galaxies are also often
present in H{\sc ii}Gs lists due to their similar optical
spectroscopic properties (for example, UM 477 and MRK 710). Thus our
sample consists of 120 starburst regions in galaxies for which we
have obtained line widths from optical high spectral resolution
spectroscopy.

\begin{figure*}[ht]
\epsscale{0.4} \plotone{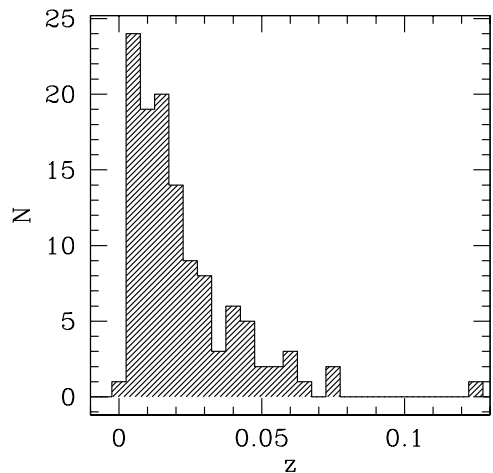} \caption{The redshift distribution
of our sample containing 120 objects. The mean of the distribution
is 0.0222, whereas the median is 0.0167.\label{fig1}}
\end{figure*}

This sample is not complete in a statistical
sense\footnote{Definition of completeness for this class of emission
line objects is rather tricky and a more detailed discussion on this
issue is given by \cite{sal89}.}. It is heterogeneous in nature
comprising four orders of magnitude in H$\alpha$ luminosity
range. Figure~\ref{fig1} shows the redshift distribution of our
sample. The mean of the distribution is 0.022, and the median is
0.017. The study including all these galaxies is fundamental to
investigate the range of the starburst magnitude for which the
$L$-$\sigma$ relation is valid. This is, to date, the
largest sample of H{\sc ii}Gs ever studied in order to test and
calibrate the local $L$-$\sigma$ relation. This has been possible
due to the private agreement between the Observat\'orio Nacional-MCT
and the European Southern Observatory (ESO) for the dedicated use of
the 1.52m and 2.2m telescopes at La Silla, Chile.

\subsection{High Spectral Resolution Spectroscopy}\label{sec2_2}

H{\sc ii}Gs are mostly compact objects and the young starburst
regions in the cores of these systems dominate the main
observational properties, i.e. emission line fluxes and their widths
\citep{tel01}. More recently, \cite{bor09} have confirmed
with a spatially resolved kinematics study of the
prototypical H{\sc ii}G II Zw 40, using 3D integral field
spectroscopy, that the line width measured in the nuclear core is
the same as the line width measured over the whole extent of the
starburst region, indicating that this kinematic core
contains information about the overall dynamics of the warm
gas.

\begin{figure*}[ht]
\epsscale{0.85} \plotone{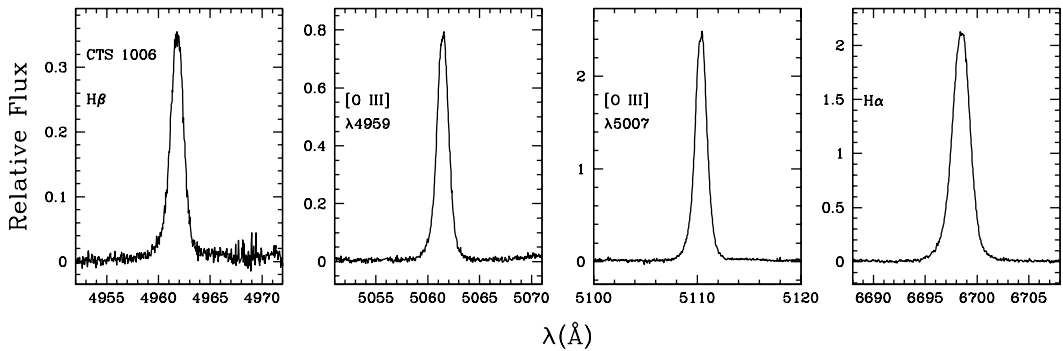}\caption{Optical wavelength
calibrated spectrum of CTS 1006 obtained with the FEROS
spectrograph. FEROS is an optical echelle spectrograph covering a
wide range in wavelength, thus the most intense lines are shown in
separated boxes.\label{fig2}}
\end{figure*}

We have decided to use the Fiber-fed Extended Range Optical
Spectrograph (FEROS) installed initially on the 1.52m and, later,
on the ESO 2.2m telescope at La Silla Observatory in Chile,
for the first time to observe galaxies, increasing our sample to the
present 120 objects. The target fiber was positioned over the
brightest region (nuclear core) of the galaxies. FEROS consists of
two fibers coupled to the Cassegrain focus of the telescope
by micro lenses, providing a spectral resolution of R = 48000
($\sigma_{\scriptsize\mbox{inst}}$ = 2.50$\pm$0.20 km s$^{-1}$,
$\sigma$ = FWHM/2.355). Each of the two fibers has a projected
2.7{\arcsec} entrance aperture and the target and sky are recorded
simultaneously.

The echelle FEROS spectrum covers the whole optical region
3560-9200{\AA}. We have observed 103 galaxies with this instrument
in five observational runs in the period between November 2000 and
April 2007. The FEROS spectra were recorded in a 2048 $\times$ 4096
15$\mu$m pixel CCD. The basic reduction, extraction and dispersion
calibration of the spectra were done by a pipeline routine in MIDAS
\citep{fra99}. It processes Bias and Flat Field calibration in a
standard way and applies dispersion calibration to the object
spectra from information of a thorium-argon-neon lamp
spectrum. The final spectrum is calibrated only in
wavelength.
Line widths for Balmer H$\beta$ (4861 {\AA}), H$\alpha$ (6563 {\AA})
and [OIII] $\lambda\lambda$4959,5007 were measured for most of the
galaxies observed. For four galaxies (Tol 0226-390, CTS 1004, Cam
08-28A and CTS 1038) it was not possible to measure H$\alpha$ widths
due to regions of bad pixels in the CCD. An example of a wavelength
calibrated FEROS spectrum is shown in Figure~\ref{fig2}.

\begin{figure*}[ht]
\epsscale{0.65} \plotone{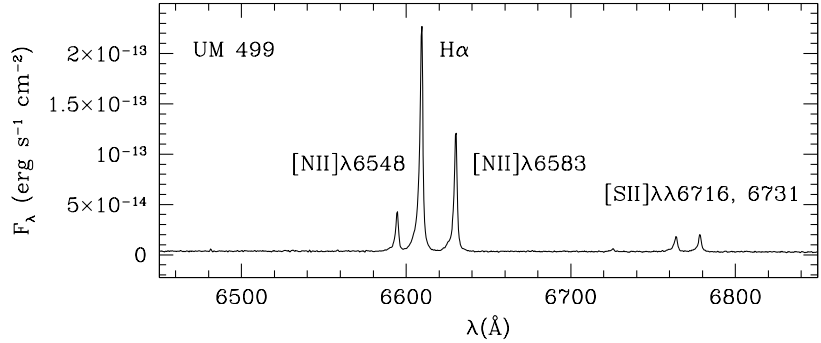} \caption{Optical calibrated
spectrum of UM 499 obtained with the Coud\'e
spectrograph.\label{fig3}}
\end{figure*}

Additional line widths were obtained
with the 1.60m telescope at Pico dos Dias Observatory (LNA/Bra\-zil)
using a Coud\'e spectrograph and the 600 l/mm diffraction
grating, resulting in a spectral resolution at 6500 {\AA}
of 0.75 {\AA} and 0.90 {\AA} (instrumental FWHM) when CCD 48 and
CCDs 101/106 were used, respectively. These correspond to
$\sigma_{\scriptsize\mbox{inst}}=14.7$ and 17.6 km s$^{-1}$,
respectively. We observed in the region 6400-6900 {\AA} to obtain
line width measurements of the H$\alpha$ line emission. The slit
used was 1{\arcsec} for all observations. We obtained data from five
observational runs between February 1997 and March 1999.

Coud\'e data were reduced in a standard procedure for slit
spectra in CCD using IRAF\footnote{IRAF is distributed by the
National Optical Astronomy Observatories, which are operated by the
Association of Universities for Research in Astronomy, Inc., under
cooperative agreement with the National Science Foundation.}. We
used the CCDRED package for Bias and Flat Field reduction and the
SPECRED package for extraction and calibration procedures. We
extracted the spectra from the brightest knot of the galaxies along
the slit. An example of a calibrated Coud\'e spectrum is shown in
Figure~\ref{fig3}. The ASCII data files of all line profiles
analyzed in this work are available at
http://www.on.br/astro/etelles/lsi\-gma.

Table~\ref{tab1} lists all observations including FEROS ones and
observations with the spectrograph.  Columns 1, 2 and 3 give the
galaxy name, and their coordinates (J2000). Columns 4, 5, 6 and 7
show the observation log for the Coud\'e spectrograph and Columns 8,
9 and 10 show the observation log for the FEROS spectrograph,
describing the total exposure times, number of exposures,
the detector used (only for Coud\'e) and date of observations,
respectively. The last column 11 shows alternative names for the
galaxy.

\begin{deluxetable}{lcccccccccl}
\rotate
\tabletypesize{\scriptsize}
\tablecaption{Journal of observations.\label{tab1}}
\tablecolumns{11}
\tablewidth{0pt}
\tablehead{
\colhead{} & \colhead{} & \colhead{} &\multicolumn{4}{c}{Coud\'e} &\multicolumn{3}{c}{FEROS}&\\
\colhead{Galaxy}&\colhead{$\alpha$(2000)} &\colhead{$\delta$(2000)} &\colhead{Total Exp.} &\colhead{Number}&\colhead{Detector}&\colhead{Obs.} &
\colhead{Total Exp.} &\colhead{Number}  &\colhead{Obs.} &\colhead{Other}\\
\colhead{}  & \colhead{}&   \colhead{} &\colhead{Time ($s$)} &\colhead{of Exp.}     &\colhead{(CCD)}    &\colhead{Date} &\colhead{Time ($s$)} &
\colhead{of Exp.} &\colhead{Date} &\colhead{Name}
}
\startdata
\object{UM 238}          &00h24m42.3s &+01d44m02s        &           &            &               &           &3600 &1    &21/07/2001 &\\
\object{MBG 00463-0239}  &00h48m53.2s &-02d22m55s        &3600       &3           &106            &13/09/1998 &     &     &           &MRK 557\\
\object{UM 304}          &01h06m54.0s &+01d56m44s        &5400       &3           &48             &28/07/1997 &5400 &2    &23/11/2000 &\\
\object{Tol 0104-388}    &01h07m02.1s &-38d31m52s        &           &            &               &           &5400 &1    &10/01/2002 &CTS 1001\\
\object{UM 306}          &01h10m35.0s &+02d06m51s        &           &            &               &           &5400 &2    &23/11/2000 &\\
\object{UM 307}          &01h11m30.7s &+01d19m16s        &1800       &1           &48             &29/07/1997 &     &     &           &\\
\object{UM 323}          &01h26m46.6s &-00d38m46s        &           &            &               &           &5400 &2    &21/11/2000 &\\
\object{Tol 0127-397}    &01h29m15.8s &-39d30m38s        &1200       &1           &106            &17/09/1998 &5400 &1    &20/11/2000 &\\
\object{Tol 0140-420}    &01h43m03.1s &-41d49m41s        &           &            &               &           &5400 &2    &24/11/2000 &\\
\object{UM 137}          &01h46m23.9s &+04d16m11s        &           &            &               &           &5400 &1    &21/07/2001 &\\
\object{UM 151}          &01h57m38.8s &+02d25m24s        &           &            &               &           &3600 &1    &21/07/2001 &MRK 1169\\
\object{UM 382}          &01h58m09.3s &-00d06m38s        &           &            &               &           &3600 &1    &22/07/2001 &\\
\object{MBG 01578-6806}  &01h59m06.0s &-67d52m13s        &1200       &1           &106            &12/09/1998 &     &     &           &NGC 802\\
\object{UM 391}          &02h03m30.4s &+02d33m59s        &6000       &5           &106            &14/09/1998 &5400 &2    &22/11/2000 &MRK 585\\
                         &            &                  &2400       &2           &106            &15/09/1998\\
\object{UM 395}          &02h06m56.8s &+01d41m52s        &           &            &               &           &5400 &2    &24/11/2000 &\\
\object{UM 396}          &02h07m26.5s &+02d56m55s        &           &            &               &           &5400 &2    &22/11/2000 &\\
\object{UM 408}          &02h11m23.4s &+02d20m30s        &           &            &               &           &5400 &2    &22/11/2000 &\\
\object{UM 417}          &02h19m30.2s &-00d59m11s        &           &            &               &           &4200 &1    &22/07/2001 &\\
\object{Tol 0226-390}    &02h28m12.3s &-38d49m20s        &2400       &2           &106            &14/09/1998 &7200 &2    &20/11/2000 &\\
\object{CTS 1003}        &02h32m43.7s &-39d34m27s        &           &            &               &           &5400 &1    &09/01/2002 &Tol 0230-397\\
\object{MBG 02411-1457}  &02h43m29.2s &-14d45m16s        &4800       &4           &106            &12/09/1998 &     &     &           &NGC 1076\\
\object{Tol 0242-387}    &02h44m37.9s &-38d34m54s        &           &            &               &           &6300 &1    &23/07/2001 &\\
\object{CTS 1004}        &03h08m43.3s &-40d24m28s        &           &            &               &           &5400 &1    &09/01/2002 &Tol 0306-405\\
                         &            &                  &           &            &               &           &4800 &2    &23/11/2000\\
\object{CTS 1005}        &03h59m08.9s &-39d06m25s        &           &            &               &           &3600 &1    &07/01/2002 &Cam 0357-3915\\
\object{Tol 0440-381}    &04h42m08.1s &-38d01m11s        &2400       &2           &106            &17/09/1998 &3600 &2    &20/11/2000 &\\
\object{CTS 1006}        &04h42m09.5s &-45d25m12s        &           &            &               &           &3600 &1    &10/01/2002 &\\
\object{CTS 1007}        &04h46m49.4s &-30d08m58s        &           &            &               &           &6000 &2    &22/11/2000 &\\
\object{CTS 1008}        &04h51m39.6s &-31d53m06s        &           &            &               &           &7200 &2    &21/11/2000 &\\
\object{Tol 0505-387}    &05h07m00.8s &-38d38m58s        &           &            &               &           &4800 &2    &23/11/2000\\
\object{Tol 0510-400}    &05h11m56.3s &-39d59m47s        &           &            &               &           &5400 &1    &09/01/2002\\
\object{Tol 0528-383}    &05h29m57.4s &-38d18m07s        &           &            &               &           &3600 &2    &20/11/2000\\
\object{II ZW 40}        &05h55m42.6s &+03d23m32s        &3600       &3           &101            &05/02/1997 &4200 &1    &21/11/2000\\
                         &            &                  &           &            &               &           &900  &1    &28/03/2001\\
\object{Tol 0559-393}    &06h00m43.9s &-39d19m07s        &           &            &               &           &3600 &1    &23/11/2000\\
\object{Tol 0610-387}    &06h12m14.2s &-38d46m23s        &           &            &               &           &5400 &2    &24/11/2000\\
\object{Tol 0614-375}    &06h16m13.8s &-37d36m37s        &           &            &               &           &3600 &1    &16/04/2007\\
\object{Tol 0633-415}    &06h35m10.2s &-41d33m42s        &           &            &               &           &4800 &2    &22/11/2000\\
                         &            &                  &           &            &               &           &3600 &1    &14/04/2007\\
\object{Tol 0645-376}    &06h46m50.1s &-37d43m22s        &1800       &1           &101            &04/02/1997 &2700 &1    &20/11/2000\\
\object{MRK 1201}        &07h25m45.7s &+29d57m10s        &           &            &               &           &3600 &1    &31/03/2001\\
\object{Cam 0840+1201}   &08h42m20.9s &+11d50m00s        &           &            &               &           &6000 &2    &24/11/2000\\
\object{Cam 0840+1044}   &08h42m36.6s &+10d33m14s        &           &            &               &           &2700 &1    &29/03/2001\\
\object{Cam 08-28A}      &08h45m33.5s &+16d05m46s        &           &            &               &           &1800 &1    &01/04/2001 &MRK 702\\
\object{MRK 710}         &09h54m49.5s &+09d16m16s        &5400       &3           &101            &05/02/1997 &1800 &1    &01/04/2001 &NGC 3049\\
\object{MRK 711}         &09h55m11.3s &+13d25m46s        &           &            &               &           &5400 &1    &07/01/2002\\
\object{Tol 0957-278}    &09h59m21.2s &-28d08m00s        &7200       &4           &101            &06/02/1997 &1800 &1    &28/03/2001 &Tol 2\\
\object{Tol 1004-296NW}  &10h06m33.1s &-29d56m09s        &1200       &2           &101            &03/02/1998 &     &     &\\
\object{Tol 1004-296SE}  &10h06m33.1s &-29d56m09s        &600        &1           &101            &03/02/1998 &     &     &\\
\object{Tol 1008-286}    &10h10m18.1s &-28d57m48s        &           &            &               &           &3600 &1    &14/04/2007 &Tol 4\\
\object{CTS 1011}        &10h19m21.2s &-22d08m35s        &           &            &               &           &2700 &1    &01/04/2001\\
\object{CTS 1012}        &10h21m21.0s &-21d36m27s        &           &            &               &           &3600 &1    &14/04/2007\\
\object{CTS 1013}        &10h25m05.9s &-19d46m57s        &           &            &               &           &2700 &1    &28/03/2001\\
\object{Tol 1025-285}    &10h27m25.5s &-28d47m33s        &           &            &               &           &5400 &2    &29/03/2001 &Tol 6\\
\object{Haro 24}         &10h27m55.4s &+19d29m26s        &           &            &               &           &3600 &1    &13/04/2007 &II Zw 47\\
\object{CTS 1014}        &10h35m05.4s &-27d20m08s        &           &            &               &           &4500 &1    &14/04/2007 &Tol 1032-2704\\
\object{CTS 1016}        &10h37m30.6s &-24d08m41s        &           &            &               &           &3600 &1    &31/03/2001\\
\object{CTS 1017}        &10h37m40.4s &-25d58m00s        &           &            &               &           &5400 &1    &09/01/2002\\
\object{CTS 1018}        &10h38m06.5s &-26d21m56s        &           &            &               &           &3600 &1    &31/03/2001\\
\object{CTS 1019}        &10h41m03.7s &-22d34m24s        &           &            &               &           &1800 &1    &01/04/2001\\
\object{CTS 1020}        &10h47m44.3s &-20d57m49s        &           &            &               &           &2700 &1    &01/04/2001\\
\object{CTS 1022}        &10h48m40.2s &-19d26m57s        &           &            &               &           &3600 &1    &29/03/2001\\
\object{$[$F80$]$ 30}    &10h56m09.1s &+06d10m22s        &           &            &               &           &1800 &1    &31/03/2001 &MRK 1271,Tol 1053+064\\
\object{MRK 36}          &11h04m58.3s &+29d08m23s        &           &            &               &           &1800 &1    &31/03/2001 &Haro 4\\
\object{UM 439}          &11h36m36.8s &+00d48m58s        &3000       &3           &101            &04/02/1998 &3600 &1    &13/04/2007\\
\object{UM 448}          &11h42m12.4s &+00d20m03s        &4800       &4           &101            &06/02/1997 &3600 &1    &13/04/2007 &MRK 1304\\
\object{Tol 1147-283}    &11h50m03.2s &-28d40m17s        &           &            &               &           &3600 &1    &28/03/2001 &Tol 17\\
\object{UM 455}          &11h50m23.8s &-00d31m41s        &           &            &               &           &2700 &1    &31/03/2001\\
\object{UM 456}          &11h50m36.3s &-00d34m03s        &           &            &               &           &3600 &1    &13/04/2007\\
\object{UM 461}          &11h51m33.3s &-02d22m22s        &           &            &               &           &1800 &1    &31/03/2001\\
\object{UM 463}          &11h52m47.5s &-00d40m08s        &           &            &               &           &3600 &1    &01/04/2001\\
\object{CTS 1026}        &12h05m59.3s &-27d00m56s        &           &            &               &           &3600 &1    &14/04/2007\\
\object{UM 477}          &12h08m11.1s &+02d52m42s        &2400       &3           &101            &03/02/1998 &3600 &1    &10/01/2002 &MRK 1466,NGC 4123\\
\object{UM 483}          &12h12m14.7s &+00d04m20s        &           &            &               &           &5400 &1    &22/07/2001 &MRK 1313\\
\object{CTS 1027}        &12h15m18.3s &+05d45m40s        &           &            &               &           &3600 &1    &16/04/2007 &Haro 6\\
\object{MRK 1318}        &12h19m09.9s &+03d51m21s        &1800       &2           &106            &14/03/1999 &3600 &1    &29/03/2001 &Haro 8\\
\object{CTS 1028}        &12h23m16.6s &+04d50m09s        &           &            &               &           &3600 &1    &01/04/2001 &Tol 1220+051,[F80] 34\\
\object{UM 499}          &12h25m42.8s &+00d34m21s        &2400       &2           &101            &04/02/1997 &     &     &\\
                         &            &                  &1200       &1           &101            &05/02/1997\\
\object{Tol 1223-359}    &12h25m46.9s &-36d14m01s        &           &            &               &           &3600 &1    &28/03/2001 &Tol 65\\
\object{Haro 30}         &12h37m41.1s &+27d07m46s        &           &            &               &           &3600 &1    &13/04/2007 &MRK 650,IC 3600\\
\object{[SC98] 01}       &13h04m15.2s &-22d52m53s        &           &            &               &           &3600 &1    &29/03/2001\\
\object{CTS 1029}        &13h06m05.1s &-22d37m22s        &           &            &               &           &6300 &2    &30/03/2001 &[SC98] 09\\
\object{[SC98] 11}       &13h06m19.3s &-22d58m49s        &           &            &               &           &4500 &1    &16/04/2007\\
\object{UM 559}          &13h17m42.8s &-01d00m01s        &           &            &               &           &3600 &1    &28/03/2001\\
\object{[SC98] 68}       &13h21m50.0s &-22d28m31s        &           &            &               &           &3600 &1    &31/03/2001\\
\object{UM 570}          &13h23m47.4s &-01d32m52s        &           &            &               &           &4500 &1    &22/07/2001\\
\object{[SC98] 88}       &13h25m33.0s &-26d02m50s        &           &            &               &           &4500 &1    &29/03/2001\\
\object{CTS 1030}        &13h25m33.3s &-25d55m33s        &           &            &               &           &3600 &1    &14/04/2007 &[SC98] 84\\
\object{POX 186}         &13h25m48.6s &-11d36m38s        &           &            &               &           &4500 &1    &01/04/2001\\
                         &            &                  &           &            &               &           &4500 &1    &13/04/2007\\
\object{CTS 1031}        &13h25m58.5s &-23d38m09s        &           &            &               &           &4500 &1    &16/04/2007 &[SC98] 91\\
\object{Tol 1345-420}    &13h48m22.2s &-42d21m15s        &600        &1           &106            &14/03/1999 &3600 &1    &30/03/2001 &Tol 111\\
\object{CTS 1033}        &13h49m44.8s &-18d11m28s        &           &            &               &           &3600 &1    &13/04/2007\\
\object{Tol 1400-397}    &14h03m05.7s &-40d02m28s        &           &            &               &           &7200 &1    &23/07/2001 &Tol 115\\
\object{UM 649}          &14h14m27.7s &-00d28m08s        &           &            &               &           &5400 &1    &21/07/2001\\
\object{CTS 1034}        &14h19m32.4s &-27d35m08s        &           &            &               &           &5400 &2    &28/03/2001\\
\object{II ZW 70}        &14h50m56.5s &+35d34m18s        &           &            &               &           &1400 &1    &28/03/2001 &MRK 829\\
\object{CTS 1035}        &14h57m19.7s &-22d23m35s        &           &            &               &           &3600 &1    &30/03/2001\\
                         &            &                  &           &            &               &           &3600 &1    &31/03/2001\\
\object{CTS 1037}        &15h15m44.0s &-18d18m52s        &           &            &               &           &4300 &1    &13/04/2007\\
\object{Cam 1543+0907}   &15h45m38.6s &+09d03m28s        &           &            &               &           &3600 &1    &31/03/2001\\
\object{Tol 1924-416}    &19h27m58.2s &-41d34m32s        &3600       &2           &48             &28/07/1997 &     &     &\\
\object{Tol 1939-419}    &19h33m32.0s &-41d50m56s        &           &            &               &           &3100 &1    &16/04/2007\\
\object{Tol 1937-423}    &19h40m58.6s &-42d15m45s        &2400       &2           &106            &16/09/1998 &5400 &1    &21/07/2001\\
\object{CTS 1038}        &19h54m52.6s &-32d56m40s        &           &            &               &           &4500 &1    &01/04/2001\\
\object{CTS 1039}        &20h05m51.3s &-45d28m42s        &           &            &               &           &3600 &1    &16/04/2007\\
\object{Tol 2010-382}    &20h14m06.4s &-38d07m41s        &5400       &3           &48             &29/07/1997 &3600 &1    &14/04/2007\\
\object{Tol 2019-405}    &20h23m06.2s &-40d20m33s        &           &            &               &           &5400 &1    &21/07/2001\\
\object{Tol 2041-394}    &20h44m50.8s &-39d13m17s        &           &            &               &           &5400 &1    &22/07/2001\\
\object{NGC 6970}        &20h52m09.4s &-48d46m40s        &2400       &2           &48             &28/07/1997 &     &     &\\
\object{MBG 20533-4410}  &20h56m43.4s &-43d59m10s        &4800       &4           &48             &28/07/1997 &     &     &  &NGC 6983\\
\object{Tol 2122-408}    &21h25m46.9s &-40d39m12s        &4800       &4           &106            &16/09/1998 &3600 &1    &21/11/2000\\
\object{Tol 2138-405}    &21h41m21.8s &-40d19m06s        &           &            &               &           &5400 &1    &23/07/2001\\
\object{Tol 2138-397}    &21h41m38.4s &-39d31m30s        &           &            &               &           &1800 &1    &16/04/2007\\
\object{Tol 2146-391}    &21h49m48.2s &-38d54m09s        &           &            &               &           &5400 &1    &21/07/2001\\
\object{MBG 21567-1645}  &21h59m26.1s &-16d30m44s        &5400       &3           &48             &29/07/1997 &     &     &  &NGC 7165\\
\object{MBG 22012-1550}  &22h03m56.3s &-15d36m00s        &5400       &3           &48             &29/07/1997 &     &     &\\
\object{IC 5154}         &22h04m30.3s &-66d06m45s        &2400       &2           &106            &12/09/1998 &     &     &\\
\object{ESO 533-G 014}   &22h19m50.6s &-26d20m30s        &3600       &3           &106            &12/09/1998 &     &     &\\
\object{MCG -01-57-017}  &22h38m13.5s &-07d02m05s        &2400       &2           &106            &14/09/1998 &     &     &\\
\object{Tol 2240-384}    &22h43m32.4s &-38d11m24s        &           &            &               &           &5400 &1    &22/07/2001\\
                         &            &                  &           &            &               &           &3600 &1    &23/07/2001\\
\object{MBG 23121-3807}  &23h14m52.3s &-37d51m20s        &8700       &8           &106            &11/09/1998 &     &     &\\
\object{Tol 2326-405}    &23h28m49.4s &-40d15m26s        &           &            &               &           &4500 &1    &23/07/2001\\
\object{UM 167}          &23h36m14.1s &+02d09m19s        &3600       &2           &48             &29/07/1997 &     &     &  &MRK 538, NGC 7714\\
\object{UM 191}          &23h56m59.6s &-02d05m02s        &5400       &3           &48             &29/07/1997 &7200 &2    &21/11/2000 &MRK 542\\
                         &               &               &2400       &2           &106            &13/09/1998\\
                         &               &               &4800       &4           &106            &14/09/1998
\enddata
\end{deluxetable}

\subsection{Spectrophotometry}\label{sec2_3}

Most of the spectrophotometric data used in this work comes from KTC
with 91 objects in common with our FEROS plus Coud\'e sample. Their
data were obtained from a Boller \& Chivens spectrograph on the
1.52m ESO telescope. The spectra cover 4000{\AA} in the optical
range centered at 5700{\AA}. They have a spectral resolution of
5{\AA} and the entrance slit was 2{\arcsec}. KTC observed
with a long slit and in some cases they extracted more than one
spectrum for a galaxy, representing different bright regions
spatially separated. Additional and complementary data were obtained
from recent works and the references will be cited below. Emission
line fluxes and equivalent widths of permitted and forbidden line
fluxes were gathered from these sources to derive the physical
parameters, such as extinction coefficient, ionization ratio,
electron temperature and density and oxygen abundance.

\begin{deluxetable}{lcccccccc}
\tabletypesize{\scriptsize}
\tablecaption{Measurements of redshifts, distances and line widths from
FEROS and Coud\'e spectra. The last column indicates the class of
the galaxies discussed in the text.\label{tab2}}
\tablecolumns{9}
\tablewidth{0pt}
\tablehead{
Galaxy &$z_{\scriptsize\mbox{hel}}$&$D$&FWHM&FWHM&FWHM&FWHM&FWHM&Class\\
              & &Mpc&H$\beta$&[OIII] 4959\AA&[OIII] 5007\AA&H$\alpha$&H$\alpha$ (Coud\'e)& 
}
\startdata
UM 238            &0.01427  & 55.3  &0.844  &0.775  &0.785  &1.158  &-      &G\\
MBG 00463-0239    &0.01328  & 51.4  &-      &-      &-      &-      &3.200  &I\\
UM 304            &0.01570  & 61.7  &3.063  &3.556  &3.444  &4.140  &4.107  &C\\
Tol 0104-388      &0.02263  & 92.4  &1.914  &1.880  &1.988  &2.618  &-      &I\\
UM 306            &0.01649  & 65.1  &0.846  &0.703  &0.736  &1.147  &-      &G\\
UM 307            &0.02249  & 90.4  &-      &-      &-      &-      &2.772  &G\\
UM 323            &0.00648  & 23.0  &0.873  &0.853  &0.786  &1.143  &-      &G$^{\prime}$\\
Tol 0127-397      &0.01735  & 70.3  &1.510  &1.420  &1.440  &1.950  &2.045  &G$^{\prime}$\\
Tol 0140-420      &0.02205  & 89.9  &1.193  &1.092  &1.104  &1.510  &-      &C\\
UM 137            &0.00591  & 20.8  &-      &-      &0.676  &0.962  &-      &G\\
UM 151            &0.01607  & 63.9  &1.196  &0.892  &1.066  &1.760  &-      &G\\
UM 382            &0.01206  & 47.0  &0.812  &0.772  &0.742  &1.028  &-      &G\\
MBG 01578-6806    &0.00490  & 19.8  &-      &-      &-      &-      &1.669  &G\\
UM 391            &0.02101  & 84.9  &2.291  &-      &3.188  &3.168  &3.320  &C\\
UM 395            &0.02234  & 90.6  &1.255  &1.281  &1.206  &1.711  &-      &G$^{\prime}$\\
UM 396            &0.02078  & 84.0  &1.134  &1.099  &1.110  &1.517  &-      &G$^{\prime}$\\
UM 408            &0.01153  & 45.0  &0.867  &0.661  &0.681  &1.165  &-      &G$^{\prime}$\\
UM 417            &0.00872  & 33.3  &0.639  &0.544  &0.606  &0.994  &-      &G$^{\prime}$\\
Tol 0226-390      &0.04771  &199.2  &3.620  &3.410  &3.346  &-      &4.219  &I\\
CTS 1003          &0.01684  & 68.9  &1.055  &1.005  &1.000  &1.452  &-      &G\\
MBG 02411-1457    &0.00686  & 26.1  &-      &-      &-      &-      &2.047  &G\\
Tol 0242-387      &0.12635  &531.5  &4.252  &4.966  &4.949  &5.664  &-      &I\\
CTS 1004          &0.04734  &198.3  &1.804  &1.780  &1.831  &-      &-      &I\\
CTS 1005          &0.07441  &313.3  &2.082  &1.963  &1.964  &2.772  &-      &I\\
Tol 0440-381      &0.04082  &172.2  &1.440  &1.308  &1.310  &1.967  &2.776  &C\\
CTS 1006          &0.02072  & 87.5  &1.492  &1.341  &1.360  &2.032  &-      &G$^{\prime}$\\
CTS 1007          &0.04130  &174.2  &1.255  &1.123  &1.108  &1.648  &-      &G$^{\prime}$\\
CTS 1008          &0.06106  &257.7  &2.030  &1.891  &1.910  &2.693  &-      &G$^{\prime}$\\
Tol 0505-387      &0.02897  &122.6  &0.967  &0.812  &0.906  &1.269  &-      &G$^{\prime}$\\
Tol 0510-400      &0.04132  &174.9  &1.425  &1.282  &1.281  &1.795  &-      &G$^{\prime}$\\
Tol 0528-383      &0.01163  & 49.8  &0.826  &0.760  &0.794  &1.188  &-      &G$^{\prime}$\\
II ZW 40          &0.00258  & 11.8  &1.329  &1.277  &1.300  &1.826  &1.951  &I\\
Tol 0559-393      &0.04478  &190.3  &2.073  &1.713  &1.929  &2.677  &-      &G\\
Tol 0610-387      &0.00575  & 25.7  &0.771  &-      &0.929  &1.199  &-      &G\\
Tol 0614-375      &0.03157  &134.8  &1.820  &2.045  &2.036  &2.646  &-      &G$^{\prime}$\\
Tol 0633-415      &0.01640  & 71.1  &1.310  &1.223  &1.251  &1.705  &-      &G$^{\prime}$\\
Tol 0645-376      &0.02579  &110.9  &1.229  &1.228  &1.198  &1.708  &1.670  &G$^{\prime}$\\
MRK 1201          &0.01857  & 80.6  &1.843  &-      &-      &2.491  &-      &G\\
Cam 0840+1201     &0.02938  &128.2  &1.510  &1.397  &1.400  &1.945  &-      &G$^{\prime}$\\
Cam 0840+1044     &0.01044  & 48.0  &0.752  &0.549  &0.573  &0.999  &-      &G$^{\prime}$\\
Cam 08-28A        &0.05304  &227.8  &2.090  &1.797  &1.856  &-      &-      &I\\
MRK 710           &0.00502  & 25.9  &2.201  &-      &2.127  &2.708  &2.697  &C\\
MRK 711           &0.01944  & 86.7  &3.685  &3.418  &3.499  &5.046  &-      &I\\
Tol 0957-278      &0.00334  & 18.8  &1.031  &1.051  &1.044  &1.503  &1.543  &I\\
Tol 1004-296NW    &0.00370  & 20.2  &-      &-      &-      &-      &1.943  &I\\
Tol 1004-296SE    &0.00359  & 19.8  &-      &-      &-      &-      &1.738  &G\\
Tol 1008-286      &0.01384  & 63.1  &1.054  &1.010  &1.008  &1.455  &-      &G$^{\prime}$\\
CTS 1011          &0.01207  & 55.9  &0.921  &0.823  &0.837  &1.241  &-      &G$^{\prime}$\\
CTS 1012          &0.01089  & 50.9  &0.787  &0.650  &0.646  &1.051  &-      &G$^{\prime}$\\
CTS 1013          &0.02688  &118.5  &1.390  &1.346  &1.380  &1.774  &-      &C\\
Tol 1025-285      &0.03073  &134.5  &2.356  &-      &2.438  &2.986  &-      &G\\
Haro 24           &0.04327  &187.3  &1.770  &1.587  &1.759  &2.565  &-      &I\\
CTS 1014          &0.05895  &253.8  &2.333  &1.778  &1.849  &2.803  &-      &I\\
CTS 1016          &0.03450  &150.6  &1.456  &1.323  &1.410  &2.203  &-      &I\\
CTS 1017          &0.03544  &154.5  &1.186  &1.176  &1.176  &1.598  &-      &G$^{\prime}$\\
CTS 1018          &0.03925  &170.6  &1.453  &1.313  &1.316  &1.977  &-      &G\\
CTS 1019          &0.06651  &285.8  &1.888  &1.895  &1.936  &2.761  &-      &G$^{\prime}$\\
CTS 1020          &0.01248  & 57.7  &1.393  &1.365  &1.399  &1.911  &-      &G$^{\prime}$\\
CTS 1022          &0.01369  & 62.9  &0.974  &0.839  &0.854  &1.427  &-      &G\\
$[$F80$]$ 30      &0.00335  & 18.2  &0.903  &0.741  &0.749  &1.205  &-      &G$^{\prime}$\\
MRK 36            &0.00212  & 13.2  &0.782  &0.687  &0.717  &1.051  &-      &G$^{\prime}$\\
UM 439            &0.00382  & 21.3  &0.805  &0.708  &0.711  &1.103  &1.444  &G$^{\prime}$\\
UM 448            &0.01834  & 82.6  &2.947  &3.043  &3.017  &4.143  &3.376  &C\\
Tol 1147-283      &0.00626  & 31.2  &0.793  &0.799  &0.724  &1.113  &-      &G$^{\prime}$\\
UM 455            &0.01306  & 60.3  &0.998  &0.746  &0.740  &1.518  &-      &I\\
UM 456            &0.00572  & 29.3  &0.757  &0.631  &0.643  &1.017  &-      &G$^{\prime}$\\
UM 461            &0.00352  & 20.0  &0.669  &0.512  &0.519  &0.912  &-      &G$^{\prime}$\\
UM 463            &0.00468  & 24.9  &0.849  &0.680  &0.676  &1.057  &-      &G$^{\prime}$\\
CTS 1026          &0.00577  & 29.2  &1.710  &1.689  &1.705  &2.284  &-      &G\\
UM 477            &0.00422  & 22.8  &2.243  &-      &2.860  &3.019  &3.038  &C\\
UM 483            &0.00792  & 38.4  &0.816  &0.776  &0.744  &1.105  &-      &I\\
CTS 1027          &0.00674  & 33.4  &0.892  &0.799  &0.821  &1.217  &-      &I\\
MRK 1318          &0.00504  & 26.2  &0.756  &0.680  &0.695  &1.052  &1.448  &G$^{\prime}$\\
CTS 1028          &0.01776  & 79.9  &1.141  &1.074  &1.098  &1.520  &-      &C\\
UM 499            &0.00707  & 34.8  &-      &-      &-      &-      &2.520  &I\\
Tol 1223-359      &0.00930  & 44.2  &0.828  &0.681  &0.707  &1.163  &-      &G$^{\prime}$\\
Haro 30           &0.01552  & 69.5  &1.746  &1.887  &1.845  &2.428  &-      &I\\
$[$SC98$]$ 01     &0.01041  & 48.4  &0.975  &0.791  &0.720  &1.185  &-      &G$^{\prime}$\\
CTS 1029          &0.03633  &157.8  &-      &-      &-      &1.787  &-      &G\\
$[$SC98$]$ 11     &0.03104  &135.5  &1.317  &1.285  &1.244  &1.801  &-      &I\\
UM 559            &0.00429  & 22.5  &0.794  &0.676  &0.687  &1.103  &-      &G$^{\prime}$\\
$[$SC98$]$ 68     &0.02377  &104.7  &1.274  &1.346  &1.300  &1.758  &-      &G\\
UM 570            &0.02249  & 99.3  &0.930  &0.884  &0.867  &1.291  &-      &G$^{\prime}$\\
$[$SC98$]$ 88     &0.01454  & 65.6  &1.066  &0.896  &0.860  &1.491  &-      &I\\
CTS 1030          &0.01505  & 67.7  &1.221  &1.213  &1.212  &1.668  &-      &C\\
POX 186           &0.00415  & 21.9  &0.715  &0.584  &0.586  &0.984  &-      &G$^{\prime}$\\
CTS 1031          &0.04525  &195.3  &1.338  &1.296  &1.386  &1.770  &-      &I\\
Tol 1345-420      &0.00807  & 37.5  &0.877  &0.743  &0.791  &1.188  &1.655  &G$^{\prime}$\\
CTS 1033          &0.01549  & 69.4  &1.921  &1.935  &1.939  &2.602  &-      &C\\
Tol 1400-397      &0.03101  &134.3  &1.352  &1.299  &1.368  &1.922  &-      &G$^{\prime}$\\
UM 649            &0.02611  &113.9  &1.152  &0.935  &1.018  &1.480  &-      &G\\
CTS 1034          &0.02292  &100.2  &-      &1.091  &1.037  &1.533  &-      &G\\
II ZW 70          &0.00406  & 19.2  &0.901  &0.794  &0.823  &1.377  &-      &G$^{\prime}$\\
CTS 1035          &0.02848  &123.2  &1.164  &1.120  &1.086  &1.409  &-      &G\\
CTS 1037          &0.02130  & 92.5  &1.562  &1.530  &1.513  &2.074  &-      &I\\
Cam 1543+0907     &0.03766  &160.8  &1.293  &1.186  &1.179  &1.716  &-      &G$^{\prime}$\\
Tol 1924-416      &0.00952  & 38.4  &-      &-      &-      &-      &2.065  &I\\
Tol 1939-419      &0.02525  &104.8  &-      &-      &0.962  &1.336  &-      &G\\
Tol 1937-423      &0.00932  & 37.4  &1.022  &0.768  &0.888  &1.262  &1.874  &G\\
CTS 1038          &0.04984  &208.0  &-      &2.123  &2.136  &-      &-      &I\\
CTS 1039          &0.04486  &187.4  &1.693  &1.704  &1.722  &2.287  &-      &I\\
Tol 2010-382      &0.02026  & 83.0  &1.506  &-      &1.480  &1.899  &2.166  &G$^{\prime}$\\
Tol 2019-405      &0.01495  & 60.6  &1.095  &0.978  &0.996  &1.300  &-      &I\\
Tol 2041-394      &0.02576  &106.0  &-      &1.178  &1.121  &1.627  &-      &G\\
NGC 6970          &0.01751  & 71.6  &-      &-      &-      &-      &2.594  &C\\
MBG 20533-4410    &0.01714  & 69.7  &-      &-      &-      &-      &3.075  &C\\
Tol 2122-408      &0.01480  & 59.4  &1.114  &1.052  &1.071  &1.416  &1.717  &G\\
Tol 2138-405      &0.05802  &241.7  &2.445  &2.468  &2.496  &3.444  &-      &C\\
Tol 2138-397      &0.01570  & 63.0  &0.996  &0.998  &0.900  &1.391  &-      &G$^{\prime}$\\
Tol 2146-391      &0.02953  &121.3  &1.154  &0.935  &0.991  &1.503  &-      &I\\
MBG 21567-1645    &0.01738  & 68.8  &-      &-      &-      &-      &5.513  &C\\
MBG 22012-1550    &0.04227  &173.9  &-      &-      &-      &-      &5.074  &C\\
IC 5154           &0.01068  & 43.7  &-      &-      &-      &-      &2.035  &G\\
ESO 533-G 014     &0.00873  & 32.6  &-      &-      &-      &-      &1.317  &G\\
MCG -01-57-017    &0.00962  & 35.6  &-      &-      &-      &-      &1.567  &G\\
Tol 2240-384      &0.07584  &316.6  &2.082  &2.112  &2.196  &2.940  &-      &C\\
MBG 23121-3807    &0.00945  & 36.2  &-      &-      &-      &-      &1.777  &G\\
Tol 2326-405      &0.05515  &229.3  &-      &-      &1.556  &2.456  &-      &I\\
UM 167            &0.00928  & 34.0  &-      &-      &-      &-      &3.925  &G\\
UM 191            &0.02427  & 97.4  &1.265  &-      &1.584  &1.798  &2.000  &G$^{\prime}$
\enddata
\end{deluxetable}

\section{Results}\label{sec3}

\subsection{Line Widths and Velocity Dispersions}\label{sec3_1}

We have measured the emission line widths from our high spectral
resolution observations by fitting single Gaussians to the observed
line profiles using the SPLOT routine of IRAF.
The H$\alpha$ line for the Coud\'e spectra were measured
and H$\alpha$, H$\beta$ and [OIII] $\lambda\lambda$4959,5007 lines
for FEROS spectra were detected and measured for almost all
galaxies. All observed FWHM, uncorrected for instrumental width, are
presented in Table~\ref{tab2}.

In several cases a single Gaussian fit did not adequately
represent the observed profile. Some present irregularities such as
prominent wings and multiple components. Different methodologies to
obtain line widths such as multiple Gaussians or Gauss-Hermite
\citep[][and references therein]{rif10} fits are viable approaches
to the problem of modeling real emission line profiles, and were
tested.
For instance, Gauss-Hermite fits provide single width measurements
that are well compared with single Gaussian fit measurements, and
they are less sensitive to small asymmetries. In the case of
profiles with a dominant broad component or with double
peaks, multiple Gaussians or Gauss-Hermite methodologies provide
better fits. However, the interpretation of the model parameters is
not obvious.
Further detailed analysis with more data is needed to resolve this
issue, but it is beyond the scope of this paper. Here, we
are interested in the simplest methodology to measure the line
widths that may provide us with a robust kinematic measurement of
the starburst region as a whole.

As a simple alternative to deal with this problem, we classified
galaxies depending on their line profiles using the following
criterion:

\begin{itemize}

\item{Gaussian Profile - Symmetrical lines well represented by a
single Gaussian fit. These profiles occur in 62\% of galaxies in our
sample.}

\item{Irregular Profile - Asymmetrical lines showing prominent
wings and generally peaked. These occur in 29\% of our sample.}

\item{Profile with Components - These clearly show more than one
component in emission, normally double-peak lines with similar
intensities, occurring in 17\% of our sample.}

\end{itemize}

\begin{figure*}[ht]
\epsscale{0.8} \plotone{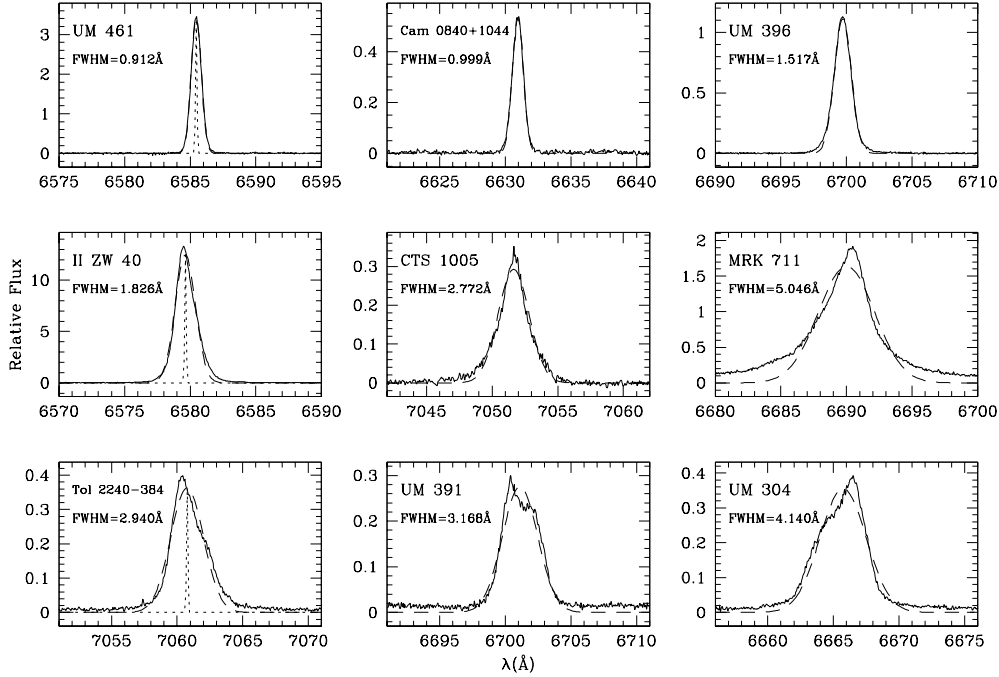} \caption{Examples of the three
classes of line profiles visually identified. (\textit{first row})
Galaxies with nearly Gaussian profiles; (\textit{second row})
galaxies with irregular profiles; (\textit{third row}) galaxies with
line profiles clearly showing components. The dotted narrow
lines in all of the left-hand boxes represent the FEROS
instrumental profile, $\sigma_{\scriptsize\mbox{inst}}=2.50$ km
s$^{-1}$. The dashed lines represent single Gaussian fits to the
observed profiles.\label{fig4}}
\end{figure*}

Our classification was done by eye comparison between the single
Gaussian fit and the observed line profile, therefore it has an
intrinsic subjectiveness even though interesting for early and
qualitative purposes in this work. Since we have for most galaxies
the four strong emission lines, we checked them all to classify the
galaxy. Figure~\ref{fig4} presents some examples of prototypical
H$\alpha$ profiles of the three classes defined. All galaxies were
classified including those showing low signal-to-noise (S/N) in
their emission line profiles.
Some galaxies may have been classified as presenting Gaussian
profiles simply because the line wings were not well
sampled. The spectra from FEROS were used in priority to Coud\'e
ones to classify those galaxies observed with both instruments. The
galaxies were assigned as letter G for Gaussian and I for irregular
profiles, and C for profiles with components. The
respective class for each galaxy is shown in the last
column of Table~\ref{tab2}. We will return to this point later in
Section~\ref{sec4_3} using a semi-quantitative analysis of line
profile classifications to select a more homogeneous
sample.

\begin{figure*}[ht]
\epsscale{0.32}
\plotone{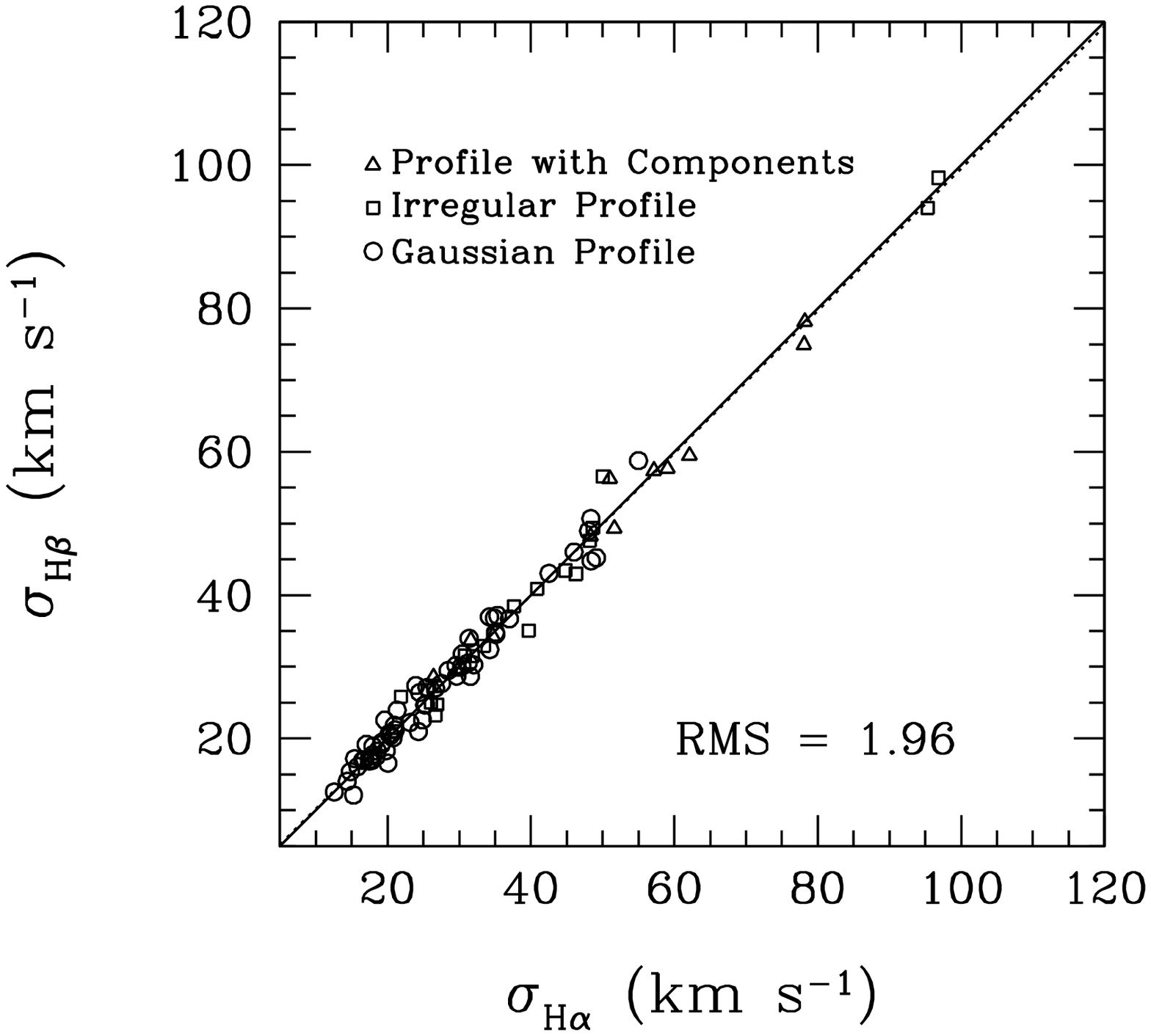}\plotone{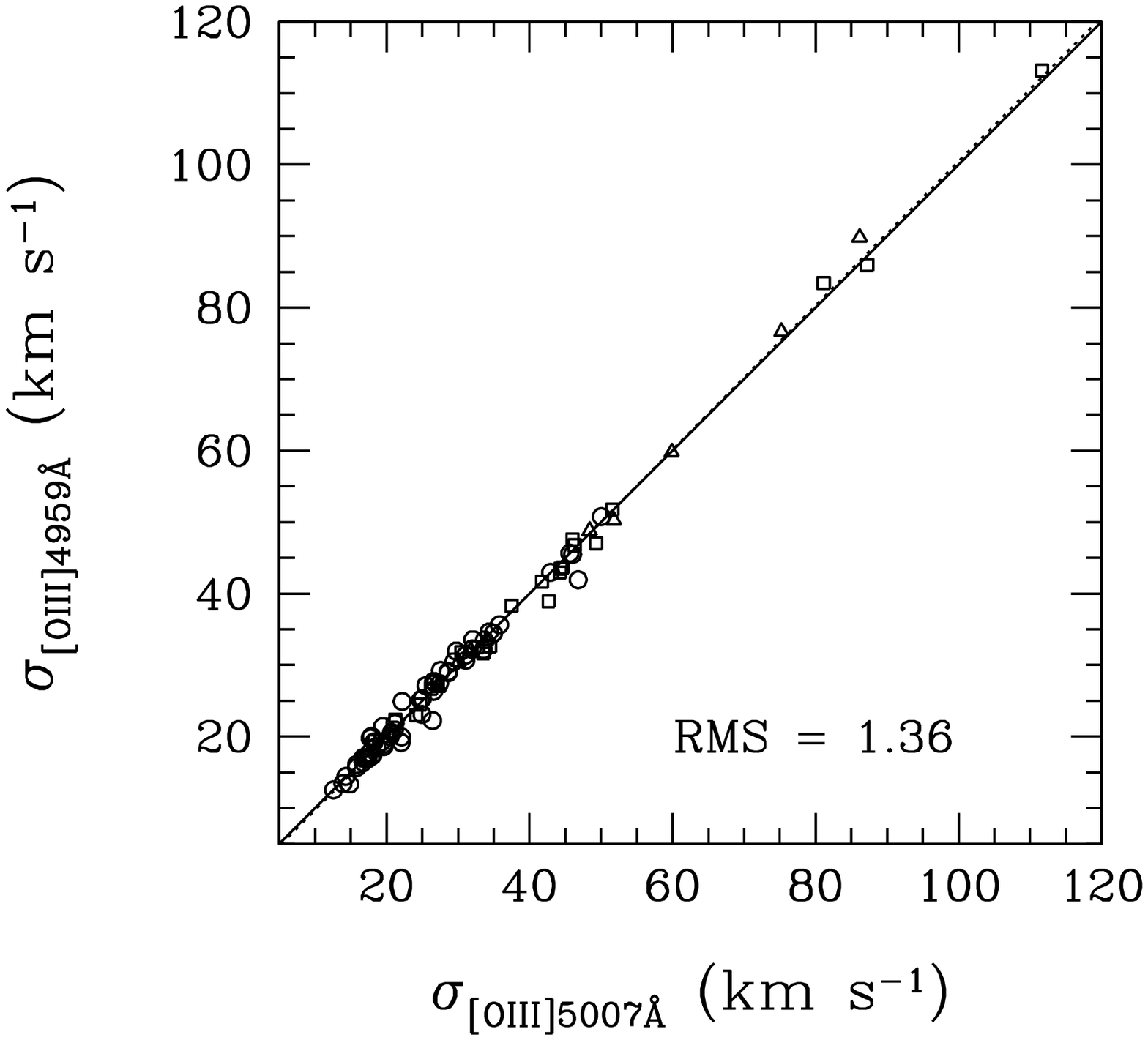}\plotone{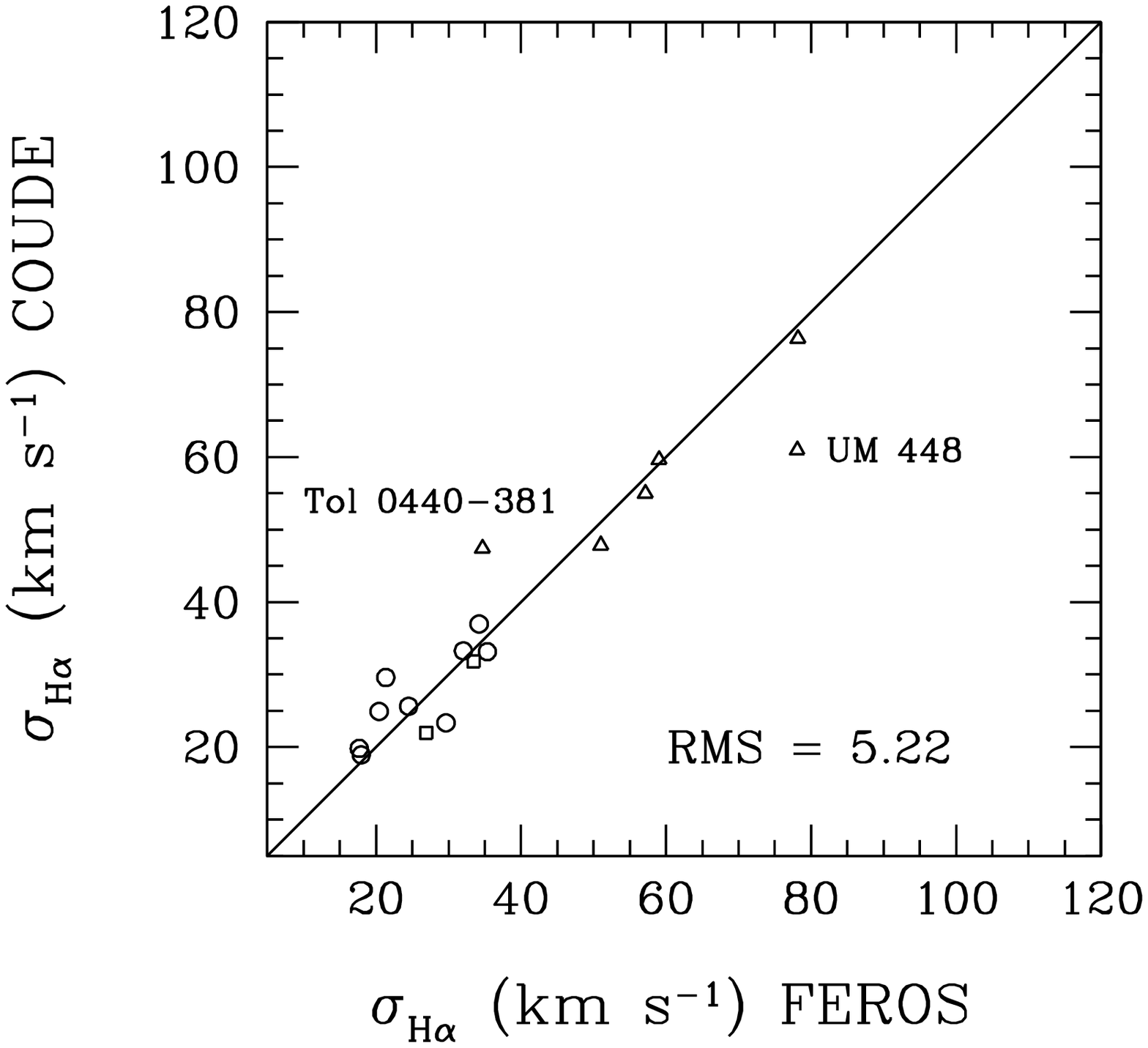}
\caption{(\textit{left}) Testing the consistency of measurements
between $\sigma$ derived from H$\alpha$ and H$\beta$ (FEROS);
(\textit{center}) $\sigma_{\scriptsize\mbox{[OIII]5007\AA}}$ and
$\sigma_{\scriptsize\mbox{[OIII]4959\AA}}$; (\textit{right})
H$\alpha$ from FEROS ($x$) and Coud\'e ($y$). Tol 0440-381 and UM
448 noted in the right plot present the most discrepant
measurements possibly due to different position covered by
observations.\label{fig5}}
\end{figure*}

We derived the radial velocity dispersions ($\sigma$) from the
observed FWHM presented in Table~\ref{tab2}. The observed velocity
dispersions ($\sigma_{\scriptsize\mbox{obs}}$) in km
s$^{-1}$ were corrected by the instrumental
($\sigma_{\scriptsize\mbox{inst}}$), and thermal broadening
($\sigma_{\scriptsize\mbox{th}}$), assuming a Maxwellian velocity
distribution of the hydrogen and oxygen atoms,
\[\sigma_{\scriptsize\mbox{th}} = \sqrt{\frac{kT_{e}}{m}},\] where $k$ is the
Boltzmann constant,
$T_{e}$ is the electronic temperature in Kelvin and $m$
is the mass of the atom.
We used $T_{e}(\mbox{OIII})$ presented in Table~\ref{tab3} to derive
$\sigma_{\scriptsize\mbox{th}}$ for all galaxies. For those galaxies
where $T_{e}$ was not directly determined or found in
literature we assumed a mean value of 14000 K. For H lines from
FEROS spectra we also corrected $\sigma_{\scriptsize\mbox{obs}}$ by
the fine structure broadening ($\sigma_{\scriptsize\mbox{fs}}$):
3.20 km s$^{-1}$ for H$\alpha$ and 2.40 km s$^{-1}$ for H$\beta$ as
adopted by \cite{gar08}. The velocity dispersion of interest here
and sometimes called ``non-thermal'' velocity dispersion, was
calculated by
\[\sigma = \sqrt{\sigma_{\scriptsize\mbox{obs}}^{2} -
\sigma_{\scriptsize\mbox{inst}}^{2} -
\sigma_{\scriptsize\mbox{th}}^{2} -
\sigma_{\scriptsize\mbox{fs}}^{2}}.\] For the FEROS,
$\sigma_{\scriptsize\mbox{inst}}$ = 2.5 km s$^{-1}$, while for
Coud\'e, $\sigma_{\scriptsize\mbox{inst}}$ = 14.7 km s$^{-1}$ and
= 17.6 km s$^{-1}$, depending on the instrumentation used (see
Section~\ref{sec2_2}). The correction due to
$\sigma_{\scriptsize\mbox{th}}$ for H lines varies between 9.5 and
12.5 km s$^{-1}$, and for the O lines, between 2 and 3 km
s$^{-1}$. We estimated the error in $\sigma$ due to uncertainties
in $T_{e}$ as being $\delta\sigma_{\scriptsize\mbox{H}} < 0.3$ km
s$^{-1}$ and $\delta\sigma_{\scriptsize\mbox{O}} < 0.1$ km
s$^{-1}$.

\begin{deluxetable}{lcccccccl}
\tabletypesize{\scriptsize}
\tablecaption{H$\alpha$ integrated fluxes, H$\beta$ extinction coefficients, H$\beta$
equivalent widths, ionization ratios [OIII]/[OII], O/H ratios and electron densities and
temperatures.\label{tab3}}
\tablecolumns{9}
\tablewidth{0pt}
\tablehead{
Galaxy &$F_{\scriptsize\mbox{H}\alpha}$     &$C_{\scriptsize\mbox{H}\beta}$&$W_{\scriptsize\mbox{H}\beta}$&[OIII]/[OII]&$N_{e}$&$T_{e}$&12+&Ref.\tablenotemark{a}\\
             &(erg s$^{-1}$ cm$^{-2}$)& &\AA & &cm$^{-3}$&$10^{4}$K&$\log(\mbox{O}/\mbox{H})$&
}
\startdata
UM 238            &2.0e-14  &0.23  & 36  & 3.31  & 867  &1.53  &7.89  &1,1,1,1,1,1,18    \\
MBG 00463-0239    &8.2e-14  &0.38  & 10  & 0.19  & 374  &1.40  &8.70  &1,1,1,1,1,22,16    \\
UM 304            &1.5e-13  &1.09  &-    &-      & 204  &1.40  &-     &14,2,-,-,22,22,-    \\
Tol 0104-388      &4.7e-14  &0.19  & 60  & 1.14  & 846  &1.49  &7.96  &1,1,1,1,1,1,18    \\
UM 306            &3.1e-14  &0.08  & 24  & 2.21  &  27  &1.16  &8.18  &1,1,1,1,1,1,18    \\
UM 307            &1.1e-13  &0.25  & 23  & 0.67  & 983  &1.40  &8.43  &1,1,1,1,1,22,16    \\
UM 323            &2.6e-14  &0.85  & 21  & 0.90  &  27  &1.76  &7.92  &1,1,1,1,1,1,18    \\
Tol 0127-397      &4.1e-14  &0.51  &-    &-      & 204  &1.40  &-     &14,2,-,-,22,22,-    \\
Tol 0140-420      &2.3e-14  &0.00  & 56  & 1.98  &  27  &1.28  &8.06  &1,1,1,1,1,1,18    \\
UM 137            &1.5e-14  &0.37  &  4  & 0.55  &  27  &1.40  &8.25  &1,1,1,1,1,22,17    \\
UM 151            &2.6e-14  &0.41  & 20  & 0.79  &  94  &1.40  &8.47  &1,1,1,15,1,22,17   \\
UM 382            &1.9e-14  &0.18  &135  &10.90  &  45  &1.62  &7.82  &6,6,6,6,6,6,6     \\
MBG 01578-6806    &-        &-     &-    &-      & 204  &1.40  &-     &-,-,-,-,22,22,-     \\
UM 391            &5.1e-14  &0.54  & 14  & 0.35  &  27  &1.40  &8.40  &1,1,1,1,1,22,16    \\
UM 395            &1.6e-14  &0.49  &  6  & 0.42  & 579  &1.40  &8.63  &1,1,1,1,1,22,17    \\
UM 396            &3.7e-14  &0.00  &153  & 5.85  &  27  &1.22  &8.18  &1,1,1,1,1,1,18    \\
UM 408            &2.0e-14  &0.06  & 33  & 3.66  & 207  &1.33  &8.02  &1,1,1,1,1,1,18    \\
UM 417            &5.8e-15  &0.45  & 46  & 7.65  &  27  &1.40  &8.04  &1,1,1,1,1,22,17    \\
Tol 0226-390      &1.0e-13  &0.31  &115  & 3.03  & 193  &1.16  &8.15  &1,1,1,1,1,1,18    \\
CTS 1003          &2.2e-14  &0.24  &-    &-      & 204  &1.40  &7.90  &3,3,-,-,22,22,3     \\
MBG 02411-1457    &4.7e-14  &0.76  &  2  & 0.15  & 132  &1.40  &8.26  &1,1,1,1,1,22,16    \\
Tol 0242-387      &1.1e-13  &0.78  &-    &-      & 204  &1.40  &8.23  &2,2,-,-,22,22,19    \\
CTS 1004          &3.5e-14  &0.00  & 77  & 3.83  & 101  &1.21  &8.14  &1,1,1,1,1,1,18    \\
CTS 1005          &4.3e-14  &0.16  &134  &11.14  & 204  &1.46  &7.91  &1,1,1,1,22,1,18    \\
Tol 0440-381      &5.5e-14  &0.11  & 29  & 1.77  &  27  &1.53  &7.96  &1,1,1,1,1,1,18    \\
CTS 1006          &1.4e-13  &0.15  & 70  & 2.88  &  27  &1.28  &8.04  &1,1,1,1,1,1,18    \\
CTS 1007          &4.3e-14  &0.01  &-    &-      & 204  &1.40  &7.83  &3,3,-,-,22,22,3     \\
CTS 1008          &5.7e-14  &0.24  &140  & 6.04  & 278  &1.21  &8.16  &1,1,1,1,1,1,18    \\
Tol 0505-387      &1.1e-14  &0.35  & 10  & 0.66  &  27  &1.40  &8.50  &1,1,1,1,1,22,17    \\
Tol 0510-400      &3.3e-14  &0.19  & 64  & 2.39  & 227  &1.40  &8.25  &1,1,1,1,1,22,16    \\
Tol 0528-383      &3.7e-14  &0.35  & 21  & 1.86  & 133  &1.48  &7.96  &1,1,1,1,1,1,18    \\
II ZW 40          &3.5e-13  &0.61  &184  &10.98  & 217  &1.31  &8.07  &1,1,1,1,1,1,18    \\
Tol 0559-393      &4.5e-14  &0.35  &-    &-      & 204  &1.40  &-     &2,2,-,-,22,22,-     \\
Tol 0610-387      &1.1e-14  &0.93  &  4  & 0.44  &  27  &1.40  &8.56  &1,1,1,2,1,22,17    \\
Tol 0614-375      &6.9e-14  &1.25  &-    &-      & 204  &1.40  &7.86  &2,2,-,-,22,22,20    \\
Tol 0633-415      &1.1e-13  &0.40  & 83  & 4.59  &  67  &1.25  &8.14  &1,1,1,21,1,21,4   \\
Tol 0645-376      &2.9e-14  &0.19  & 28  & 1.58  &  27  &1.78  &7.77  &1,1,1,1,1,1,18    \\
MRK 1201          &3.9e-14  &0.52  &  8  & 0.50  &1467  &1.40  &9.36  &1,1,1,1,1,22,16    \\
Cam 0840+1201     &1.1e-13  &0.03  &105  & 3.53  &  27  &1.32  &7.98  &1,1,1,1,1,1,18    \\
Cam 0840+1044     &2.1e-14  &0.25  & 44  & 7.85  &  27  &1.58  &7.73  &1,1,1,1,1,1,18    \\
Cam 08-28A        &1.5e-13  &0.28  & 37  & 1.60  &  97  &1.11  &8.13  &1,1,1,1,1,1,18    \\
MRK 710           &4.5e-13  &0.50  & 29  & 0.23  & 184  &1.40  &8.95  &1,1,1,1,1,22,16    \\
MRK 711           &1.9e-13  &0.54  & 28  & 1.61  & 460  &1.40  &8.79  &1,1,1,2,1,22,17    \\
Tol 0957-278      &2.0e-13  &0.17  & 36  & 1.92  &  74  &1.24  &8.02  &1,1,1,1,1,1,18    \\
Tol 1004-296NW    &7.4e-13  &0.40  & 62  & 3.50  & 122  &1.04  &8.28  &1,1,1,1,1,1,18    \\
Tol 1004-296SE    &5.0e-13  &0.30  & 52  & 2.69  &  69  &1.08  &8.20  &1,1,1,1,1,1,18    \\
Tol 1008-286      &5.1e-14  &1.05  &123  & 9.55  & 395  &1.30  &8.17  &1,1,1,21,1,21,4   \\
CTS 1011          &6.2e-14  &0.34  & 93  & 3.80  & 233  &1.28  &8.18  &1,1,1,1,1,1,18    \\
CTS 1012          &6.3e-14  &0.01  &-    &-      & 204  &1.40  &8.41  &3,3,-,-,22,22,3     \\
CTS 1013          &1.7e-14  &0.00  & 38  & 3.40  & 185  &1.29  &8.08  &1,1,1,1,1,1,18    \\
Tol 1025-285      &4.0e-14  &0.79  &  9  & 0.31  &  27  &1.40  &8.71  &1,1,1,2,1,22,17    \\
Haro 24           &4.8e-14  &0.57  & 11  & 0.88  &  27  &1.40  &8.23  &1,1,1,1,1,22,17    \\
CTS 1014          &2.2e-14  &0.01  &-    &-      & 204  &1.40  &7.98  &3,3,-,-,22,22,3     \\
CTS 1016          &1.7e-14  &0.19  & 26  & 1.24  &  27  &1.40  &8.36  &1,1,1,1,1,22,3     \\
CTS 1017          &2.2e-14  &0.22  &161  & 6.68  & 247  &1.46  &7.98  &1,1,1,1,1,1,18    \\
CTS 1018          &1.8e-14  &0.16  & 58  & 2.43  & 140  &1.40  &7.97  &1,1,1,1,1,1,18    \\
CTS 1019          &4.2e-14  &0.22  & 90  & 3.87  &  27  &1.11  &8.22  &1,1,1,1,1,1,18    \\
CTS 1020          &1.5e-13  &0.33  &109  & 2.94  & 101  &1.12  &8.25  &1,1,1,1,1,1,18    \\
CTS 1022          &2.6e-14  &0.43  & 57  & 1.53  & 147  &1.33  &8.09  &1,1,1,1,1,1,18    \\
$[$F80$]$ 30      &2.9e-13  &0.00  & 97  & 5.07  & 215  &1.41  &7.99  &1,1,1,7,1,7,7     \\
MRK 36            &1.5e-13  &0.08  & 62  & 3.48  &  96  &1.38  &7.89  &1,1,1,1,1,1,18    \\
UM 439            &1.2e-13  &0.05  & 49  & 4.30  &  27  &1.39  &8.01  &1,1,1,1,1,1,18    \\
UM 448            &7.3e-13  &0.40  & 48  & 1.16  & 151  &1.08  &8.17  &1,1,1,1,1,1,18    \\
Tol 1147-283      &5.6e-14  &0.20  & 40  & 1.08  &  79  &1.51  &7.88  &1,1,1,1,1,1,18    \\
UM 455            &2.0e-14  &0.48  & 29  & 4.32  &  27  &1.73  &7.74  &1,1,1,1,1,1,18    \\
UM 456            &8.9e-14  &0.06  & 44  & 3.09  &  27  &1.41  &7.95  &1,1,1,1,1,1,18    \\
UM 461            &1.1e-13  &0.05  &155  & 9.43  & 115  &1.66  &7.77  &1,1,1,1,1,1,18    \\
UM 463            &3.3e-14  &0.17  & 74  & 6.08  & 102  &1.32  &7.92  &1,1,1,20,1,11,11  \\
CTS 1026          &1.0e-12  &0.33  &-    &-      & 204  &1.40  &8.30  &3,3,-,-,22,22,3     \\
UM 477            &2.6e-13  &0.88  & 17  & 0.34  & 979  &1.40  &9.12  &1,1,1,1,1,22,16    \\
UM 483            &4.7e-14  &0.45  & 19  & 0.84  & 133  &1.71  &7.85  &1,1,1,11,1,11,11  \\
CTS 1027          &1.5e-13  &0.08  & 50  & 1.88  &  63  &1.01  &8.35  &8,8,8,8,8,8,8     \\
MRK 1318          &2.2e-13  &0.27  & 68  & 1.70  &  78  &1.01  &8.27  &1,1,1,1,1,1,18    \\
CTS 1028          &4.5e-14  &0.57  & 82  & 3.90  & 344  &1.40  &8.05  &1,1,1,1,1,1,18    \\
UM 499            &6.1e-13  &0.55  & 24  & 0.45  & 611  &1.40  &8.82  &1,1,1,1,1,22,16    \\
Tol 1223-359      &7.5e-14  &0.16  &129  & 7.18  &  27  &1.73  &7.54  &1,1,1,12,1,12,12  \\
Haro 30           &6.1e-14  &0.00  & 28  & 0.93  & 103  &1.54  &7.67  &1,1,1,1,1,1,18    \\
$[$SC98$]$ 01     &1.8e-14  &0.14  & 34  & 1.39  &  45  &1.40  &8.27  &1,1,1,1,1,22,17    \\
CTS 1029          &1.9e-14  &0.29  & 35  & 1.05  & 489  &1.40  &8.42  &1,1,1,1,1,22,3     \\
$[$SC98$]$ 11     &-        &-     &-    &-      & 204  &1.40  &-     &-,-,-,-,22,22,-     \\
UM 559            &5.0e-14  &0.00  &535  & 4.92  & 204  &1.58  &7.72  &1,1,1,1,22,9,9     \\
$[$SC98$]$ 68     &2.1e-14  &0.52  & 22  & 0.76  & 343  &1.40  &8.49  &1,1,1,1,1,22,17    \\
UM 570            &3.1e-14  &0.00  &180  &43.00  & 204  &1.83  &7.71  &1,1,1,9,22,9,9     \\
$[$SC98$]$ 88     &1.9e-14  &0.29  & 20  & 1.24  & 339  &1.41  &8.02  &1,1,1,1,1,1,18    \\
CTS 1030          &7.0e-14  &0.01  &-    &-      & 204  &1.40  &8.25  &3,3,-,-,22,22,3     \\
POX 186           &7.2e-14  &0.01  &274  &19.26  & 342  &1.66  &7.74  &1,1,1,1,1,1,18    \\
CTS 1031          &3.4e-14  &0.22  &-    &-      & 204  &1.40  &8.23  &3,3,-,-,22,22,3     \\
Tol 1345-420      &8.7e-14  &0.27  & 51  & 2.89  &  71  &1.07  &8.26  &1,1,1,1,1,1,18    \\
CTS 1033          &7.4e-14  &0.24  & 59  & 8.69  & 155  &1.42  &8.01  &1,1,1,1,1,1,18    \\
Tol 1400-397      &2.3e-14  &0.20  &-    &-      & 204  &1.40  &-     &2,2,-,-,22,22,-     \\
UM 649            &1.2e-14  &0.00  &-    &-      & 204  &1.40  &-     &2,2,-,-,22,22,-     \\
CTS 1034          &1.3e-14  &0.28  & 20  & 1.53  &1288  &1.47  &7.96  &1,1,1,1,1,1,18    \\
II ZW 70          &2.7e-13  &0.30  & 49  & 1.90  &  96  &1.21  &8.07  &5,13,10,10,13,13,4\\
CTS 1035          &1.5e-14  &0.12  & 62  & 2.91  & 510  &1.40  &8.01  &1,1,1,1,1,22,3     \\
CTS 1037          &4.0e-13  &0.29  &-    &-      & 204  &1.40  &8.21  &3,3,-,-,22,22,3     \\
Cam 1543+0907     &5.9e-14  &0.05  &192  & 8.96  &  74  &1.68  &7.71  &1,1,1,1,1,1,18    \\
Tol 1924-416      &1.0e-12  &0.11  &100  & 4.86  & 131  &1.35  &8.01  &1,1,1,1,1,1,18    \\
Tol 1939-419      &1.3e-14  &0.00  &-    &-      & 204  &1.40  &-     &2,2,-,-,22,22,-     \\
Tol 1937-423      &2.0e-14  &0.70  &  5  & 0.49  &  39  &1.40  &8.48  &1,1,1,1,1,22,17    \\
CTS 1038          &1.9e-13  &0.53  &-    &-      & 204  &1.40  &7.82  &3,3,-,-,22,22,3     \\
CTS 1039          &9.7e-14  &0.01  &-    &-      & 204  &1.40  &7.70  &3,3,-,-,22,22,3     \\
Tol 2010-382      &8.2e-14  &0.76  &-    &-      & 204  &1.40  &-     &14,2,-,-,22,22,-    \\
Tol 2019-405      &1.5e-14  &0.10  & 11  & 1.90  &  76  &1.45  &7.99  &1,1,1,1,1,1,18    \\
Tol 2041-394      &1.9e-14  &0.00  &-    &-      & 204  &1.40  &-     &2,2,-,-,22,22,-     \\
NGC 6970          &9.7e-14  &0;00  &-    &-      & 204  &1.40  &-     &14,2,-,-,22,22,-    \\
MBG 20533-4410    &1.0e-13  &0.81  &  8  & 0.20  &  90  &1.40  &8.88  &1,1,1,1,1,22,16    \\
Tol 2122-408      &2.6e-14  &0.41  & 14  & 4.47  &  97  &1.40  &8.49  &1,1,1,1,1,22,16    \\
Tol 2138-405      &8.9e-14  &0.19  &208  & 7.08  & 204  &1.38  &7.98  &9,9,9,9,22,9,9     \\
Tol 2138-397      &1.8e-14  &0.12  & 33  & 2.49  &  27  &1.86  &7.64  &1,1,1,1,1,1,18    \\
Tol 2146-391      &2.8e-14  &0.09  &246  & 7.70  &  47  &1.59  &7.78  &1,1,1,1,1,1,18    \\
MBG 21567-1645    &3.2e-14  &1.36  &  2  & 0.08  & 281  &1.40  &8.93  &1,1,1,1,1,22,17    \\
MBG 22012-1550    &3.9e-14  &0.84  &  7  & 0.70  &  27  &1.40  &8.20  &1,1,1,1,1,22,16    \\
IC 5154           &4.9e-14  &0.38  & 10  & 0.79  & 250  &1.40  &8.52  &1,1,1,1,1,22,17    \\
ESO 533-G 014     &5.7e-14  &0.45  &  6  & 0.41  &  27  &1.51  &7.90  &1,1,1,1,1,1,18    \\
MCG -01-57-017    &6.5e-14  &0.10  & 11  & 0.23  &  27  &1.40  &8.37  &1,1,1,1,1,22,16    \\
Tol 2240-384      &3.4e-14  &0.37  &165  & 8.72  & 204  &1.53  &7.85  &1,1,1,1,22,1,18    \\
MBG 23121-3807    &2.4e-14  &0.71  &  4  & 0.11  &  27  &1.40  &8.75  &1,1,1,1,1,22,17    \\
Tol 2326-405      &4.1e-14  &0.22  &-    &-      & 204  &1.40  &8.03  &2,2,-,-,22,22,19   \\
UM 167            &1.2e-12  &0.26  &-    &-      & 204  &1.40  &-     &14,2,-,-,22,22,-    \\
UM 191            &3.6e-14  &0.43  &  7  & 0.26  & 108  &1.40  &8.30  &1,1,1,1,1,22,16    
\enddata
\tablenotetext{a}{References.- (1) \cite{keh04}; (2) T91; (3)
\cite{pen91}; (4) \cite{den02}; (5) \cite{tel01}; (6)
\cite{kni01}; (7) \cite{izo98}; (8) \cite{vil03}; (9)
\cite{pap06}; (10) \cite{mas99}; (11) \cite{kni04}; (12)
\cite{izo01}; (13) \cite{keh08}; (14) Coud\'e spectrophotometry;
(15) \cite{pus02}; (16) O/H derived from p-method \citep{pil00};
(17) O/H derived from N2 calibrator \citep{den02}; (18) O/H
derived from $T_{e}$-method; (19) \cite{mel88}; (20) \cite{mas94};
(21) \cite{cam86}; (22) mean values for $T_{e}$ and $N_{e}$.}
\end{deluxetable}

The internal errors were determined as a function of the S/N
calculated for the emission lines and defined as the ratio between
the peak intensity and the adjacent continuum RMS. Figure~\ref{fig5}
shows the comparisons between $\sigma$ derived from the same ion
considering all data. The FEROS measurements are very consistent.
The RMS of a linear fit for all points are 1.96 for H{\sc ii} and
1.36 for [OIII] lines in km s$^{-1}$ (Figure~\ref{fig5} left and
center). The comparison between FEROS and Coud\'e measurements also
shows good agreement except for two objects (Tol 0440-381 and UM
448) that present components in their line profiles. The observation
of these objects is more sensitive to the position of the slit and
fiber over the galaxy. The RMS = 5.22 km s$^{-1}$ from a linear fit
is also shown inside the box (Figure~\ref{fig5} right). For
different ranges of S/N we selected samples of galaxies for their
$\sigma$ values to be compared. The selection was based on the S/N
of the weaker lines of each ion (i.e. H$\beta$ and
[OIII]$\lambda$4959) and their $\sigma$ values were plotted in $y$
axis against the $\sigma$ values derived from the more intense line
in $x$. The errors were estimated by taking the RMS of a direct
least square fit for each data set. In a similar procedure, we
estimated the errors for Coud\'e
$\sigma_{\scriptsize\mbox{H}\alpha}$ comparing sets of two ranges of
S/N with $\sigma_{\scriptsize\mbox{H}\alpha}$ of galaxies observed
also with FEROS. Table~\ref{tab4} shows the estimated errors in
$\sigma$ as a function of S/N for H and O lines. The $\sigma$ values
and their respective errors for each line are shown in
Table~\ref{tab5}.

\begin{table}[ht]
\begin{center}
\caption{Errors in $\sigma$ as a function of signal-to-noise ratio
of the line-emission.\label{tab4}}
\begin{tabular}{lclc}\\ \tableline\tableline
S/N             &$\delta\sigma_{\scriptsize\mbox{H}}$ &S/N  &$\delta\sigma_{\scriptsize\mbox{O}}$ \\
H lines               &km s$^{-1}$        &O lines    &km s$^{-1}$
\\ \tableline
FEROS\\
S/N $<$ 10            &2.9                &S/N $<$ 10            &2.1                \\
10 $<$ S/N $<$ 20     &2.2                &10 $<$ S/N $<$ 20     &1.4                \\
20 $<$ S/N $<$ 30     &1.8                &20 $<$ S/N $<$ 30     &0.9                \\
30 $<$ S/N $<$ 40     &1.2                &30 $<$ S/N $<$ 50     &0.7                \\
40 $<$ S/N $<$ 75     &0.7                &50 $<$ S/N $<$ 150    &0.5                \\
S/N $>$ 75            &0.4                &S/N $>$ 150 &0.2
\\ \tableline
Coud\'e\\
S/N $<$ 110           &4.6 \\
S/N $>$ 110           &3.5
\\ \tableline
\end{tabular}
\end{center}
\end{table}

\subsection{Physical Conditions}\label{sec3_2}

Before we use line fluxes to derive physical conditions we need to
infer the amount of extinction for each galaxy. Dust in starburst
regions is responsible for extinction of light in the line
of sight due to absorption and scattering. In optical wavelengths,
the amount of extinction can be reasonably well estimated from H
recombination lines through the Balmer Decrement method. To derive
H$\beta$ extinction coefficient ($C_{\scriptsize\mbox{H}\beta}$) for
galaxies with KTC spectrophotometry we used the theoretical ratios
H$\alpha$/H$\beta$ = 2.87 and H$\gamma$/H$\beta$ = 0.466 for case B
optically thick with T = 10$^{4}$ K (Osterbrock 1989). In cases
where H$\alpha$/H$\beta$ was smaller than the theoretical
value 2.87, we calculated $C_{\scriptsize\mbox{H}\beta}$ using the
ratio H$\gamma$/H$\beta$. When ratios were H$\alpha$/H$\beta$ $<$
2.87 and H$\gamma$/H$\beta$ $>$ 0.466 simultaneously we adopted the
zero value for $C_{\scriptsize\mbox{H}\beta}$. Dereddened
fluxes $I_{\lambda}$ were thus calculated by
\[I_{\lambda}=F_{\lambda}\exp[C_{\scriptsize\mbox{H}\beta}(1+f_{\lambda})],\]
where $F_{\lambda}$ is the published flux corrected by
atmospheric extinction and $f_{\lambda}$ is the interstellar
reddening function normalized at H$\beta$. We adopted $f_{\lambda}$
from \cite{whi58} as normalized by \cite{leq79}.

H$\beta$ equivalent widths ($W_{\scriptsize\mbox{H}\beta}$) were
taken directly from KTC for 91 objects. From the same work
ionization ratios [OIII]$\lambda\lambda$4959+5007/[OII]
$\lambda$3727 (hereafter [OIII]/[OII]) were directly determined for
80 objects from dereddened fluxes of oxygen lines.

We derived oxygen abundances for 51 objects adopting the
$T_{e}-$method and the standard model for a two-zone photoionized
H{\sc ii} region. This number was limited by the number of objects
with all oxygen lines available in KTC, i.e.
[OIII]$\lambda\lambda$4959, 5007, [OIII]$\lambda$4363 and
[OII]$\lambda$3727 lines. We did not consider those galaxies in
which [OII]$\lambda$3727 fluxes were indirectly determined by KTC.
Temperatures for low- and high-ionization zones were derived
according to \cite{pag92} using the [OIII] ($\lambda\lambda$4959 +
5007)/$\lambda$4363 ratio and electron densities, $N_{e}$(SII).
$N_{e}$(SII) was derived for 87 galaxies from the [SII]
$\lambda$6717/$\lambda$6731 ratio, using TEMDEN from NEBULAR package
of IRAF, based on five-level atom calculations developed by
\cite{sha95}. All [SII] $\lambda$6717/$\lambda$6731 ratios higher
than 1.4 were fixed at this value, corresponding to a minimum
$N_{e}$(SII) of 27 cm$^{-3}$ in TEMDEN task. For those galaxies
where $N_{e}$ was not directly determined or found in
literature we assumed a mean value of 204 cm$^{-3}$. Oxygen ionic
abundances could be derived using \cite{pag92} expressions for
O$^{++}$/H$^{+}$ and O$^{+}$/H$^{+}$ to obtain the total oxygen
abundance
\[\frac{\mbox{O}}{\mbox{H}}=\frac{\mbox{O}^{+}}{\mbox{H}^{+}}+\frac{\mbox{O}^{++}}{\mbox{H}^{+}}.\]
Typical errors in oxygen abundances derived by the
$T_{e}$-method and provided by Monte Carlo simulations were
$\delta$(O/H)=0.05-0.06.

For galaxies in KTC that could not have their abundances determined
from the $T_{e}-$method, or even were not present in their sample,
we compiled some recent results found in the literature
using also the $T_{e}-$method. From most of these other works the
values for $W_{\scriptsize\mbox{H}\beta}$, [OIII]/[OII],
$N_{e}$(SII) and $T_{e}$(OIII) were also taken. For those galaxies
where we could not find oxygen abundances determined from
the $T_{e}-$method, mainly due to the absence of
the auroral [OIII] $\lambda$4363 line, we derived O/H empirically
using the p-method \citep{pil00} and N2 calibrator \citep{den02}.

Firstly, we calculated both high- and low-abundance values for O/H
\citep[equations (4) and (6)]{pil00}. We only considered those
values in agreement with the respective abundance regimes of the
Pilyugin's best fits, i.e. 12+log(O/H)$_{P3} < 7.95$ or
12+log(O/H)$_{P2} > 8.15$. In order to break the degeneracy
when the values were acceptable, we adopted a criterion
similar to the one described in \cite{van98}. Those galaxies with
log [NII]($\lambda\lambda$6548 + 6584)/[OII]($\lambda$3727 blended)
$> -0.8$ should have their high-abundance values assigned (P2),
whereas those with log [NII]/[OII] $< -1.05$ should have their
low-abundance values assigned (P3). With this criterion we
assigned 14 empirical abundance values for 12+log(O/H) and all from
the high-abundance side. Instead of using the calibrator
[NII]/H$\alpha$ provided in \cite{van98} for the turnover region,
and when the Pilyugin's values were not calculated or acceptable, we
used the N2 calibrator from \cite{den02},
\[12+\log(\mbox{O}/\mbox{H})=9.12+0.73\times\mbox{N2},\] where N2 is
defined as N2=log([NII]$\lambda$6584/H$\alpha$). For this, we also
used [NII] dereddened fluxes from KTC. Other 15 O/H ratios
were calculated by the N2 calibrator. The uncertainties in oxygen
abundances derived by both empirical methods (p-method and N2) were
estimated to be $\delta$(O/H)=0.14, corresponding to the RMS of
least square fits between $T_{e}$-method abundances and empirical
abundances derived independently.

Table~\ref{tab3} presents all physical parameters discussed in this
section. Columns 1-5 show the galaxy name, the observed flux of
H$\alpha$ ($F_{\scriptsize\mbox{H}\alpha}$), the derived logarithmic
reddening parameter ($C_{\scriptsize\mbox{H}\beta}$), the equivalent
width of H$\beta$ ($W_{\scriptsize\mbox{H}\beta}$), the ionization
ratio [OIII]/[OII], the derived electron density ($N_e$) and
temperature ($T_e$), and oxygen abundance (O/H), and finally,
the last column shows the references to the sources of the
data in the same order they appear in the table.

\subsection{Distance and H$\alpha$ Luminosity}\label{sec3_3}

The distances ($D$) to all galaxies in Mpc were derived using the
Hubble's Law, $D=cz_{\scriptsize\mbox{H}}/H_{0}$, where
$z_{\scriptsize\mbox{H}}$ is the cosmological redshift of the
galaxies and $H_{0}$ is the Hubble Constant in km s$^{-1}$
Mpc$^{-1}$.
Heliocentric redshifts ($z_{\scriptsize\mbox{hel}}$)
were derived from the observed redshift
($z_{\scriptsize\mbox{obs}}=\Delta\lambda/\lambda$)
by removing the earth's rotational and orbital motions using the
resultant velocity component from RVCORRECT routine of IRAF. This
correction is smaller than 30 km s$^{-1}$ in modulus and it was only
applied to the redshifts from FEROS spectra. The uncertainty $\delta
z_{\scriptsize\mbox{hel}}=
5\times 10^{-5}$ was estimated for FEROS spectra's redshifts and
$\delta z_{\scriptsize\mbox{hel}}=1.5\times10^{-4}$ for Coud\'e
spectra's redshifts. Heliocentric redshifts from FEROS were
used in priority to Coud\'e to derive distances for those galaxies
observed with both instruments.
We found $z_{\scriptsize\mbox{H}}$ from
$z_{\scriptsize\mbox{hel}}$ by removing the solar motion with
respect to 3K Cosmic Microwave Background (CMB).
We use the NASA/IPAC Extragalactic Database -
NED\footnote{http://nedwww.ipac.caltech.edu/} to obtain the
resultant velocity correction
(heliocentric to 3K background) for all galaxies of our sample.
 Heliocentric redshifts
and distances for all galaxies
are presented in columns 2 and 3 of Table~\ref{tab2}.

H$\alpha$ luminosities ($L_{\scriptsize\mbox{H}\alpha}$) were
therefore derived from dereddened fluxes
($I_{\scriptsize\mbox{H}\alpha}$) and $D$
($L_{\scriptsize\mbox{H}\alpha}=4\pi
D^{2}I_{\scriptsize\mbox{H}\alpha}$). We have used H$\alpha$ rather
than H$\beta$ in the $L$-$\sigma$ relation since it is more
intense and relatively less affected by extinction and underlying
absorption.
The last column of Table~\ref{tab5} shows the derived H$\alpha$
luminosities for 118 galaxies for which we had reliable
spectrophotometry.

\begin{deluxetable}{lcccccc}
\tabletypesize{\small}
\tablecaption{Velocity dispersion and H$\alpha$
luminosities.\label{tab5}}
\tablecolumns{7}
\tablewidth{0pt}
\tablehead{
Galaxy &\multicolumn{5}{c}{$\sigma$ (km s$^{-1}$)} &\\
&\multicolumn{4}{c}{FEROS} &Coud\'e &$\log$ $L$ (erg s$^{-1}$)\\
&H$\beta$ &$\lambda 4959$ &$\lambda 5007$ &H$\alpha$ &H$\alpha$
&H$\alpha$
}
\startdata
UM 238            &18.3$\pm$1.8   & 19.2$\pm$0.5   & 19.3$\pm$0.5   &18.6$\pm$0.4   &-              &40.02\\
MBG 00463-0239    &-              &-               &-               &-              & 57.7$\pm$3.5  &40.67\\
UM 304            &78.2$\pm$2.2   & 89.8$\pm$2.1   & 86.1$\pm$1.4   &78.2$\pm$0.4   & 76.3$\pm$4.6  &41.55\\
Tol 0104-388      &47.6$\pm$1.8   & 47.0$\pm$0.7   & 49.3$\pm$0.5   &48.2$\pm$0.4   &-              &40.81\\
UM 306            &19.1$\pm$1.2   & 17.4$\pm$0.5   & 18.1$\pm$0.5   &19.1$\pm$0.4   &-              &40.25\\
UM 307            &-              &-               &-               &-              & 49.4$\pm$3.5  &41.20\\
UM 323            &18.9$\pm$2.2   & 21.4$\pm$0.9   & 19.5$\pm$0.5   &18.0$\pm$0.4   &-              &39.78\\
Tol 0127-397      &37.2$\pm$0.7   & 35.6$\pm$0.5   & 35.8$\pm$0.5   &35.4$\pm$0.4   & 33.2$\pm$3.5  &40.72\\
Tol 0140-420      &28.6$\pm$2.2   & 27.2$\pm$2.1   & 27.2$\pm$2.1   &26.4$\pm$0.4   &-              &40.35\\
UM 137            &-              &-               & 16.7$\pm$2.1   &14.5$\pm$2.2   &-              &39.13\\
UM 151            &28.7$\pm$2.9   & 22.2$\pm$2.1   & 26.4$\pm$2.1   &31.6$\pm$1.8   &-              &40.39\\
UM 382            &17.2$\pm$2.9   & 19.2$\pm$2.1   & 18.2$\pm$1.4   &15.4$\pm$2.2   &-              &39.83\\
MBG 01578-6806    &-              &-               &-               &-              & 24.8$\pm$4.6  &-    \\
UM 391            &57.7$\pm$2.2   &-               & 79.3$\pm$1.4   &59.1$\pm$0.4   & 59.6$\pm$4.6  &41.00\\
UM 395            &30.1$\pm$2.2   & 32.0$\pm$2.1   & 29.8$\pm$1.4   &30.3$\pm$0.4   &-              &40.52\\
UM 396            &27.1$\pm$2.2   & 27.4$\pm$0.2   & 27.4$\pm$0.5   &26.7$\pm$0.4   &-              &40.49\\
UM 408            &19.5$\pm$1.2   & 16.4$\pm$0.5   & 16.7$\pm$0.5   &19.3$\pm$0.4   &-              &39.73\\
UM 417            &12.1$\pm$2.9   & 13.4$\pm$2.1   & 14.8$\pm$0.7   &15.3$\pm$1.2   &-              &39.18\\
Tol 0226-390      &89.9$\pm$0.7   & 83.5$\pm$0.5   & 81.1$\pm$0.2   &-              & 75.5$\pm$3.5  &41.89\\
CTS 1003          &24.7$\pm$1.8   & 25.1$\pm$0.5   & 24.7$\pm$0.7   &25.2$\pm$0.4   &-              &40.10\\
MBG 02411-1457    &-              &-               &-               &-              & 33.7$\pm$4.6  &40.09\\
Tol 0242-387      &98.2$\pm$2.9   &113.1$\pm$2.1   &111.7$\pm$0.9   &96.9$\pm$2.9   &-              &43.11\\
CTS 1004          &43.8$\pm$2.9   & 43.5$\pm$2.1   & 44.3$\pm$0.2   &-              &-              &41.21\\
CTS 1005          &49.4$\pm$1.8   & 46.8$\pm$2.1   & 46.3$\pm$0.5   &48.6$\pm$0.4   &-              &41.81\\
Tol 0440-381      &34.3$\pm$0.7   & 32.0$\pm$0.5   & 31.8$\pm$0.2   &34.7$\pm$0.4   & 47.4$\pm$4.6  &41.36\\
CTS 1006          &36.7$\pm$0.4   & 33.5$\pm$0.2   & 33.7$\pm$0.2   &37.0$\pm$0.4   &-              &41.22\\
CTS 1007          &29.5$\pm$1.8   & 27.4$\pm$0.5   & 26.8$\pm$0.5   &28.5$\pm$0.4   &-              &41.20\\
CTS 1008          &49.0$\pm$1.8   & 45.6$\pm$0.5   & 45.6$\pm$0.7   &48.0$\pm$1.2   &-              &41.81\\
Tol 0505-387      &21.8$\pm$2.9   & 19.9$\pm$2.1   & 22.1$\pm$0.7   &21.0$\pm$1.8   &-              &40.55\\
Tol 0510-400      &34.0$\pm$1.8   & 31.4$\pm$0.9   & 31.1$\pm$0.9   &31.4$\pm$0.7   &-              &41.21\\
Tol 0528-383      &18.0$\pm$2.2   & 18.9$\pm$0.9   & 19.6$\pm$1.4   &19.5$\pm$0.7   &-              &40.27\\
II ZW 40          &32.9$\pm$0.4   & 32.5$\pm$0.2   & 32.8$\pm$0.2   &33.5$\pm$0.4   & 31.8$\pm$3.5  &40.17\\
Tol 0559-393      &50.7$\pm$2.2   & 41.9$\pm$2.1   & 46.8$\pm$0.7   &48.3$\pm$1.2   &-              &41.53\\
Tol 0610-387      &16.6$\pm$2.9   &-               & 23.2$\pm$0.7   &20.1$\pm$0.7   &-              &39.55\\
Tol 0614-375      &44.8$\pm$1.8   & 50.8$\pm$2.1   & 50.0$\pm$0.9   &48.4$\pm$0.4   &-              &42.01\\
Tol 0633-415      &32.0$\pm$0.7   & 30.7$\pm$0.5   & 31.1$\pm$0.2   &30.6$\pm$0.4   &-              &41.09\\
Tol 0645-376      &28.7$\pm$2.2   & 30.5$\pm$0.9   & 29.4$\pm$0.5   &29.6$\pm$0.7   & 23.3$\pm$3.5  &40.76\\
MRK 1201          &46.0$\pm$2.2   &-               &-               &46.0$\pm$0.7   &-              &40.83\\
Cam 0840+1201     &36.8$\pm$1.8   & 34.6$\pm$0.7   & 34.4$\pm$0.2   &34.9$\pm$0.4   &-              &41.37\\
Cam 0840+1044     &15.4$\pm$1.8   & 13.4$\pm$1.4   & 13.9$\pm$0.2   &14.8$\pm$0.4   &-              &39.94\\
Cam 08-28A        &51.0$\pm$2.2   & 43.7$\pm$1.4   & 44.7$\pm$0.5   &-              &-              &42.14\\
MRK 710           &56.2$\pm$1.8   &-               & 53.7$\pm$1.4   &51.0$\pm$0.4   & 47.8$\pm$3.5  &40.89\\
MRK 711           &94.0$\pm$0.7   & 86.0$\pm$0.7   & 87.2$\pm$0.5   &95.3$\pm$0.4   &-              &41.59\\
Tol 0957-278      &24.7$\pm$2.2   & 26.7$\pm$1.4   & 26.2$\pm$0.5   &26.9$\pm$0.4   & 22.0$\pm$3.5  &40.05\\
Tol 1004-296NW    &-              &-               &-               &-              & 31.9$\pm$3.5  &40.83\\
Tol 1004-296SE    &-              &-               &-               &-              & 27.1$\pm$3.5  &40.57\\
Tol 1008-286      &24.9$\pm$0.4   & 25.3$\pm$0.2   & 25.0$\pm$0.2   &25.5$\pm$0.4   &-              &41.08\\
CTS 1011          &21.2$\pm$0.7   & 20.6$\pm$0.5   & 20.7$\pm$0.5   &21.1$\pm$0.4   &-              &40.59\\
CTS 1012          &17.0$\pm$0.7   & 16.1$\pm$0.5   & 15.8$\pm$0.2   &16.6$\pm$0.4   &-              &40.29\\
CTS 1013          &33.7$\pm$2.9   & 33.5$\pm$1.4   & 34.0$\pm$0.5   &31.6$\pm$1.8   &-              &40.45\\
Tol 1025-285      &58.8$\pm$2.9   &-               & 60.0$\pm$1.4   &55.0$\pm$1.2   &-              &41.47\\
Haro 24           &43.0$\pm$2.2   & 38.9$\pm$1.4   & 42.7$\pm$0.7   &46.3$\pm$0.7   &-              &41.69\\
CTS 1014          &56.6$\pm$2.2   & 42.9$\pm$0.9   & 44.2$\pm$0.7   &50.0$\pm$1.8   &-              &41.23\\
CTS 1016          &35.1$\pm$2.9   & 32.6$\pm$2.1   & 34.5$\pm$1.4   &39.7$\pm$1.8   &-              &40.79\\
CTS 1017          &27.7$\pm$2.2   & 28.9$\pm$0.7   & 28.6$\pm$1.4   &27.5$\pm$0.7   &-              &40.94\\
CTS 1018          &34.8$\pm$1.8   & 32.2$\pm$0.5   & 32.0$\pm$0.5   &35.1$\pm$0.7   &-              &40.91\\
CTS 1019          &45.2$\pm$2.9   & 45.5$\pm$0.7   & 46.0$\pm$0.5   &49.1$\pm$0.7   &-              &41.76\\
CTS 1020          &34.5$\pm$0.4   & 34.4$\pm$0.5   & 35.0$\pm$0.2   &35.1$\pm$0.4   &-              &40.98\\
CTS 1022          &22.6$\pm$1.8   & 20.9$\pm$0.7   & 21.1$\pm$0.7   &24.9$\pm$0.4   &-              &40.37\\
$[$F80$]$ 30      &20.6$\pm$0.4   & 18.6$\pm$0.2   & 18.6$\pm$0.2   &20.2$\pm$0.4   &-              &40.07\\
MRK 36            &17.1$\pm$1.2   & 17.2$\pm$0.5   & 17.8$\pm$0.2   &16.8$\pm$0.4   &-              &39.55\\
UM 439            &17.7$\pm$0.4   & 17.7$\pm$0.2   & 17.6$\pm$0.2   &18.0$\pm$0.4   & 18.9$\pm$4.6  &39.86\\
UM 448            &75.1$\pm$0.7   & 76.6$\pm$0.7   & 75.3$\pm$0.5   &78.3$\pm$0.4   & 61.2$\pm$3.5  &42.04\\
Tol 1147-283      &17.0$\pm$2.2   & 20.0$\pm$0.9   & 17.9$\pm$0.5   &17.8$\pm$0.4   &-              &39.95\\
UM 455            &22.6$\pm$2.2   & 18.5$\pm$0.7   & 18.2$\pm$0.9   &26.2$\pm$1.2   &-              &40.25\\
UM 456            &16.1$\pm$0.7   & 15.7$\pm$0.5   & 15.8$\pm$0.2   &15.9$\pm$0.4   &-              &40.00\\
UM 461            &12.5$\pm$0.4   & 12.5$\pm$0.2   & 12.6$\pm$0.2   &12.5$\pm$0.4   &-              &39.77\\
UM 463            &19.2$\pm$2.2   & 17.0$\pm$0.9   & 16.7$\pm$0.2   &17.0$\pm$0.7   &-              &39.50\\
CTS 1026          &43.1$\pm$0.4   & 43.0$\pm$0.2   & 42.9$\pm$0.2   &42.5$\pm$0.4   &-              &41.02\\
UM 477            &57.4$\pm$1.2   &-               & 72.3$\pm$1.4   &57.2$\pm$0.4   & 55.0$\pm$3.5  &40.80\\
UM 483            &17.2$\pm$0.7   & 19.4$\pm$0.5   & 18.4$\pm$0.2   &17.1$\pm$0.4   &-              &40.22\\
CTS 1027          &21.0$\pm$0.4   & 20.1$\pm$0.5   & 20.5$\pm$0.2   &21.2$\pm$0.4   &-              &40.29\\
MRK 1318          &17.1$\pm$1.2   & 17.0$\pm$0.5   & 17.2$\pm$0.2   &17.7$\pm$0.4   & 19.8$\pm$4.6  &40.44\\
CTS 1028          &27.1$\pm$1.2   & 26.8$\pm$0.5   & 27.2$\pm$0.5   &26.6$\pm$0.4   &-              &40.92\\
UM 499            &-              &-               &-               &-              & 44.0$\pm$3.5  &41.31\\
Tol 1223-359      &17.5$\pm$2.2   & 16.9$\pm$0.9   & 17.4$\pm$0.2   &18.4$\pm$0.4   &-              &40.35\\
Haro 30           &43.4$\pm$2.2   & 47.6$\pm$2.1   & 46.0$\pm$0.9   &44.8$\pm$0.7   &-              &40.55\\
$[$SC98$]$ 01     &22.6$\pm$2.9   & 19.8$\pm$1.4   & 17.7$\pm$0.7   &19.6$\pm$0.7   &-              &39.80\\
CTS 1029          &-              &-               &-               &31.4$\pm$1.8   &-              &40.94\\
$[$SC98$]$ 11     &31.5$\pm$1.8   & 31.8$\pm$0.9   & 30.5$\pm$0.5   &31.9$\pm$0.4   &-              &-    \\
UM 559            &16.9$\pm$0.7   & 16.9$\pm$0.5   & 17.0$\pm$0.2   &17.5$\pm$0.4   &-              &39.48\\
$[$SC98$]$ 68     &30.6$\pm$2.9   & 33.6$\pm$2.1   & 32.1$\pm$0.9   &31.3$\pm$0.7   &-              &40.79\\
UM 570            &20.1$\pm$2.2   & 21.8$\pm$0.9   & 21.2$\pm$0.5   &20.8$\pm$0.4   &-              &40.57\\
$[$SC98$]$ 88     &25.1$\pm$2.2   & 22.4$\pm$1.4   & 21.2$\pm$1.4   &26.1$\pm$1.2   &-              &40.19\\
CTS 1030          &29.4$\pm$0.4   & 30.5$\pm$0.5   & 30.1$\pm$0.2   &29.7$\pm$0.4   &-              &40.59\\
POX 186           &14.1$\pm$0.7   & 14.4$\pm$0.5   & 14.3$\pm$0.2   &14.4$\pm$0.4   &-              &39.62\\
CTS 1031          &31.6$\pm$2.2   & 31.6$\pm$1.4   & 33.5$\pm$0.7   &30.8$\pm$0.4   &-              &41.19\\
Tol 1345-420      &20.4$\pm$2.2   & 18.6$\pm$0.7   & 19.6$\pm$0.2   &20.4$\pm$0.4   & 24.9$\pm$4.6  &40.35\\
CTS 1033          &48.2$\pm$0.4   & 48.8$\pm$0.2   & 48.4$\pm$0.2   &48.3$\pm$0.4   &-              &40.79\\
Tol 1400-397      &32.4$\pm$2.2   & 32.1$\pm$1.4   & 33.5$\pm$0.5   &34.3$\pm$0.7   &-              &40.84\\
UM 649            &27.1$\pm$2.2   & 23.1$\pm$1.4   & 25.0$\pm$0.5   &25.5$\pm$1.8   &-              &40.27\\
CTS 1034          &-              & 27.1$\pm$2.1   & 25.5$\pm$1.4   &26.6$\pm$1.8   &-              &40.38\\
II ZW 70          &21.0$\pm$1.2   & 20.0$\pm$0.5   & 20.5$\pm$0.2   &24.3$\pm$0.4   &-              &40.27\\
CTS 1035          &27.4$\pm$2.9   & 27.7$\pm$2.1   & 26.6$\pm$0.5   &23.9$\pm$2.2   &-              &40.53\\
CTS 1037          &38.4$\pm$0.4   & 38.3$\pm$0.2   & 37.5$\pm$0.2   &37.7$\pm$0.4   &-              &41.61\\
Cam 1543+0907     &30.2$\pm$0.4   & 29.1$\pm$0.2   & 28.6$\pm$0.2   &29.5$\pm$0.4   &-              &41.29\\
Tol 1924-416      &-              &-               &-               &-              & 35.4$\pm$3.5  &41.34\\
Tol 1939-419      &-              &-               & 23.6$\pm$0.9   &22.5$\pm$1.2   &-              &40.22\\
Tol 1937-423      &24.0$\pm$2.2   & 19.2$\pm$1.4   & 22.1$\pm$0.7   &21.3$\pm$0.7   & 29.6$\pm$4.6  &40.00\\
CTS 1038          &-              & 51.8$\pm$2.1   & 51.6$\pm$0.7   &-              &-              &41.98\\
CTS 1039          &40.9$\pm$0.4   & 41.7$\pm$0.2   & 41.7$\pm$0.2   &40.9$\pm$0.4   &-              &41.61\\
Tol 2010-382      &37.0$\pm$1.8   &-               & 36.7$\pm$1.4   &34.2$\pm$0.4   & 37.0$\pm$3.5  &41.34\\
Tol 2019-405      &25.8$\pm$1.8   & 24.5$\pm$0.5   & 24.7$\pm$0.7   &21.9$\pm$0.4   &-              &39.89\\
Tol 2041-394      &-              & 29.3$\pm$2.1   & 27.5$\pm$0.7   &28.5$\pm$1.8   &-              &40.41\\
NGC 6970          &-              &-               &-               &-              & 46.0$\pm$4.6  &40.77\\
MBG 20533-4410    &-              &-               &-               &-              & 55.8$\pm$4.6  &41.31\\
Tol 2122-408      &26.4$\pm$1.8   & 26.4$\pm$0.7   & 26.6$\pm$0.5   &24.5$\pm$0.4   & 25.6$\pm$4.6  &40.32\\
Tol 2138-405      &59.5$\pm$2.2   & 59.8$\pm$0.9   & 59.9$\pm$0.5   &62.1$\pm$1.8   &-              &41.92\\
Tol 2138-397      &22.2$\pm$2.2   & 24.9$\pm$1.4   & 22.2$\pm$0.7   &23.1$\pm$0.7   &-              &40.01\\
Tol 2146-391      &26.8$\pm$1.8   & 23.0$\pm$0.5   & 24.2$\pm$0.2   &25.6$\pm$0.4   &-              &40.75\\
MBG 21567-1645    &-              &-               &-               &-              &103.5$\pm$4.6  &41.17\\
MBG 22012-1550    &-              &-               &-               &-              & 92.7$\pm$4.6  &41.71\\
IC 5154           &-              &-               &-               &-              & 33.2$\pm$3.5  &40.30\\
ESO 533-G 014     &-              &-               &-               &-              & 14.5$\pm$3.5  &40.16\\
MCG -01-57-017    &-              &-               &-               &-              & 22.0$\pm$3.5  &40.06\\
Tol 2240-384      &49.3$\pm$2.2   & 50.3$\pm$0.7   & 51.8$\pm$0.2   &51.6$\pm$0.4   &-              &41.85\\
MBG 23121-3807    &-              &-               &-               &-              & 27.3$\pm$4.6  &40.04\\
Tol 2326-405      &-              &-               & 37.3$\pm$0.7   &43.7$\pm$2.2   &-              &41.56\\
UM 167            &-              &-               &-               &-              & 73.2$\pm$3.5  &41.39\\
UM 191            &30.3$\pm$2.2   &-               & 39.2$\pm$0.9   &32.0$\pm$0.4   & 33.3$\pm$3.5  &40.90
\enddata
\end{deluxetable}

The error in luminosity is very difficult to evaluate precisely
since it depends on several factors. Firstly, slit spectrophotometry
for extended objects suffers intrinsically from the aperture effect
(see Section~\ref{sec4_2}). In addition, individual
spectrophotometric errors were not provided by KTC, instead, they
provided the random error of the flux measurements of the line
emissions as being in the range $5\%-40\%$ for $F>10^{-15}$ erg
s$^{-1}$ cm$^{-2}$, decreasing for high intensity lines. However,
they compared their results with those in T91 showing good agreement
with their spectrophotometry. A point in favor of these new
spectrophotometric data is their homogeneity on technique and
instrumentation. In view of this, to obtain quantitative results in
the analysis sections we will only consider a sample of homogeneous
spectrophotometric data, i.e. luminosities derived from KTC data and
only a few newly available data with similar techniques.

\subsection{The $L$-$\sigma$ Relation}\label{sec3_4}

\begin{figure*}[ht]
\epsscale{0.35} \plotone{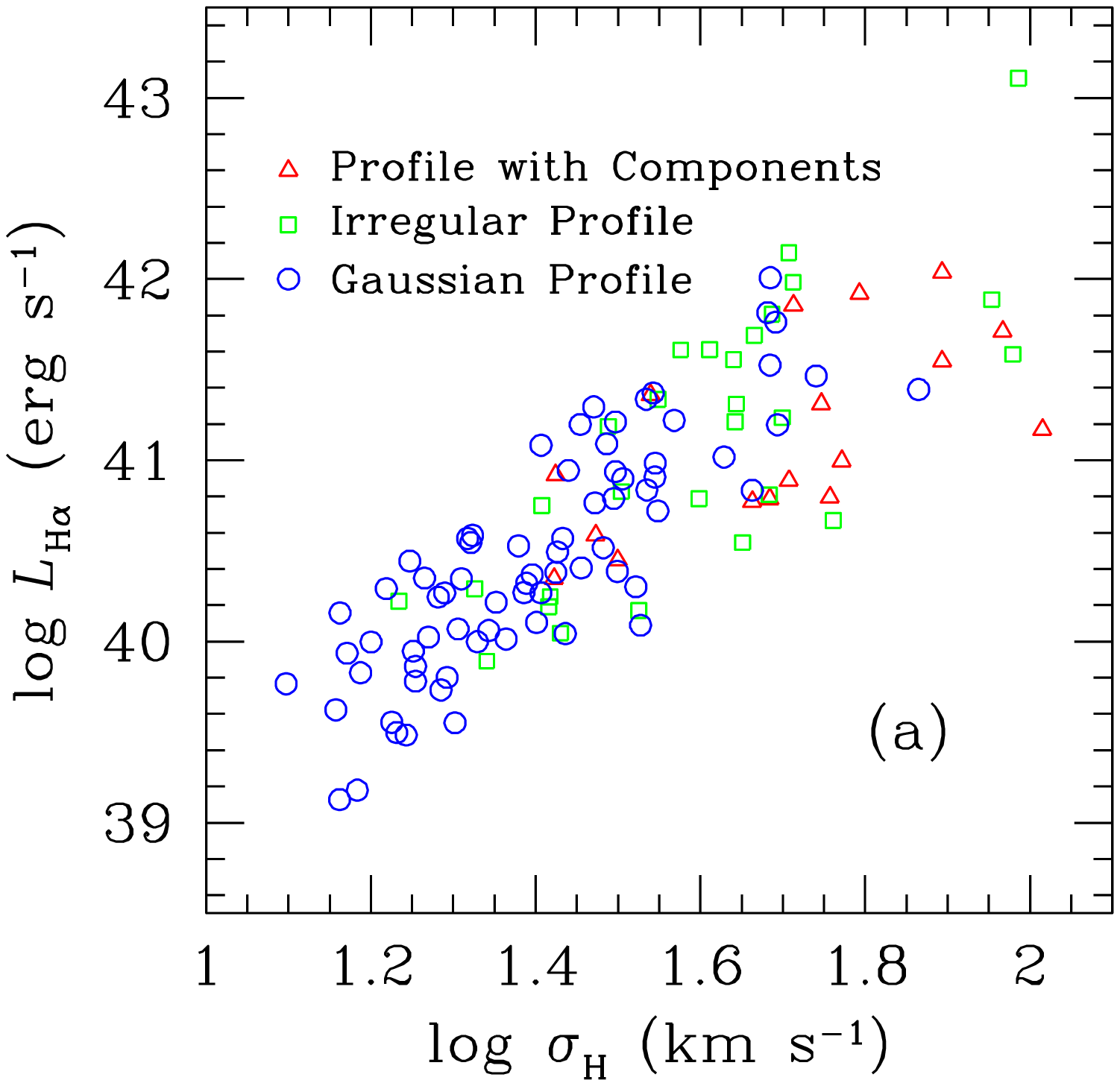}\\
\epsscale{0.6}\plottwo{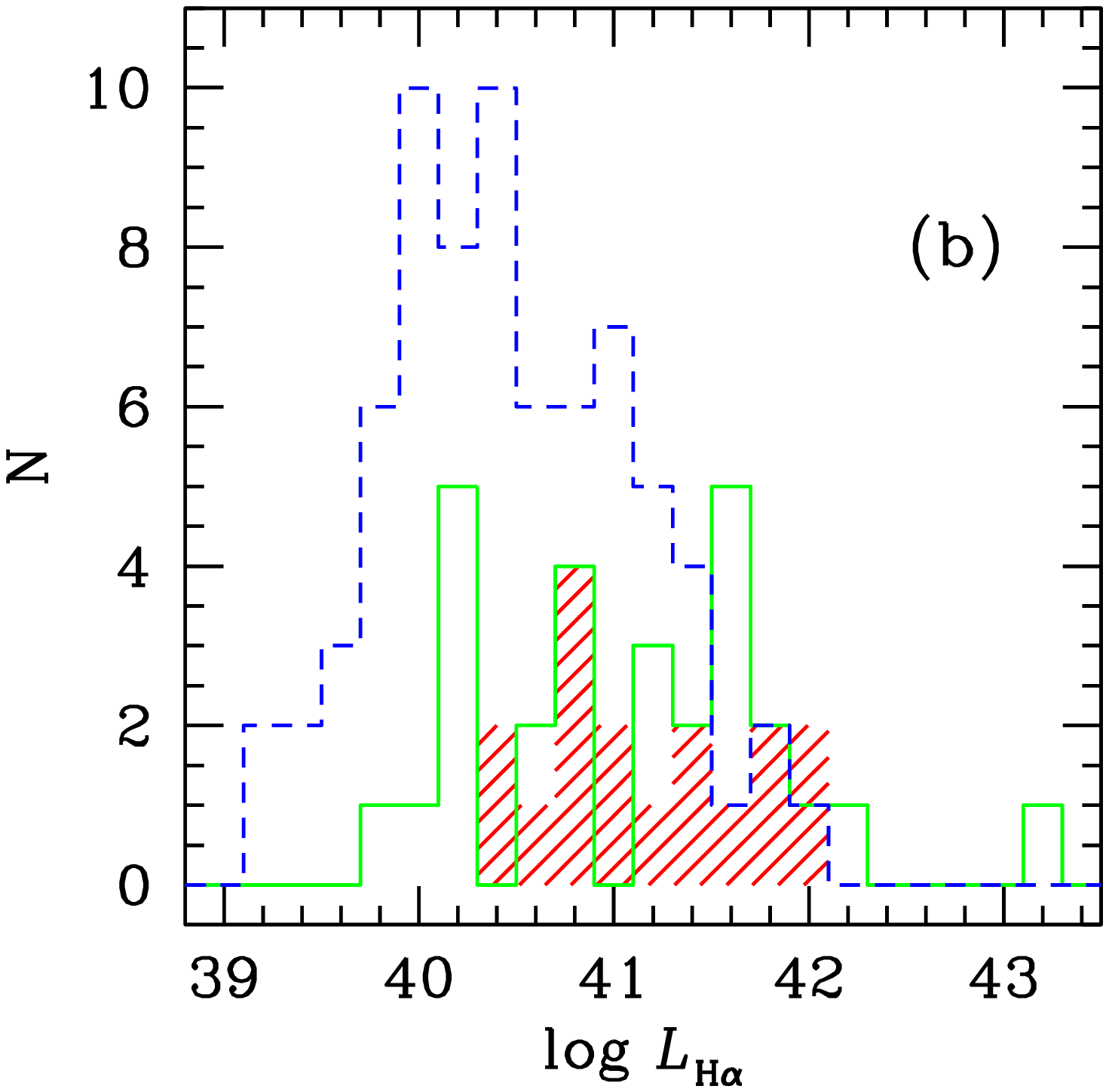}{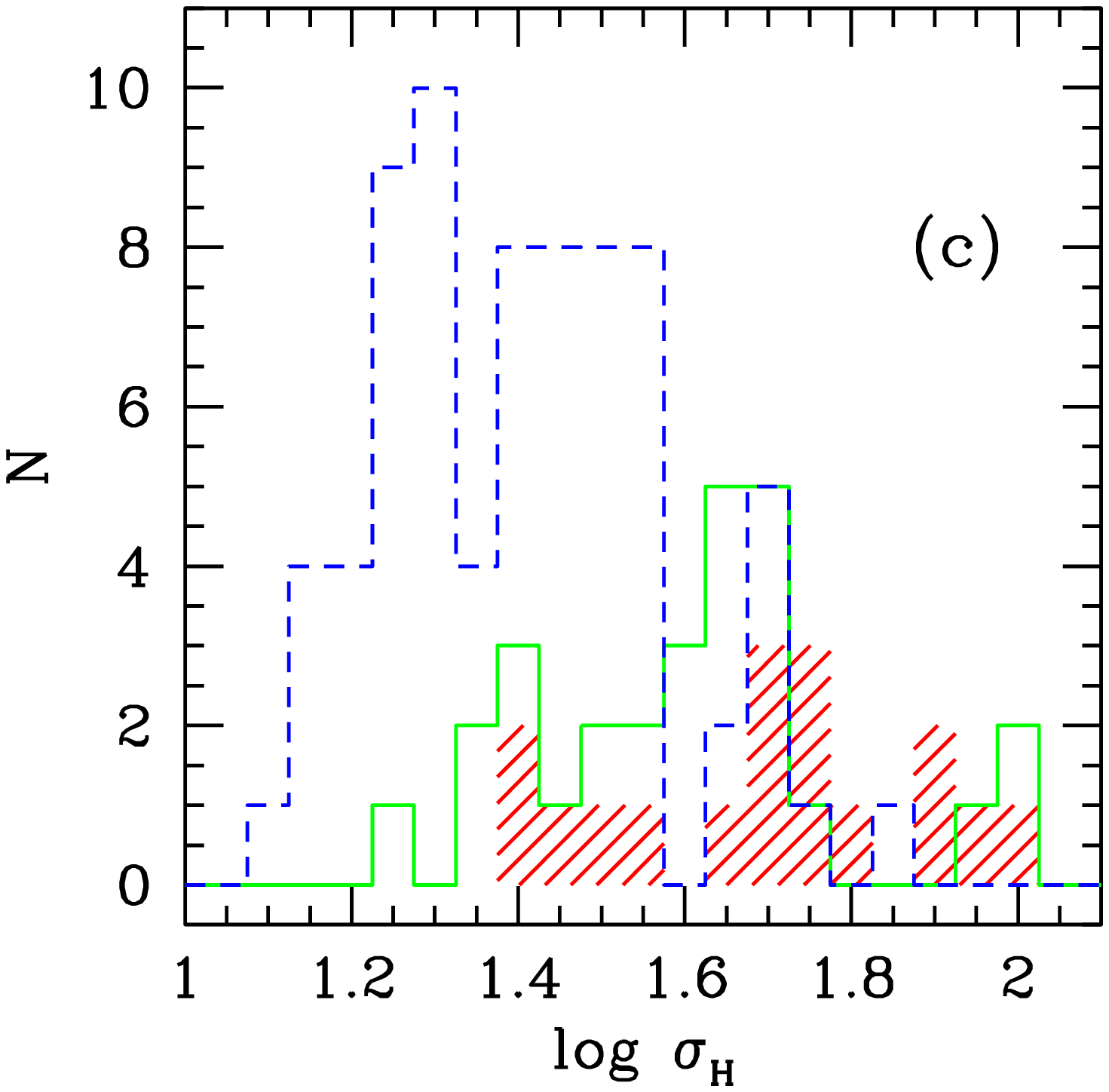} \caption{(a) The
$L_{\scriptsize\mbox{H}\alpha}$-$\sigma_{\scriptsize\mbox{H}}$ for
117 galaxies with spectrophotometry obtained; (b) the luminosity
distribution; and (c) the velocity dispersion distribution. The
colour codes for each class of galaxy, (G, I and C) are as given in
the label of plot (a). \label{fig6}}
\end{figure*}

Figure~\ref{fig6} (a) shows the $L$-$\sigma$ relation in logarithmic
units for 117 galaxies all observed with FEROS and Coud\'e
spectrographs. We also show the $L_{\scriptsize\mbox{H}\alpha}$
distribution in (b) and $\sigma$ in (c) for each subsample of line
profile class (G, I and C), which will help the visualization of the
subsample properties highlighted below. We have plotted in
Figure~\ref{fig6} (a) all galaxies for which we have obtained
spectrophotometric data. It shows the strong correlation between the
nebular luminosity and its velocity dispersion. It confirms, once
more, the existence of this relation for H{\sc ii}Gs, but now for a
sample of more than one hundred galaxies, doubling the old samples
studied in the past. Galaxies which present irregular profiles and
especially those showing profiles with components are systematically
concentrated in the high velocity dispersion (log $\sigma$ $>$ 1.6)
and luminosity (log $L_{\scriptsize\mbox{H}\alpha}$ $>$ 41) regions
of the $L$-$\sigma$ plot (top right). On the other hand, galaxies
showing Gaussian profile are more concentrated in the regime log
$\sigma$ $<$ 1.6 (bottom left), but they span from typical values of
$\sigma$ found for single GH{\sc ii}Rs, $12-30$ km s$^{-1}$
($1.1<\log\sigma<1.5$), to $\sim60$ km s$^{-1}$ ($\log\sigma \sim
1.8$). It is clear in Figure~\ref{fig6} (a) that galaxies showing
irregularities and multiple components in their emission line
profiles contribute to flattening the $L$-$\sigma$ relation
resulting in its curved shape toward high $L$ and $\sigma$
values. This behavior was in fact predicted by MTM but they only
identified two galaxies that clearly disagree of the mean line (Tol
0226-390 and Tol 0242-387) due with their sample size. These two
galaxies are also presented in our sample and were classified as I.
MTM restricted their analysis to those objects which present
$W_{\scriptsize\mbox{H}\beta}>30$\AA. We will show below that this
criterion seems to be also efficient to select galaxies with the
most Gaussian line profiles.

In order to minimize the uncertainties due to a heterogeneous data
set, we further analyze the $L$-$\sigma$ relation for those galaxies
that have line-widths measured from FEROS data and spectrophotometry
compiled from KTC's work. An additional 4 objects (UM 382,
CTS 1027, II Zw 70 and Tol 2138-405) that have good new
spectrophotometry are included.

\begin{figure*}[ht]
\epsscale{0.75} \plotone{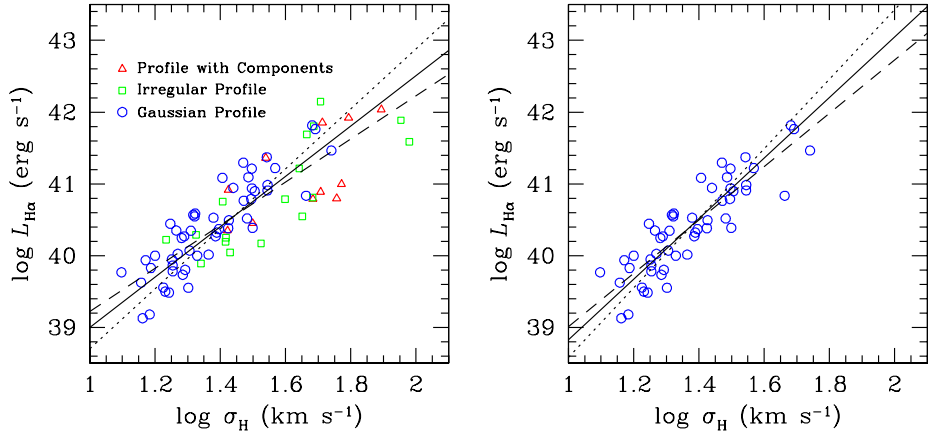} \caption{The
$L_{\scriptsize\mbox{H}\alpha}$-$\sigma_{\scriptsize\mbox{H}}$
relation for all galaxies with homogeneous spectrophotometry (81
objects G, I and C, \textit{left}) and only for those showing
regular Gaussian profiles (53 objects G, \textit{right}). The dashed
line represents OLS(Y$|$X), the dotted line represents OLS(X$|$Y)
and the solid line represents the bisector fit. All regression
coefficients are presented in Table~\ref{tab6}. \label{fig7}}
\end{figure*}

Figure~\ref{fig7} shows the $L$-$\sigma$ relation for the
homogeneous sample described above including galaxies with
close to Gaussian emission line profiles (81 objects).
The regression fits for the total sample and only for the G
subsample (53 objects) are presented in Table~\ref{tab6}. The class
of ordinary least-square fits (OLS) used in this work is appropriate
to problems where the intrinsic scatter dominates the errors arising
from the measurement process \citep{iso90}. We argue that the
OLS(Y$|$X) is the most appropriate to describe the $L$-$\sigma$
relation for our data set and is the best to be compared with
previous calibrations \citep[MTM,][]{tel01}. A second point is that
the uncertainties in $\sigma$ are much smaller than in luminosity,
which justify the first to be treated as an independent parameter in
a direct linear regression. Nevertheless the other fits give us
an idea of maximum limits of the regression coefficients (slope
and zero point).

\begin{figure}[hd]
\epsscale{0.4} \plotone{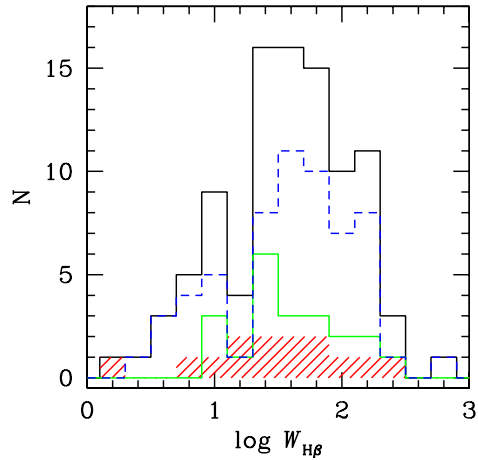} \caption{The H$\beta$ equivalent
width distributions (in log) for each class of galaxies namely G
(53), I (17) and C (11) presented in Figure~\ref{fig7}. The
solid line represents the distribution including all subsamples
(81).\label{fig8}}
\end{figure}

It is clear that the $L$-$\sigma$ relation including only G galaxies
is tighter and steeper than the one including the whole sample of
H{\sc ii}Gs. It suggests that the Gaussianity of the emission line
profiles in these systems may be associated with the nature of the
$L$-$\sigma$ relation. Figure~\ref{fig8} shows the
$W_{\scriptsize\mbox{H}\beta}$ distribution for each class G, I and
C. Note that most G galaxies are concentrated in the region of high
$W_{\scriptsize\mbox{H}\beta}$ ($>30$\AA). Thus a sample selection
criterion based on high $W_{\scriptsize\mbox{H}\beta}$
would be efficient to select also galaxies with the most Gaussian
profiles.
The consequences of this result to the nature of the $L$-$\sigma$
relation are profound. Since $W_{\scriptsize\mbox{H}\beta}$ is an
age indicator of the current starburst, resultant Gaussian line
profiles may be associated with the youngest systems. If the
$L$-$\sigma$ relation for H{\sc ii}Gs has in fact an upper
envelope described by those galaxies with zero-age and maximum
$W_{\scriptsize\mbox{H}\beta}$ values, it should be populated only
by H{\sc ii}Gs showing the most Gaussian line profiles.
In Section~\ref{sec4_3} we will be more rigorous in selecting a
subsample of G galaxies with the most Gaussian line profiles by a
semi-quantitative criterion in order to investigate if it
has additional consequences on the slope and scatter of the
$L$-$\sigma$ relation.

\begin{table}[ht]
\begin{center}

\caption{Regressions for log $L_{\scriptsize\mbox{H}\alpha}$
versus log $\sigma_{\scriptsize\mbox{H}}$.\label{tab6}}

\begin{tabular}{lccc}\\ \tableline\tableline
Linear Regression        &Intercept   &Slope    &RMS\\
                      &(A)          &(B)                    &\\
\tableline
\multicolumn{3}{l}{All galaxies (81 objects)}\\
\multicolumn{2}{l}{Pearson correlation coefficient ($r$) = 0.85}\\
&\multicolumn{2}{c}{log $L$ = A + B $\times$ log $\sigma$}\\
OLS(Y$|$X) ..............           &36.21 $\pm$ 0.32    &3.01 $\pm$ 0.23   &0.37\\
OLS(X$|$Y) ..............           &34.52 $\pm$ 0.38    &4.18 $\pm$ 0.27   &\\
OLS Bisector .........              &35.49 $\pm$ 0.32    &3.51 $\pm$ 0.23   &\\
\tableline


\multicolumn{3}{l}{Galaxies with Gaussian profiles (53 objects)}\\
\multicolumn{2}{l}{$r$ = 0.88}\\
OLS(Y$|$X) ..............           &35.29 $\pm$ 0.42    &3.72 $\pm$ 0.31   &0.31\\
OLS(X$|$Y) ..............           &33.73 $\pm$ 0.47    &4.85 $\pm$ 0.34   &\\
OLS Bisector .........              &34.61 $\pm$ 0.41    &4.22 $\pm$ 0.30   &\\
\tableline
\multicolumn{3}{l}{More restrictive subsample (37 objects)}\\
\multicolumn{2}{l}{$r$ = 0.90}\\
OLS(Y$|$X) ..............           &34.80 $\pm$ 0.41    &4.14 $\pm$ 0.29   &0.29\\
OLS(X$|$Y) ..............           &33.45 $\pm$ 0.53    &5.13 $\pm$ 0.38   &\\
OLS Bisector .........              &34.19 $\pm$ 0.43    &4.58 $\pm$ 0.30   &\\

\tableline

\end{tabular}

\end{center}

\end{table}

\section{Data Analysis}\label{sec4}

\subsection{Aperture Effects}\label{sec4_2}

The $L$-$\sigma$ relation may be in principle affected by the
aperture effect in two ways. Firstly, if the observation comes from
fixed-slit spectroscopy ($\sim$ 2{\arcsec}) as in this work, nearby
objects should have their fluxes underestimated and hence their
luminosities too. Furthermore, extended and more compact objects,
both at the same distance, may suffer differentially from
this effect. It may also have implications in determining their
physical conditions \citep{kew05}.
An aperture correction factor may be applied to our calibration by
using the available data in \cite{lag07}. They observed a sample of
H{\sc ii}Gs with H$\beta$ narrow band imaging and analyzed the
surface photometry as compared with the spectrophotometry of
\cite{keh04}. Many of the galaxies in their sample are also
part of our present sample.
Taking into account the observational and calibration errors in
their analysis, there is a constant offset of the order of the
observational scatter in the comparison that can be added to our
derived $L_{\scriptsize\mbox{spec}}$: $\Delta
L_{\scriptsize\mbox{H}\beta} \sim 0.25$ with an RMS = 0.2,
comparable to the RMS of our $L$-$\sigma$ relation. We have not done
this a priori, since this correction would only introduce additional
observational scatter, masking the contribution of the physical
parameters in the analysis of the manifold of H{\sc ii}Gs. This
effect is expected to be negligible for galaxies beyond $> 100$ Mpc
according to this analysis, and an improvement in this calibration
will only come when this relation is verified for a sample of H{\sc
ii}Gs beyond the local supercluster with new spectrophotometric data
sets and new high resolution spectroscopy.
We have also verified that a correlation of the narrow band surface
photometry of the main star forming knot seems to be tighter than
with the narrow band surface photometry of the integrated galaxy.
This may be due to the mixed morphology of the galaxies as a
function of luminosity \citep[see][]{lag07,tmt97}, but also due to
the very nature of the $L$-$\sigma$ relation, in the sense that it
may be more closely related to the local gravitational potential of
the starburst, rather than the global dynamics. The relation still
holds for the determination of galactic masses if we assume that
more massive galaxies host more massive starbursts in a homologous
way. These issues remain to be investigated in more detail.

Multiplicity of star-forming regions and the aperture effect can
also introduce a bias in $\sigma$ determination | a single
observation integrates light from more than one starburst region in
the same galaxy. This could introduce a systematic velocity
component in the integrated line profiles since Super Star Clusters
and their associated GH{\sc ii}Rs may present relative radial
velocities that would add light in a non-trivial way. Although
multiplicity does not necessarily preclude Gaussianity, it seems to
be usually associated with asymmetric line profiles \citep{bor09}.
We have also found very compact objects presenting line profiles
with multiple components (e.g. CTS 1033 shown in
Figure~\ref{fig12}). However, most of the galaxies classified as C
are not compact, but extended systems frequently associated to
nuclear starburst galaxies, where the systemic rotational component
dominates the features in emission line profiles.

The multiplicity effect seems to be inherent in the $L$-$\sigma$
relation causing no strong bias, otherwise the $L$-$\sigma$ relation
would be not verified, even in the short redshift range of our
sample. The physical sizes covered by the observations span from
a few hundred parsecs in nearby objects (e.g. UM 461 and
MRK 36), characterizing sizes of single GH{\sc ii}Rs, to a
few kiloparsecs (e.g. CTS 1008 and Cam 08-28A). In addition,
multiplicity and aperture effects can be greatly reduced by
selecting only the galaxies showing the most Gaussian line profiles.

\subsection{Gaussian Profile Galaxies}\label{sec4_3}

In order to find a more homogeneous sample and, in addition, test
the visual classification presented in Section~\ref{sec3_1}, we have
further adopted a semi-quantitative criterion to search for
galaxies with the most Gaussian emission line profiles. These seem
to be the ones that show the lower scatter in the $L$-$\sigma$
plane, as suggested in Figure~\ref{fig7}, and therefore deserve
special attention. The quantitative estimators that we used were the
skewness ($\xi$) and kurtosis ($\kappa$) of the emission line
profiles \citep*[see][for the use of these estimators in one-point
velocity statistics of star-forming regions]{mie99}.
These estimators are related to the third ($m_{3}$) and fourth
($m_{4}$) moments of the distribution through the formulae
\[\xi = \frac{m_{3}}{s^{3}}\] and
\[\kappa = \frac{m_{4}}{s^{4}},\] where $s$ is the
standard deviation of the distribution and $m_{3}$ and $m_{4}$ their
higher moments\footnote{We used $s$ to denote the standard
deviation in order not to be confused with the velocity dispersion
derived ($\sigma$).}. For practical cases concerning real data
spectra, where $\lambda$ is the wavelength and $y$ is the flux, the
formulae for these estimators can be written by:
\[\mbox{mean}=\mu=\frac{\sum \lambda \cdot y}{\sum y},\]
\[\mbox{standard deviation}=s=\sqrt{\frac{1}{\sum
y}\sum (\lambda-\mu)^{2}\cdot y},\]
\[\mbox{skewness}=\xi=\frac{1}{\sum y}\frac{\sum (\lambda-\mu)^{3}\cdot
y}{s^{3}},\]
\[\mbox{kurtosis}=\kappa=\frac{1}{\sum y}\frac{\sum (\lambda-\mu)^{4}\cdot
y}{s^{4}}.\] Before we proceed the calculation of $\xi$ and $\kappa$
estimators we smoothed all FEROS spectra with a box of 11 pixels
(FEROS's scale is 0.03\AA/pix) using SPLOT. Since our aim was to
estimate the global shape of the emission lines, the smooth
procedure is very useful to eliminate high frequency noise. The most
Gaussian profile galaxies were selected as the ones which met the
criterion of symmetry\footnote{$\xi$=0 and $\kappa$=3 for
the Gaussian distribution.}: $\mid\xi\mid$ $<$ 0.1 and 2.9 $<$
$\kappa$ $<$ 3.1. The estimators were calculated in a window
centered at the mean, obtained from the single Gaussian fit
(centroid), and defined as the one in which the line intensity falls
to 20\% of its peak value. This methodology ensures that the
estimators show a good consistency between them, however the size of
integration window over $\lambda$ interval is somewhat a matter of
definition which should be based on the quality of the data.

\begin{figure*}[ht]
\epsscale{0.32} \plotone{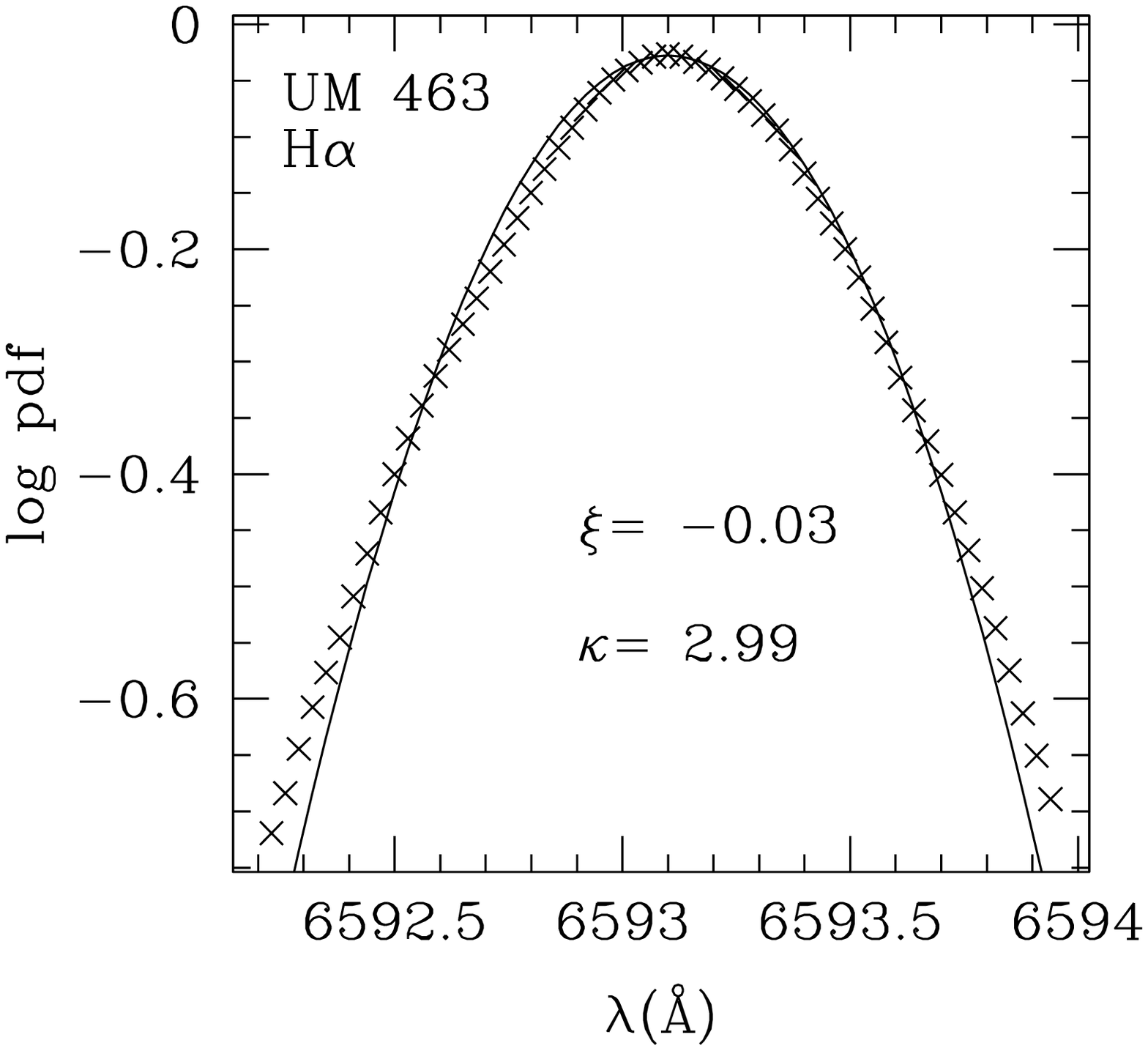}\plotone{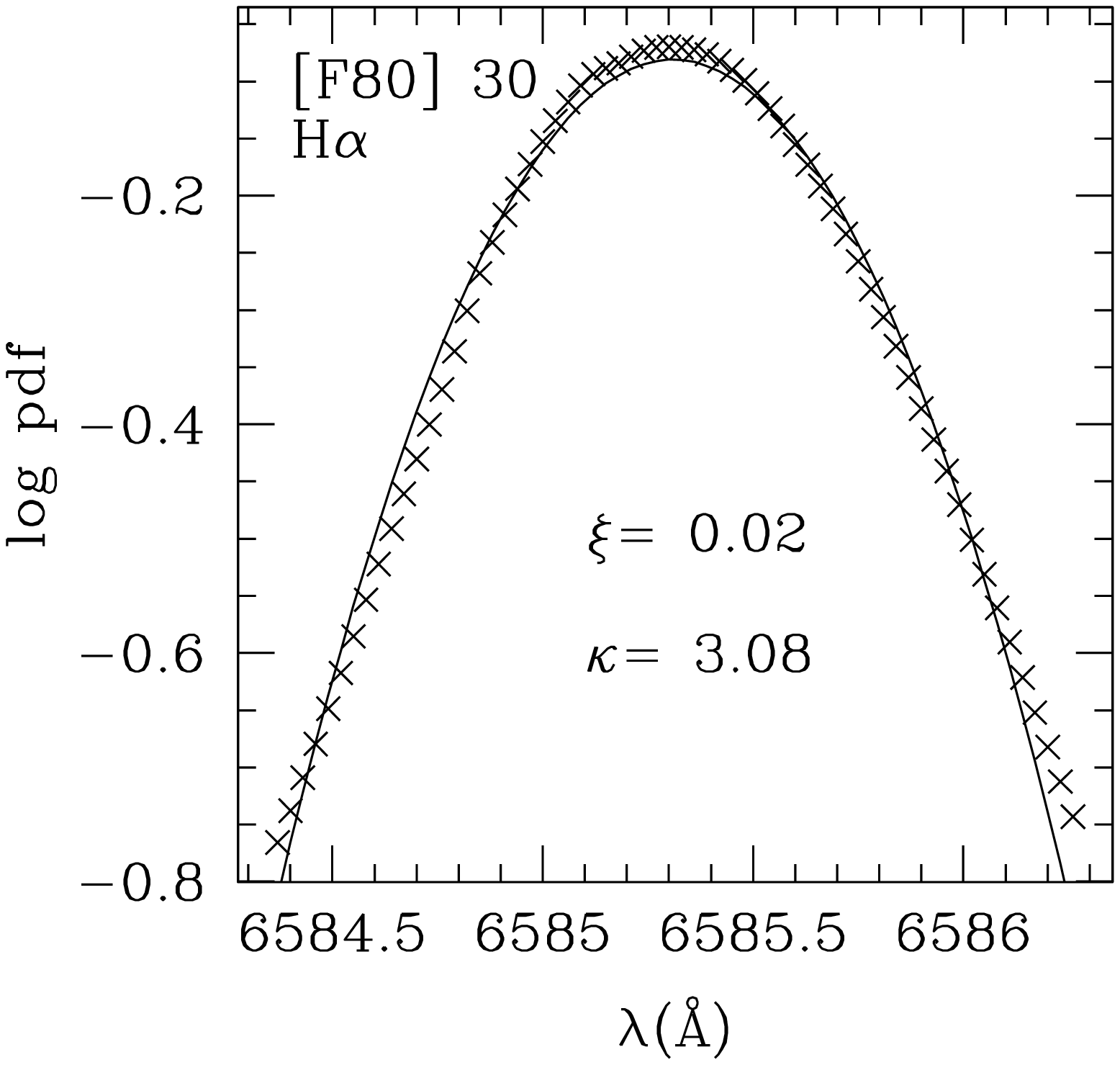}\\
\plotone{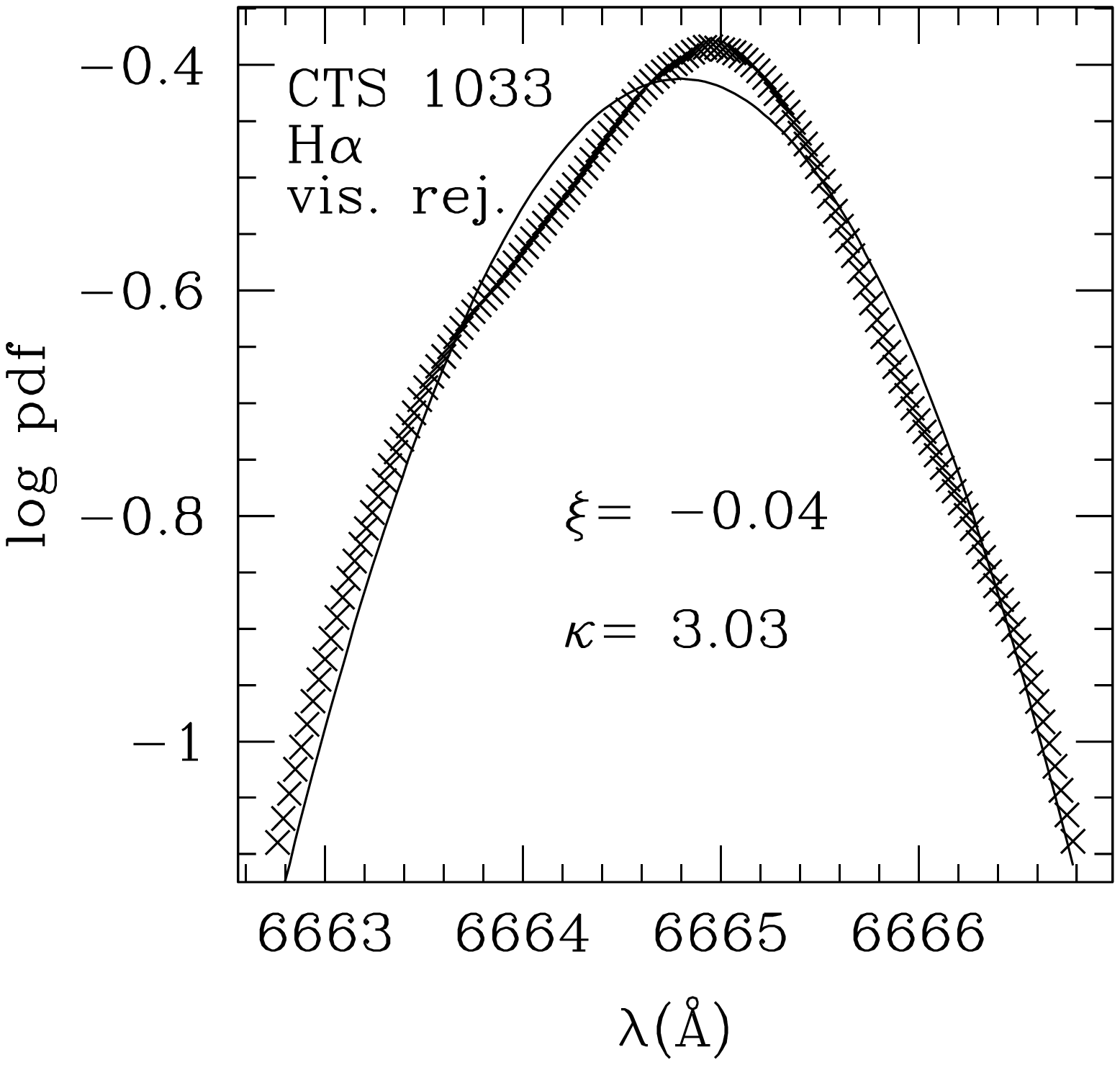}\plotone{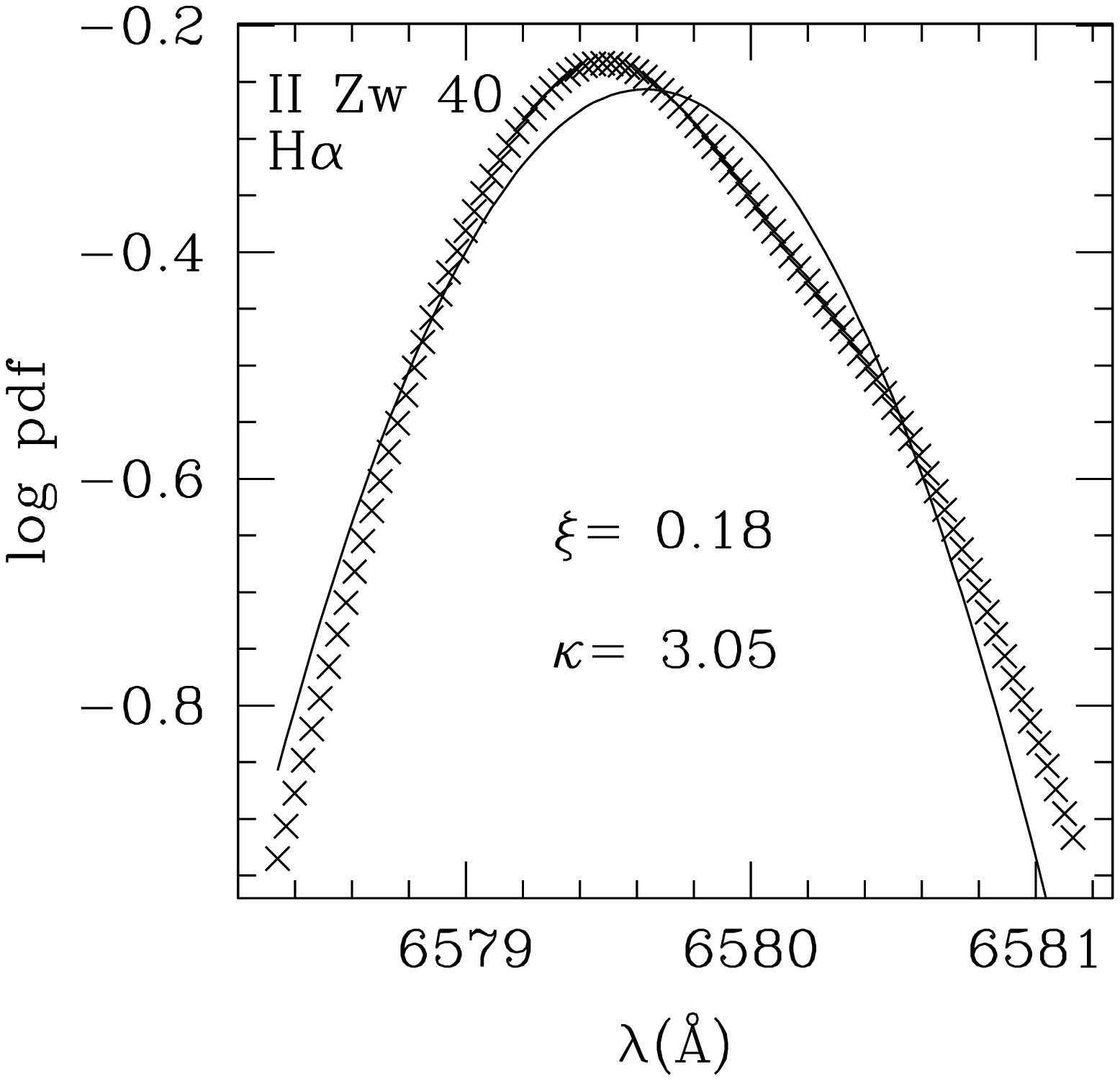}\caption{Examples of emission
line profiles in linear-logarithmic axes. UM 463 (upper left) and
[F80] 30 (upper right) met the criterion of symmetry described in
the text (G$^{\prime}$), (bottom left) CTS 1033 was qualitatively
visually rejected by presenting multiple components and (bottom
right) II Zw 40 does not meet the symmetry criterion. Measurements
of $\xi$ and $\kappa$ estimators are shown in each panel.
Solid lines represent single Gaussian fits to the observed profiles
(crosses). \label{fig12}}
\end{figure*}

Only FEROS galaxies with S/N in the line greater than 30 were
analyzed by this method. Emission line profiles of 49 galaxies met
the above criterion of symmetry. Two examples (UM463 and
[F80] 30) are presented in Figure~\ref{fig12} (upper panels),
including their shape estimators $\xi$ and $\kappa$. We also present
the H$\alpha$ emission line profile of II Zw 40 ($\xi$=0.18,
$\kappa$=3.05) as an example in which the criterion of
symmetry was efficient to reject profiles with strong asymmetries
(Figure~\ref{fig12} bottom right). Most estimators were calculated
using the H$\alpha$ line and only few using [OIII]
5007{\AA} due to the presence of bad pixels or significantly lower
S/N in H$\alpha$. A comparison with the G subsample classified
visually (Table~\ref{tab2}) showed good agreement except for six
galaxies that were previously classified as I or C, namely Tol
0104-388, Tol 0957-278, CTS 1030, CTS 1033, CTS 1037 and CTS 1039.
These objects were thus visually rejected. We kept the visual
classification for these galaxies since
the first two objects present two intense components which are
identified in the other emission lines observed, while CTS 1033
presents at least three emission components (Figure~\ref{fig12}
bottom left). The last is a very special case in which asymmetries
in the integration window are compensated resulting in estimators
that mimic a symmetric profile ($\xi$=-0.04, $\kappa$=3.03). The
other three objects present intense broad components not
well evaluated in the integration window though they are symmetric.
Thus 43 galaxies compose a more restrictive subsample of the
galaxies showing the most Gaussian line profiles and they were
assigned as G$^{\prime}$ in Table~\ref{tab2}.

The need for a qualitative analysis of the emission line profiles is
in fact a limitation of the method applied to our data. Since the
S/N ratio is relatively low in the wings of the line, we need to
limit the integration window to a fixed proportion of the
peak value (see above), in order to keep the consistency between the
values from very different S/N lines. Nevertheless, the method
provides a way to determine a homogeneous sample of galaxies based
on their kinematic properties\footnote{Complete data from this
analysis is available at http://www.on.br/astro/etelles/lsigma.}.

Table~\ref{tab6} shows the regression fits for the $L$-$\sigma$
relation considering the more restrictive subsample G$^{\prime}$.
From 43 objects identified as G$^{\prime}$, 37 had homogeneous
observational data as described in Section~\ref{sec3_4} and were
used to derive the calibration coefficients.
Although this sample presents few objects in the range 1.6 $<$ $\log
\sigma$ $<$ 1.8, the OLS fits show that the $L$-$\sigma$ is steeper
than the one for the whole G subsample (53 objects). We conclude
that the galaxies showing the most Gaussian line profiles show the
$L$-$\sigma$ relation that could be identified as an upper envelope
$L \propto \sigma^{4}$ with minimum scatter $\delta\log
L_{\scriptsize\mbox{H}\alpha}=0.29$.

\subsection{The Second Parameter}\label{sec4_4}

\cite{ter81} and \cite{mel88} have suggested that the dependent
parameter $L_{\scriptsize\mbox{Balmer}}$ can be predicted by
$\sigma$ and O/H, which should define the manifold of H{\sc ii}Gs.
\cite{tel93} also showed that the structural parameter namely radius
of the burst should be considered as a second parameter in
$L$-$\sigma$, acting remarkably similarly to effective radius
in the fundamental plane of elliptical galaxies. It is
crucial to investigate the existence of an independent second
parameter (or a third) in the $L$-$\sigma$ relation not only to
obtain a precise distance indicator but also to understand the
physics behind this scaling relation. If the underlying relation is
in fact that between mass and luminosity, as proposed by these
authors, then $\sigma$ can be used properly to obtain dynamical
masses of these systems with great relevance for the high redshift
studies of H{\sc ii}Gs counterparts, such as the Lyman Break
Galaxies \citep[][and references therein]{low09}.

\begin{table}[ht]



\caption{Correlation Matrix.\label{tab7}}

\begin{tabular}{lccccc}\\
\multicolumn{6}{l}{PCA $L_{\scriptsize\mbox{H}\alpha}$,
$\sigma_{\scriptsize\mbox{H}}$, O/H,
$W_{\scriptsize\mbox{H}\beta}$ and [OIII]/[OII] (95 objects)}\\
\hline
\\
$\log$ $L_{\scriptsize\mbox{H}\alpha}$   &1       &      &       &       &       \\
$\log$ $\sigma_{\scriptsize\mbox{H}}$    &0.82    &1     &       &       &       \\
$\log$ (O/H)                             &0.22    &0.47  &1      &       &       \\
$\log$ $W_{\scriptsize\mbox{H}\beta}$    &0.13    &-0.15 &-0.58  &1      &       \\
$\log$ [OIII]/[OII]                      &-0.03   &-0.32 &-0.69  &0.86   &1      \\
&$\log$ $L_{\scriptsize\mbox{H}\alpha}$  &$\log$ $\sigma_{\scriptsize\mbox{H}}$  &$\log$ (O/H)  &$\log$ $W_{\scriptsize\mbox{H}\beta}$ &$\log$ [OIII]/[OII]\\
\\
\hline
\end{tabular}

\end{table}

\begin{table}[hd]


\centering

\caption{Eigenvectors and eigenvalues.\label{tab8}}

\begin{tabular}{lccc}\\ \hline\hline
&\multicolumn{3}{c}{Principal component}\\
&I &II &III\\ 
\hline
\\
$\log$ $L_{\scriptsize\mbox{H}\alpha}$    & 0.39   &-0.88     &-0.11\\  
$\log$ $\sigma_{\scriptsize\mbox{H}}$     & 0.66   &-0.70     &-0.08\\  
$\log$ (O/H)                              & 0.86   & 0.07     & 0.51\\  
$\log$ $W_{\scriptsize\mbox{H}\beta}$     &-0.77   &-0.54     & 0.25\\  
$\log$ [OIII]/[OII]                       &-0.88   &-0.38     & 0.16\\  
\\\hline
Eigenvalues                               &2.69    &1.70     &0.37\\   
\% variance                               &53.9\%  &34.0\%   &7.3\%\\  
\hline

\end{tabular}

\end{table}

We have used the Principal Component Analysis (PCA) to investigate
the relative dependence of the $L$-$\sigma$ relation on a possible
second parameter based on our new data set, following the early
analysis by MTM. Let us first analyze the largest sample possible
considering the 5 parameters (variables) |
$L_{\scriptsize\mbox{H}\alpha}$, $\sigma_{\scriptsize\mbox{H}}$,
O/H, $W_{\scriptsize\mbox{H}\beta}$ and [OIII]/[OII]. The sample
includes $n=95$ objects with all these quantities known
(Tables~\ref{tab3} and~\ref{tab5}). All quantities were considered
in logarithmic scales. Table~\ref{tab7} shows the lower half
triangle of the correlation matrix. Visual analysis of
Table~\ref{tab7} shows that strong correlations are found
between $W_{\scriptsize\mbox{H}\beta}$ and [OIII]/[OII] ($r=0.86$)
and $L_{\scriptsize\mbox{H}\alpha}$ and
$\sigma_{\scriptsize\mbox{H}}$ ($r=0.82$). It is also worth noting
that the correlations between $L_{\scriptsize\mbox{H}\alpha}$ and
[OIII]/[OII], $L_{\scriptsize\mbox{H}\alpha}$ and
$W_{\scriptsize\mbox{H}\beta}$ and $\sigma_{\scriptsize\mbox{H}}$
and $W_{\scriptsize\mbox{H}\beta}$ are near zero. The correlations
of O/H and $W_{\scriptsize\mbox{H}\beta}$ and O/H and [OIII]/[OII]
are negative, while O/H and $L_{\scriptsize\mbox{H}\alpha}$ and O/H
and $\sigma_{\scriptsize\mbox{H}}$ are positive.

The first three principal components (PCs) with their
loadings\footnote{The loadings represent correlations between the
PCs and the original variables. PCs are the new set of variables
uncorrelated by definition and written as linear combinations of the
original variables.} obtained from the correlation matrix, as well
as the eigenvalues ($l$) and the respective individual percentages
($l/5$) of total variance (5) are presented in Table~\ref{tab8}.
Nearly 88\% of the variance among the 95 sample points lies in only
two dimensions. The number 2 of dimensions of the manifold can be
easily verified due to the bimodal behavior of the variance
\citep{jol02}. The first two principal components (PCI and PCII)
present eigenvalues well above 1 (or 1/5=20\%), while the others lie
well below this value. In addition, one can estimate the confidence
interval for the eigenvalue 1 using the formula
$\sqrt{2l^{2}/(n-1)}$ which provides $l=1.00\pm0.15$ (or
$20\pm3\%$), showing that PCII with 34\% and PCIII with 7.3\% are
far from $l=20\pm3\%$. This test is justified since in PCA all
variables have (0,1)-normalization, so any PC with an eigenvalue
less than 1 is not worth consideration. This criterion to
decide the dimensionality of the manifold on the space parameters,
also called ``eigenvalue-one'', has gained almost universal
acceptance (see \cite{bro73} and \cite{buj81} for the earliest
papers applying the PCA technique to galaxy samples).

Since PCI and PCII contain no more than 88\% of the total variance
and other PCs present very small eigenvalues, we show that the
addition of O/H, $W_{\scriptsize\mbox{H}\beta}$ and [OIII]/[OII]
does not contribute to explain the total variance of the space
parameters and hence they can not be used simultaneously to reduce
the scatter in the $L$-$\sigma$ relation. These three well known
indicators of the physical conditions in star-forming regions are
strongly correlated with each other and the correlations found here
are in good agreement with early findings \citep[e.g.][]{cam86}. The
interrelationships between them are known and of intense debate in
literature \citep[][and references therein]{ter04,hoy06}. It
introduces a problem commonly faced by the PCA technique and
multiple regression known as collinearity (or multicollinearity).
Multicollinearities are often indicated by large correlations
between subsets of the variables which can be seen in the
correlation matrix for variables O/H, $W_{\scriptsize\mbox{H}\beta}$
and [OIII]/[OII] (Table~\ref{tab7}). These three parameters are the
ones that have high loadings of PCI, hence they are variables which
are most closely related to PCI (Table~\ref{tab8}). On the other
hand, the high loadings in PCII identify
$L_{\scriptsize\mbox{H}\alpha}$ and $\sigma_{\scriptsize\mbox{H}}$
as the parameters most closely related to PCII.

The main result of the above analysis is that PCI can be thought as
a measure of physical conditions in H{\sc ii}Gs, whereas PCII is a
measure of the strength of the $L$-$\sigma$ relation. Note that we
do not invoke any prior knowledge about the physical origin of the
$L$-$\sigma$ relation. As PCs are pairwise uncorrelated, the
conclusion is that physical conditions are responsible for part of
the observed scatter in the $L$-$\sigma$ relation, though
the three parameters studied here can not be used simultaneously to
explain that. Note that $\sigma_{\scriptsize\mbox{H}}$ presents its
highest loading in PCII, which suggests that it is not a primary
consequence of internal physical conditions (measured by these three
parameters), and is possibly not controlled by subsequent mechanical
feedback processes due to massive stellar evolution. However, the
weaker but real correlation between $\sigma_{\scriptsize\mbox{H}}$
and O/H ($r=0.47$, see Table~\ref{tab7}) is responsible for the
non-negligible loading for $\sigma_{\scriptsize\mbox{H}}$ in PCI
(0.66). It shows that O/H introduces some degree of
multicollinearity as an additional independent parameter in
the $L$-$\sigma$ relation. Although a high degree of
multicollinearity does not violate the assumptions of the regression
model, it influences the variance of the estimated regression
coefficients (partial slopes and intercept). We will present the
$\sigma$-$Z$ relation in the context of the discussion about the
internal dynamics of H{\sc ii}Gs in Section~\ref{sec5_1}.

For now, we are not interested in fully predicting the
dependent variable $L_{\scriptsize\mbox{H}\alpha}$ based on the
other parameters studied here.
They present a high degree of multicollinearity mainly due
to the $W_{\scriptsize\mbox{H}\beta}$-[OIII]/[OII]
relation. Instead, we want to identify which physical condition
parameter explains more efficiently the scatter as a second
independent parameter in the $L$-$\sigma$ relation.
A possible way to further investigate the variance of the space
parameters, overcoming the problem of multicollinearity in PCA, is
by using only a subset of parameters, where the subset is chosen so
that it does not contain or it is intended to minimize
multicollinearities\footnote{Principal Component Regression
(PCR) is also a multivariate analysis appropriate to investigate
problems in which multicollinearity is present \citep{jol02}}.

Table~\ref{tab9} presents the results of the PCA table for the same
previous sample containing 95 objects, undistinguished by their
emission line profiles or $W_{\scriptsize\mbox{H}\beta}$, for the
three parameter spaces:
[$L_{\scriptsize\mbox{H}\alpha}$,$\sigma_{\scriptsize\mbox{H}}$,O/H],
[$L_{\scriptsize\mbox{H}\alpha}$,$\sigma_{\scriptsize\mbox{H}}$,$W_{\scriptsize\mbox{H}\beta}$]
and
[$L_{\scriptsize\mbox{H}\alpha}$,$\sigma_{\scriptsize\mbox{H}}$,[OIII]/[OII]].
The results for
[$L_{\scriptsize\mbox{H}\alpha}$,$\sigma_{\scriptsize\mbox{H}}$,O/H]
show that in addition to $L_{\scriptsize\mbox{H}\alpha}$ and
$\sigma_{\scriptsize\mbox{H}}$, O/H has a relatively high loading in
the PCI (0.61), which shows its strong relevance to explain the
variance in the PCI. It results from the relatively higher
correlation of O/H with $\sigma_{\scriptsize\mbox{H}}$ as mentioned
above. Therefore PCII only explains 27\% of the total variance. On
the other hand, the analysis with the subset
[$L_{\scriptsize\mbox{H}\alpha}$,$\sigma_{\scriptsize\mbox{H}}$,$W_{\scriptsize\mbox{H}\beta}$]
shows that $W_{\scriptsize\mbox{H}\beta}$ is uncorrelated with
$\sigma_{\scriptsize\mbox{H}}$ and contributes efficiently to
explain 35\% of the variance in the PCII. The ionization ratio
[OIII]/[OII] also presents a higher contribution to PCII
than O/H.

\begin{table}[ht]


\centering

\caption{Testing the second parameter through individual principal
component analysis - 95 objects undistinguished by their emission
line profiles.}\label{tab9}

\begin{tabular}{lccc}\\ \hline\hline
&\multicolumn{3}{c}{Principal Component}\\
Parameter &I &II &III\\ \hline
$\log$ $L_{\scriptsize\mbox{H}\alpha}$ & 0.88 &-0.42 & 0.23\\
$\log$ $\sigma_{\scriptsize\mbox{H}}$  & 0.96 &-0.12 &-0.27\\
$\log$ (O/H)                           & 0.61 & 0.79 & 0.09\\
Eigenvalues                            &68.5\% &27.0\% &4.5\%\\
\hline
$\log$ $L_{\scriptsize\mbox{H}\alpha}$ & 0.95 & 0.18 &-0.25\\
$\log$ $\sigma_{\scriptsize\mbox{H}}$  & 0.96 &-0.15 & 0.25\\
$\log$ $W_{\scriptsize\mbox{H}\beta}$  &-0.03 & 0.99 & 0.08\\
Eigenvalues                            &60.8\% &34.9\% &4.3\%\\
\hline
$\log$ $L_{\scriptsize\mbox{H}\alpha}$ &-0.90 &-0.36 & 0.24\\
$\log$ $\sigma_{\scriptsize\mbox{H}}$  &-0.97 &-0.03 &-0.25\\
$\log$ [OIII]/[OII]                    & 0.38 &-0.92 &-0.08\\
Eigenvalues                            &63.2\% &32.6\% &4.2\%\\
\hline
\end{tabular}

\end{table}

The residual variance of 4-5\% in PCIII (Table~\ref{tab9}) accounts
for observational errors, but a fraction of it can be explained by
the fact that we have analyzed a heterogeneous sample. We have not
distinguished galaxies by their emission line profiles. In
fact, part of the large scatter in the high $L$ and $\sigma$ regime
populated mostly by I and C galaxies (see Figure~\ref{fig6}) may not
be explained at all, since single Gaussian fits seem not to well
represent the internal kinematics in these systems or the resultant
line widths are not well correlated with $L$. In addition, there
must be outliers also due to photometric errors propagated to
parameters that may be perturbing the PCA results. PCA is a powerful
technique that can be also used to detect outliers in data sets
providing robust results. In order to obtain the distance estimators
presented in next section, we have investigated and detected
outliers in the context of multiple regression.

We conclude by saying that $W_{\scriptsize\mbox{H}\beta}$ as an age
estimator of the starburst in H{\sc ii}Gs can not be neglected in a
regression model in order to obtain a distance indicator based on
the $L$-$\sigma$ relation. The ionization ratio
[OIII]/[OII] acts in a similar way as $W_{\scriptsize\mbox{H}\beta}$
and can be explicitly considered in the absence of
$W_{\scriptsize\mbox{H}\beta}$. Finally, the results shown above do
not mean that O/H can not be used as a second parameter, instead we have verified that
its efficiency to reduce the scatter of the $L$-$\sigma$
relation may be real but it seems to be smaller than the one using
$W_{\scriptsize\mbox{H}\beta}$. Additionally, O/H introduces a
higher degree of collinearity due to its relationship with
$\sigma$.

\subsection{Empirical Relations}\label{sec4_5}

Primary luminosity dependence on $\sigma$ has been evaluated by the
$L$-$\sigma$ relation considering only galaxies showing nearly
Gaussian emission line profiles. For a given velocity dispersion
$\sigma$ and metallicity (O/H), the galaxies with the largest
H$\beta$ equivalent width $W_{\scriptsize\mbox{H}\beta}$ or
ionization ratio [OIII]/[OII] should present also the largest Balmer
luminosities $L_{\scriptsize\mbox{H}\alpha}$. On the other hand, for
a given $\sigma$ and $W_{\scriptsize\mbox{H}\beta}$ or [OIII]/[OII],
galaxies with the lowest O/H ratio should also present the largest
$L_{\scriptsize\mbox{H}\alpha}$. The results found here for
the $L$-$\sigma$ relation (Table~\ref{tab6}) and PCA
(Section~\ref{sec4_4}), as well as in early studies
\citep[][MTM]{ter81}, corroborate the same empirical model for H{\sc
ii}Gs. Therefore, we need to obtain precise regression coefficients
through multiple regressions.

\begin{table*}[ht]

\scriptsize

\centering

\caption{Multiple regressions for $L_{\scriptsize\mbox{H}\alpha}$,
$\sigma_{\scriptsize\mbox{H}}$, O/H, $W_{\scriptsize\mbox{H}\beta}$
and [OIII]/[OII].}

\begin{tabular}{lcc}\\ \hline\hline
Parameters  &Regressions for $\log L_{\scriptsize\mbox{H}\alpha}$ &RMS\\
\hline
\multicolumn{2}{l}{Galaxies with Gaussian profiles (53 objects$^{1}$)}\\
\\
$L$-$\sigma$ ......................... &\multicolumn{1}{l}{(35.29$\pm$0.42) + (3.72$\pm$0.31)log $\sigma$}  &0.312\\
$L$-$\sigma$-O/H ................. &\multicolumn{1}{l}{(38.22$\pm$1.02) + (4.19$\pm$0.31)log $\sigma$ $-$ (0.44$\pm$0.14)log(O/H)} &0.289\\
$L$-$\sigma$-$W_{\scriptsize\mbox{H}\beta}$ ................ &\multicolumn{1}{l}{(34.60$\pm$0.39) + (3.84$\pm$0.25)log $\sigma$ + (0.32$\pm$0.08)log $W_{\scriptsize\mbox{H}\beta}$} &0.274\\
$L$-$\sigma$-[OIII]/[OII] ...... &\multicolumn{1}{l}{(34.69$\pm$0.38) + (4.06$\pm$0.27)log $\sigma$ + (0.34$\pm$0.09)log [OIII]/[OII]} &0.275\\
$L$-$\sigma$-O/H-$W_{\scriptsize\mbox{H}\beta}$ ........ &\multicolumn{1}{l}{(35.63$\pm$1.43) + (3.96$\pm$0.31)log $\sigma$ $-$ (0.14$\pm$0.18)log(O/H) + (0.27$\pm$0.11)log $W_{\scriptsize\mbox{H}\beta}$} &0.275\\
$L$-$\sigma$-O/H-[OIII]/[OII] &\multicolumn{1}{l}{(35.95$\pm$1.01) + (4.17$\pm$0.22)log $\sigma$ $-$ (0.17$\pm$0.17)log(O/H) + (0.27$\pm$0.11)log [OIII]/[OII]} &0.275\\
\\
\hline

\end{tabular}

\begin{tabular}{lcc}\\ \hline\hline
Parameters  &Regressions for $\log L_{\scriptsize\mbox{H}\alpha}$ &RMS\\
\hline
\multicolumn{2}{l}{Galaxies with Gaussian profiles (45 objects) without outliers.}\\
\\
$L$-$\sigma$ ......................... &\multicolumn{1}{l}{(35.26$\pm$0.38) + (3.76$\pm$0.27)log $\sigma$}  &0.270\\
$L$-$\sigma$-O/H ................. &\multicolumn{1}{l}{(39.94$\pm$0.99) + (4.33$\pm$0.25)log $\sigma$ $-$ (0.68$\pm$0.14)log(O/H)} &0.217\\
$L$-$\sigma$-$W_{\scriptsize\mbox{H}\beta}$ ................ &\multicolumn{1}{l}{(34.58$\pm$0.30) + (3.78$\pm$0.20)log $\sigma$ + (0.39$\pm$0.06)log $W_{\scriptsize\mbox{H}\beta}$} &0.201\\
$L$-$\sigma$-[OIII]/[OII] ...... &\multicolumn{1}{l}{(34.64$\pm$0.30) + (4.09$\pm$0.21)log $\sigma$ + (0.39$\pm$0.07)log [OIII]/[OII]} &0.204\\
$L$-$\sigma$-O/H-$W_{\scriptsize\mbox{H}\beta}$ ........ &\multicolumn{1}{l}{(36.64$\pm$1.38) + (4.01$\pm$0.25)log $\sigma$ $-$ (0.27$\pm$0.18)log(O/H) + (0.29$\pm$0.09)log $W_{\scriptsize\mbox{H}\beta}$} &0.197\\
$L$-$\sigma$-O/H-[OIII]/[OII] &\multicolumn{1}{l}{(37.04$\pm$1.29) + (4.26$\pm$0.22)log $\sigma$ $-$ (0.32$\pm$0.17)log(O/H) + (0.28$\pm$0.09)log [OIII]/[OII]} &0.198\\
\\
\hline

\end{tabular}

\quote 1.- The same subsample with 53 G objects used in
Table~\ref{tab6}.

\label{tab10}

\end{table*}

\begin{figure*}
\epsscale{0.4} \plotone{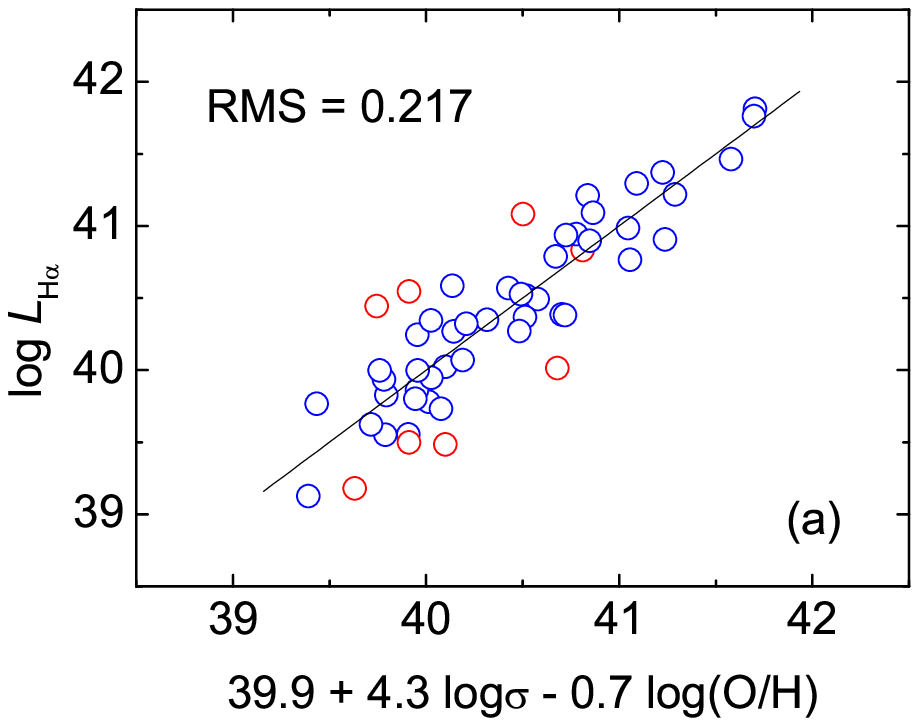}\plotone{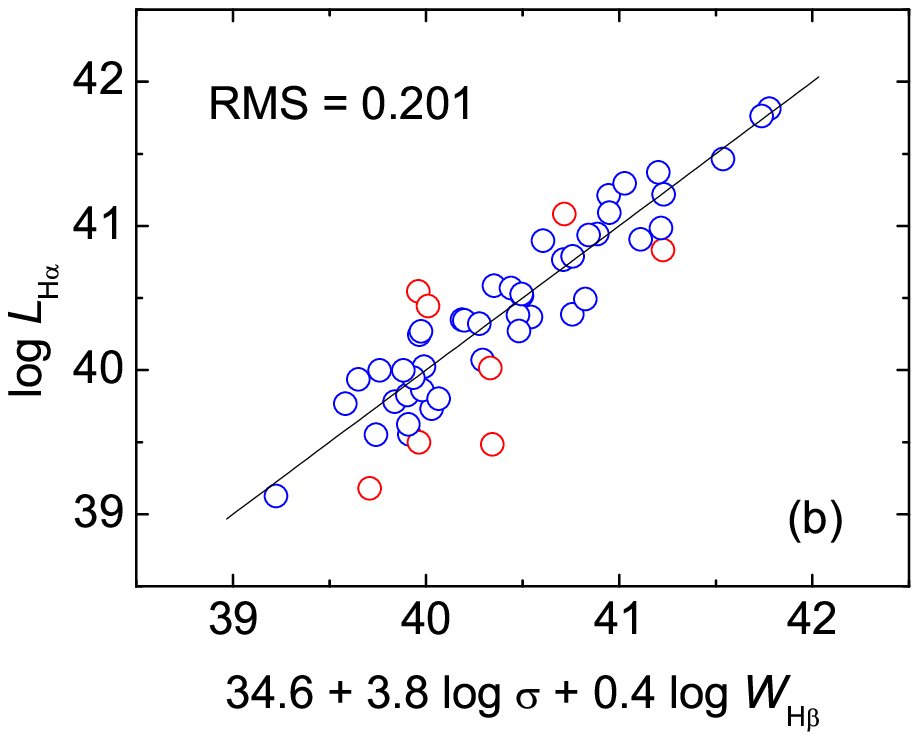}
\plotone{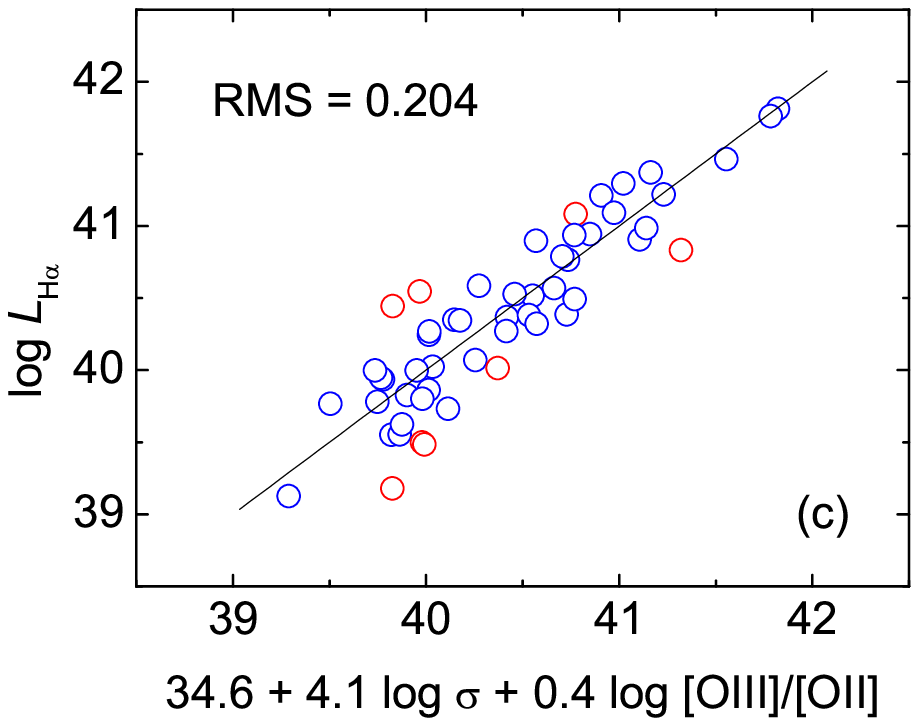}\plotone{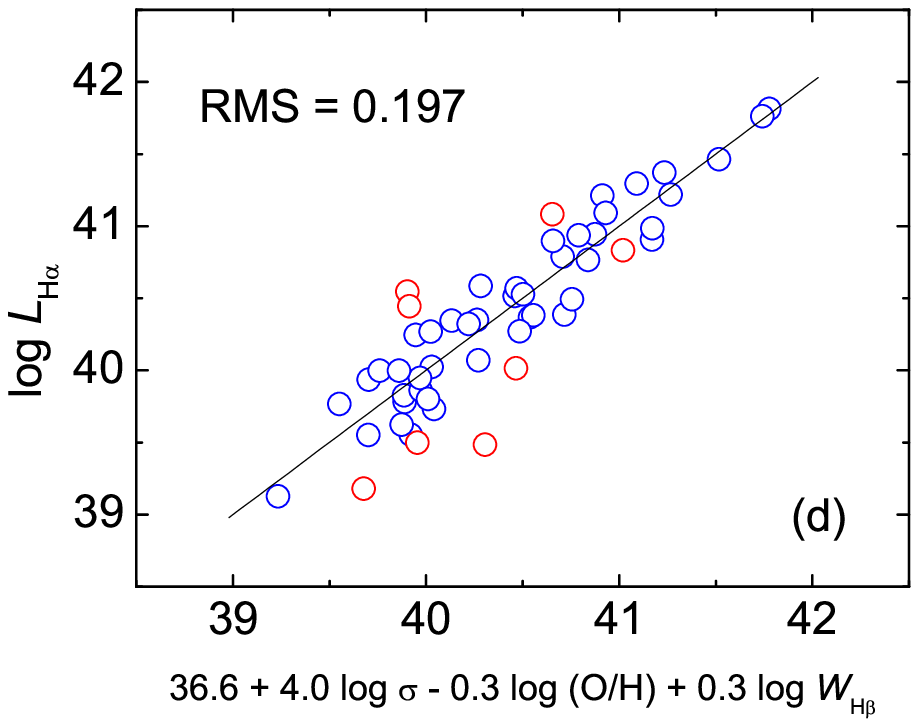} \plotone{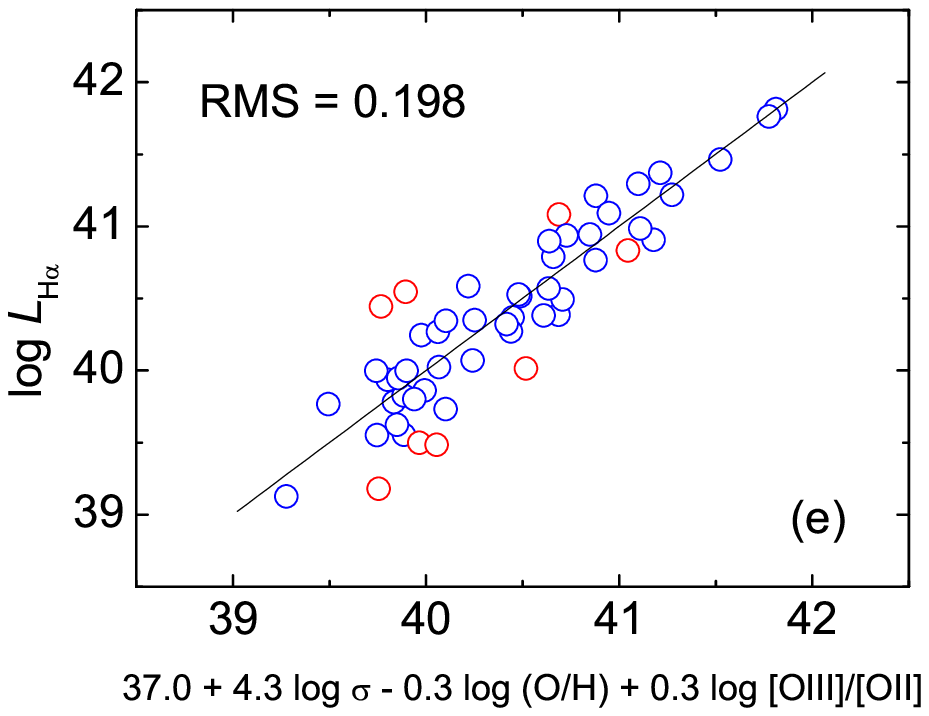}
\caption{Regression models plotted against observed H$\alpha$
luminosities for 53 objects. The functional forms are presented in
the $x$ axis of each graph. The solid lines represent the 1
to 1 lines fitted to the data. The calibration galaxies are plotted
as open blue circles, whereas the outliers are plotted as open red
circles. \label{fig10a}}
\end{figure*}

\begin{figure*}
\epsscale{0.4} \plotone{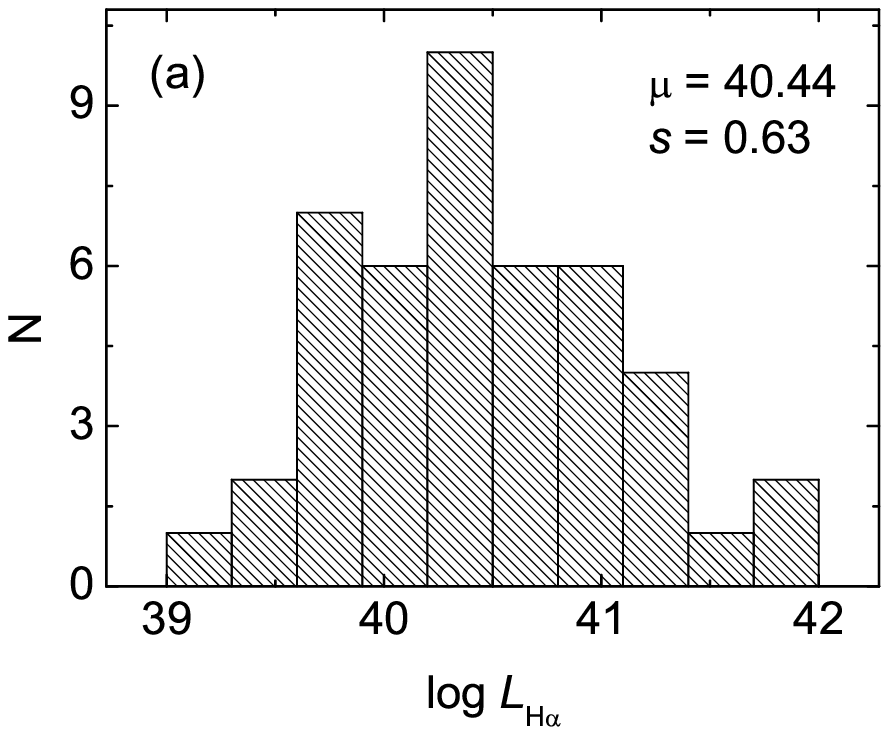}\plotone{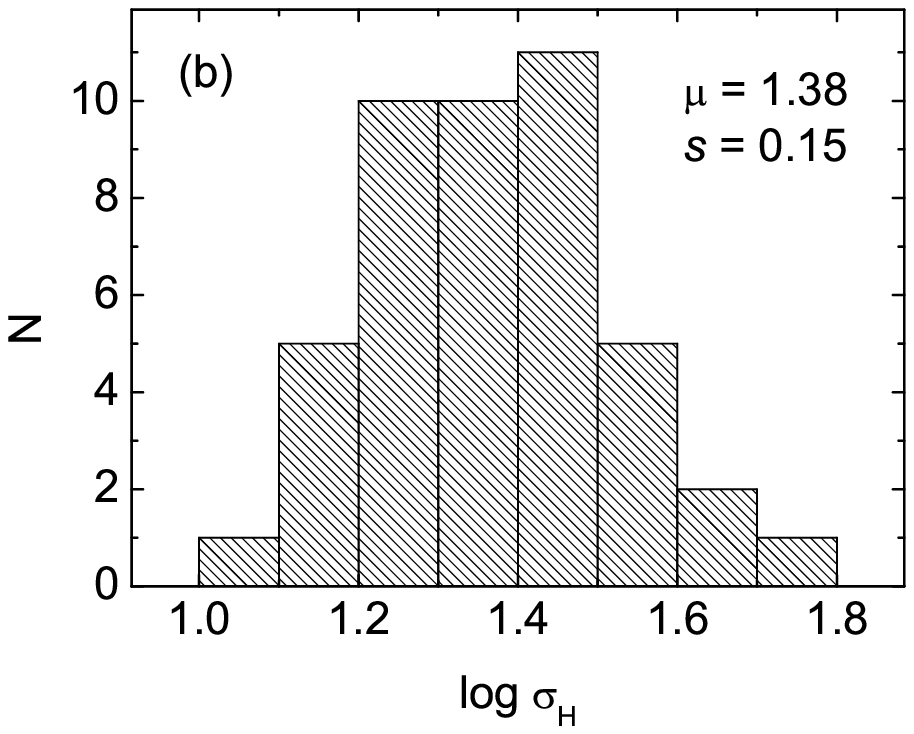}
\plotone{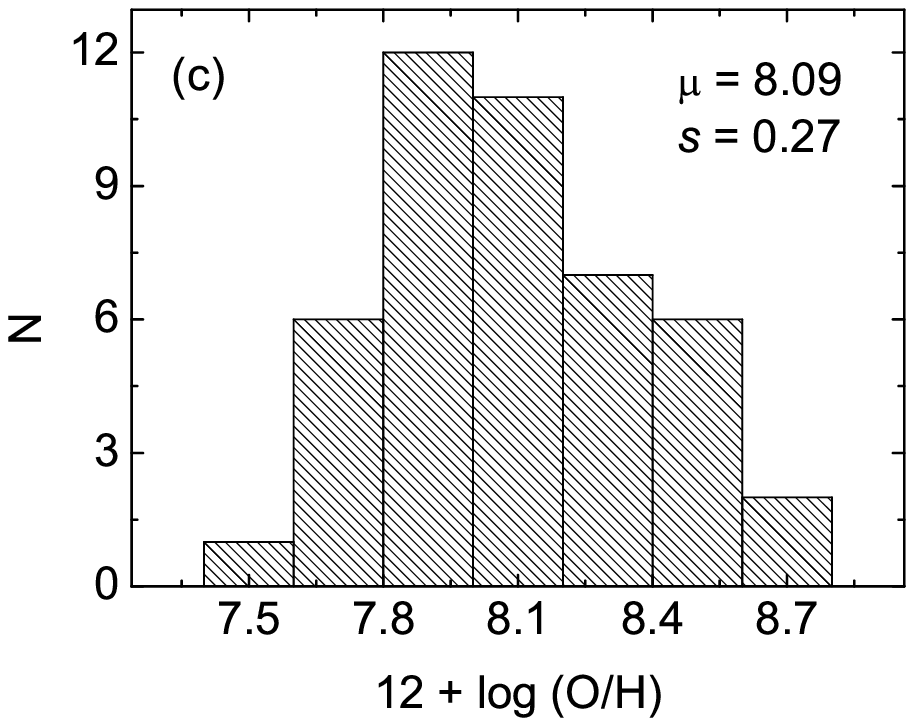}\plotone{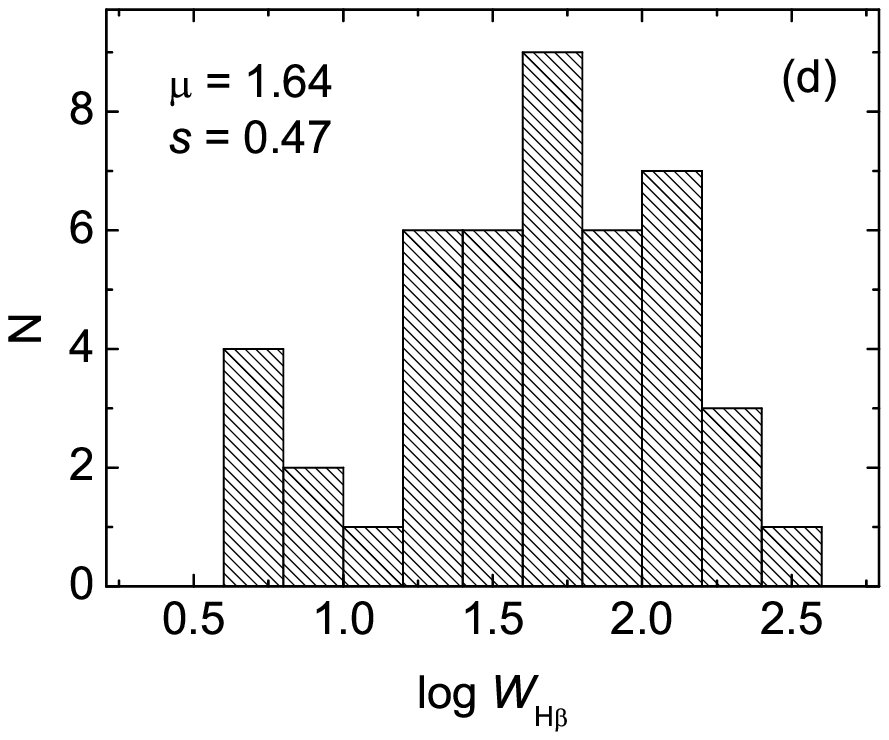}
\plotone{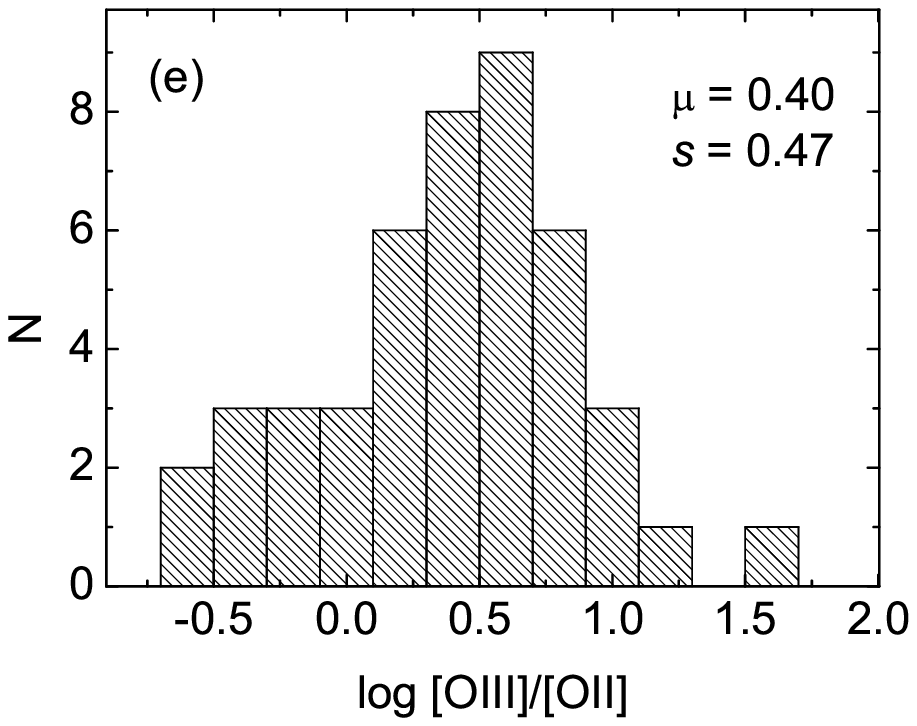}\caption{Histograms for all parameters
in the calibration sample (45 objects) in logarithmic
units. Mean ($\mu$) and standard deviations ($s$) of the
distributions are shown in each plot. \label{fig11a}}
\end{figure*}

Let us derive a set of empirical relations that can provide distance
indicators based on the homogeneous subsample of 53 objects showing
nearly Gaussian line profiles (visually classified as G) that have
shown the tightest $L$-$\sigma$ relation (Table~\ref{tab6}). It is
possible that some I or even C galaxies would fit in a regression
model increasing the statistical significance of the regression
coefficients, but in general, as shown in Section~\ref{sec3_4}, they
contribute to flatten the $L$-$\sigma$ relation increasing its
scatter especially in the regime of high $L$ and $\sigma$ values. It
suggests that some of them do not share the same physical
properties as most of the G galaxies or the single fit
procedure to derive their resultant $\sigma$ is not appropriate.
These objects may also suffer differentially from aperture effects
(Section~\ref{sec4_2}).
The sample containing 53 objects was chosen instead of the more
restrictive one (37 objects), selected by the semi-quantitative
criterion (Section~\ref{sec4_3}), since the later are not
well represented in the range 1.6 $<$ $\log \sigma$ $<$ 1.8 km
s$^{-1}$.

Multiple regression fits for 53 objects are shown in
Table~\ref{tab10}. Linear fits to the $L$-$\sigma$ relation for the
same sample were also presented in Table~\ref{tab6}. Note that the
scatter given by the RMS of the multiple regression fits did not
reduce significantly by the inclusion of a second or a third
independent parameter. It is mainly due to the presence of outliers.
Outliers may not introduce a significant problem in the
$L$-$\sigma$ relation (shown in Figure~\ref{fig7} and
Table~\ref{tab6}) because $L$ and $\sigma$ outlying measurements do
not surpass the intrinsic scatter (RMS $\sim$ 0.30). We have
included O/H, $W_{\scriptsize\mbox{H}\beta}$ and [OIII]/[OII], each
carrying a potential source of error. Some data were in fact
compiled from different works and they may have not been obtained at
exact same region in the galaxies. A specialized statistics used to
detect outliers in the context of multiple regressions is through
the computation of leverage, discrepancy (studentized
residual)\footnote{Also called ``Externally Studentized Residual''
or ``Studentized Deleted Residual''.} and influence indices
(Cook's $D$)\footnote{DFFITS is another global measure of influence
very closely related to Cook's $D$.}. Conceptually, influence
represents the product of leverage and discrepancy. These are known
as the three characteristics of potentially errant data points
\citep[see][for a didactic presentation and concepts]{coh03}. Using
this methodology eight data points were deleted from the sample
containing 53 objects. We have verified that the presence of these
outliers is mainly due to inclusion of objects with an uncertain
nature (MRK 1201 and MRK 1318), with data considered from different
sources (Tol 1008-286 and UM 463), and unusually high uncertainties
in derived parameters due to low S/N spectra (UM 417, Tol 0505-387,
UM 559 and Tol 2138-397).
Thus, Table~\ref{tab10} also shows a set of empirical relations
based on a sample of 45 objects without outliers. For RMS comparison
with the ones from regression models with two or three independent
parameters, we also presented the linear fit OLS(Y$|$X) to
$L$-$\sigma$ relation for the sample containing 45 objects free from
outliers. We reproduced a new version of the early empirical model
$L$-$\sigma$-O/H used by MTM to derive their distance
indicator with the same RMS scatter of $\delta\log
L_{\scriptsize\mbox{H}\alpha}=0.22$.
Independent functional forms for
$L$-$\sigma$-$W_{\scriptsize\mbox{H}\beta}$ and
$L$-$\sigma$-[OIII]/[OII] produce virtually the same result and they
are both more efficient than O/H as a second independent parameter
in the $L$-$\sigma$ relation reducing the RMS scatter to
$\delta\log L_{\scriptsize\mbox{H}\alpha}=0.20$. In addition, we
present the ``true'' empirical model for H{\sc ii}Gs,
$L$-$\sigma$-O/H-$W_{\scriptsize\mbox{H}\beta}$ (R1) and
$L$-$\sigma$-O/H-[OIII]/[OII] (R2).The scatter has not reduced
significantly by the inclusion of a third independent parameter |
O/H in $L$-$\sigma$-$W_{\scriptsize\mbox{H}\beta}\sim 2\%$ and O/H
in $L$-$\sigma$-[OIII]/[OII] $\sim3\%$ | but the residual is
compatible with the expected error in luminosity, which prevents us
from determining what fraction is intrinsic.

It is an important result that $W_{\scriptsize\mbox{H}\beta}$ can
account alone for a significant fraction of the scatter in
the $L$-$\sigma$ relation ($\sim25\%$) with a minimum
degree of collinearity. $W_{\scriptsize\mbox{H}\beta}$ has the
advantage of being a simpler parameter to obtain from the
spectra, though it depends on continuum detection. Alternatively,
if the intense [OIII]$\lambda\lambda$4959, 5007 and
[OII]$\lambda$3727 lines are detected one can use, with the same
precision, the $L$-$\sigma$-[OIII]/[OII] regression model. On the
other hand, the $L$-$\sigma$-O/H empirical model as distance
indicator would be more difficult to apply at great distances, since
it requires the detection of the [OIII]$\lambda4363$
auroral line to precisely calculate oxygen abundances by
the $T_{e}$-method.

Figure~\ref{fig10a} presents all five regression models, including
those with two (a, b and c panels) and three (d and e panels)
independent parameters plotted against the observed
$L_{\scriptsize\mbox{H}\alpha}$. The outliers
are plotted as open red circles in all panels.
Their predicted luminosity values were computed from the respective
regression models for 45 objects presented in Table~\ref{tab10}.
Figure~\ref{fig11a} presents the histograms of all five parameters
used in regression models, $L_{\scriptsize\mbox{H}\alpha}$ (a),
$\sigma_{\scriptsize\mbox{H}}$ (b), O/H (c),
$W_{\scriptsize\mbox{H}\beta}$ (d) and [OIII]/[OII] (e), with their
mean values and standard deviation in logarithmic units. These
histograms characterize our calibration sample. Note that we have
included some galaxies with $W_{\scriptsize\mbox{H}\beta}<30${\AA}
in our sample, which were not included by MTM in their calibration
sample. Our calibration sample with
45 local objects is a good representative sample of H{\sc ii}Gs in
general. The mean, standard deviation and median (8.03) of the
distribution of metallicity (Figure~\ref{fig11a} panel c) are
consistent with studies that characterize the population of H{\sc
ii}Gs from much larger samples \citep{kni04}.

\section{Discussion}\label{sec5}

\begin{figure*}[ht]
\epsscale{0.35}\plotone{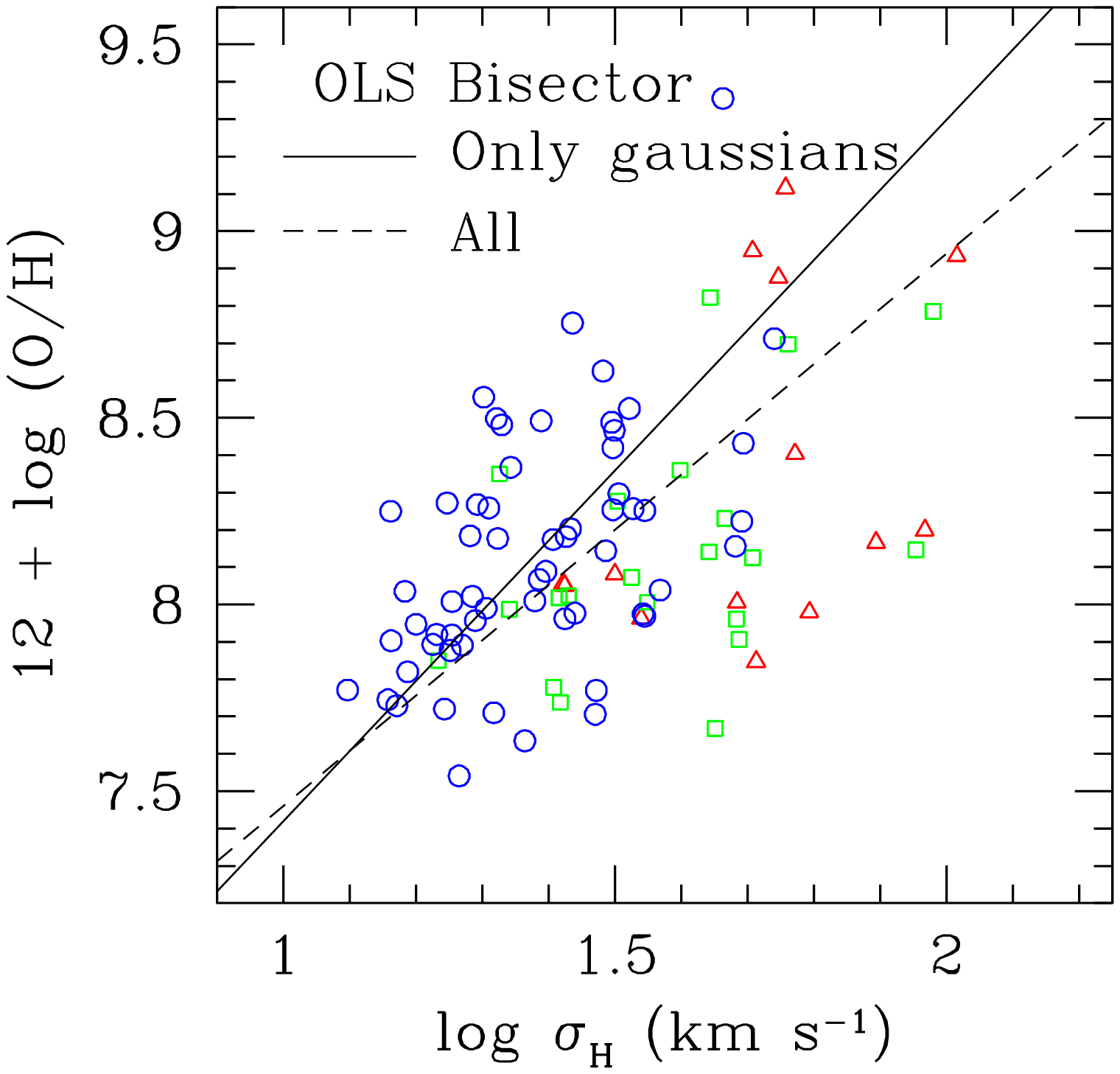}\plotone{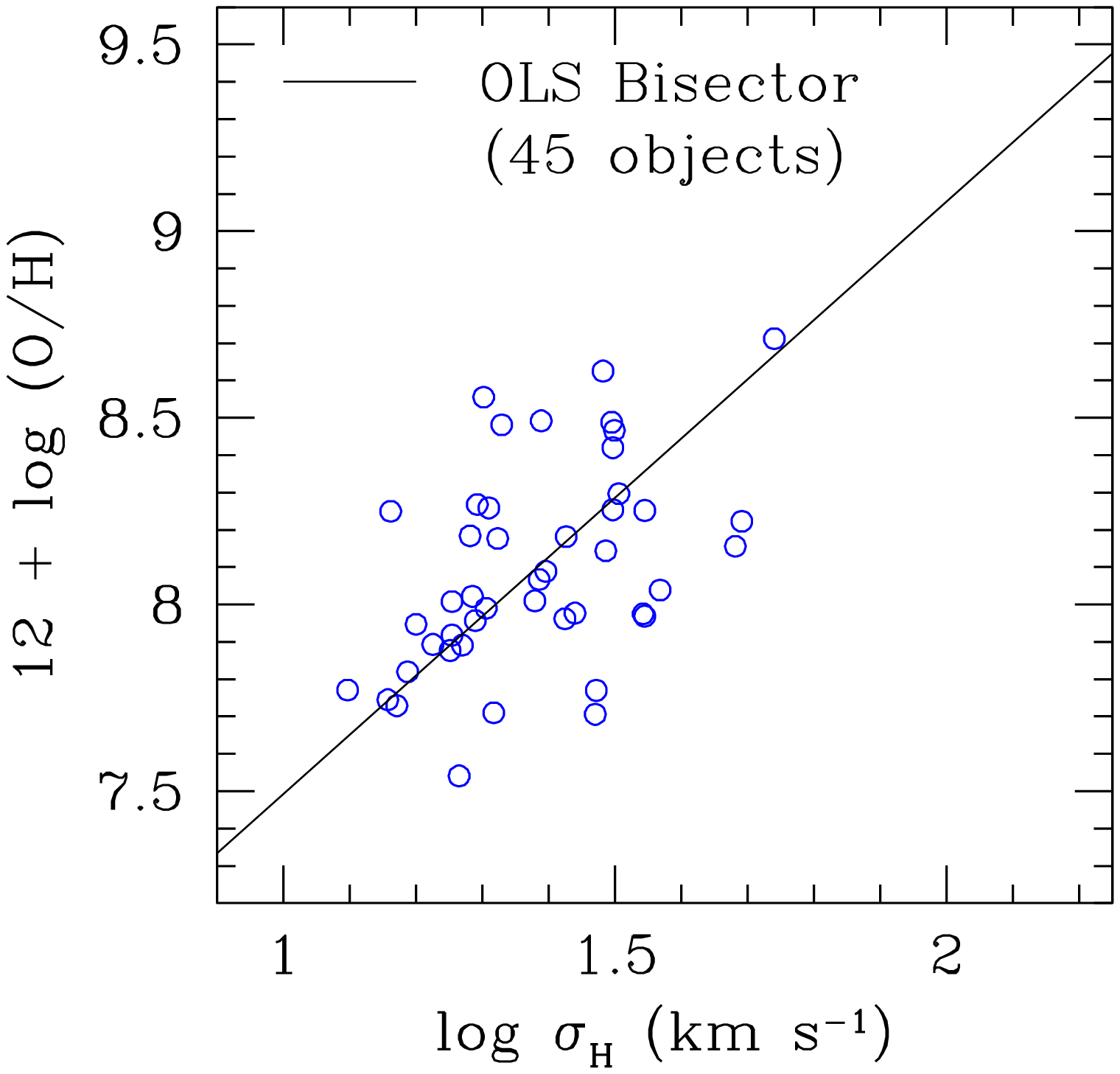}\caption{(\textit{left
panel}) The $\sigma$-$Z$ relation presented by 95 objects included
in PCA. The solid and dashed lines represent the OLS bisector fits
to the entire sample,
$12+\log$(O/H)$=(5.98\pm0.19)+(1.48\pm0.14)\log
\sigma_{\scriptsize\mbox{H}}$, and to the G sample,
$12+\log$(O/H)$=(5.54\pm0.34)+(1.88\pm0.26)\log
\sigma_{\scriptsize\mbox{H}}$, 60 objects), respectively.
(\textit{right panel}) The $\sigma$-$Z$ relation for the 45 objects
used in the calibration sample without outliers
(Table~\ref{tab10}). The solid line represents the OLS bisector fit,
$12+\log$(O/H)$=(5.91\pm0.23)+(1.59\pm0.17)\log
\sigma_{\scriptsize\mbox{H}}$.\label{fig12a}}
\end{figure*}

\subsection{Internal Dynamics}\label{sec5_1}

The existence of the $L$-$\sigma$ relation for GH{\sc ii}Rs
and H{\sc ii}Gs poses an intriguing question about the origin of the
supersonic velocity widths of the emission line profiles. Several
works have tried to answer this question through different
approaches, but the hypothesis that these systems are
gravitationally bound at least in the early stages of their
evolution persists as a paradigm \citep{ter81}. Figure~\ref{fig12a}
presents the $\sigma$-$Z$ relation for the whole sample presented in
PCA (Section~\ref{sec4_4}, 95 objects) and for those galaxies
presented in the calibration sample (Section~\ref{sec4_5}, 45
objects). If $\sigma$ correlates with metallicity in a similar
manner as in gravitationally bound systems, the underlying
relation could be the one between mass and metallicity. We present
the bisector regression fits for different samples in the caption of
Figure~\ref{fig12a}. In a recent work considering a large sample of
H{\sc ii}Gs, \cite{sal05} have shown the existence of the $L$-$Z$
relation for these systems. The Near-infrared (NIR) $L^{0.20}\propto
Z$ relation found by the authors, which more directly reflects the
underlying relationship between mass and metallicity, could be
compared with our results in order to check whether the
gravitational interpretation of the $L$-$\sigma$ relation of H{\sc
ii}Gs is acceptable. However, the NIR luminosity comes mostly from
the old stellar population, which may not be directly associated
with the current mass of the starburst. Another option would be to
compare the B band $L$-$Z$ with our $Z$-$\sigma$ relation. In fact,
if we use the scaling relation $L^{0.28}\propto Z$ derived for the B
band by \cite{sal05} and consider the $Z\propto\sigma^{1.6}$ found
here, we get closer to the $L\propto\sigma^{4}$. It still can not
constrain much about the gravitational interpretation, since we have
to explain how $\sigma$ of gas from the current starburst correlates
with metallicity. A simple explanation would be that metal rich
galaxies produce more massive starbursts, and this should be
verified by observations. These interesting questions are above the
scope of this work and should be further investigated.

\begin{figure*}[ht]
\epsscale{0.4}\plotone{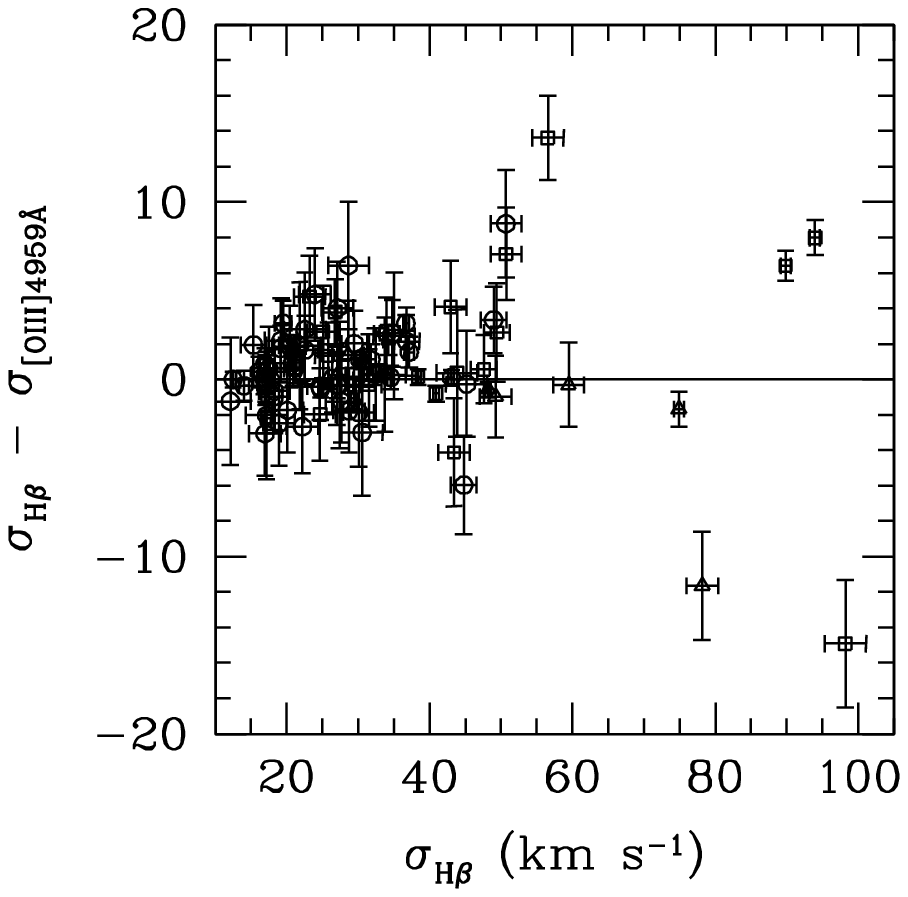}\plotone{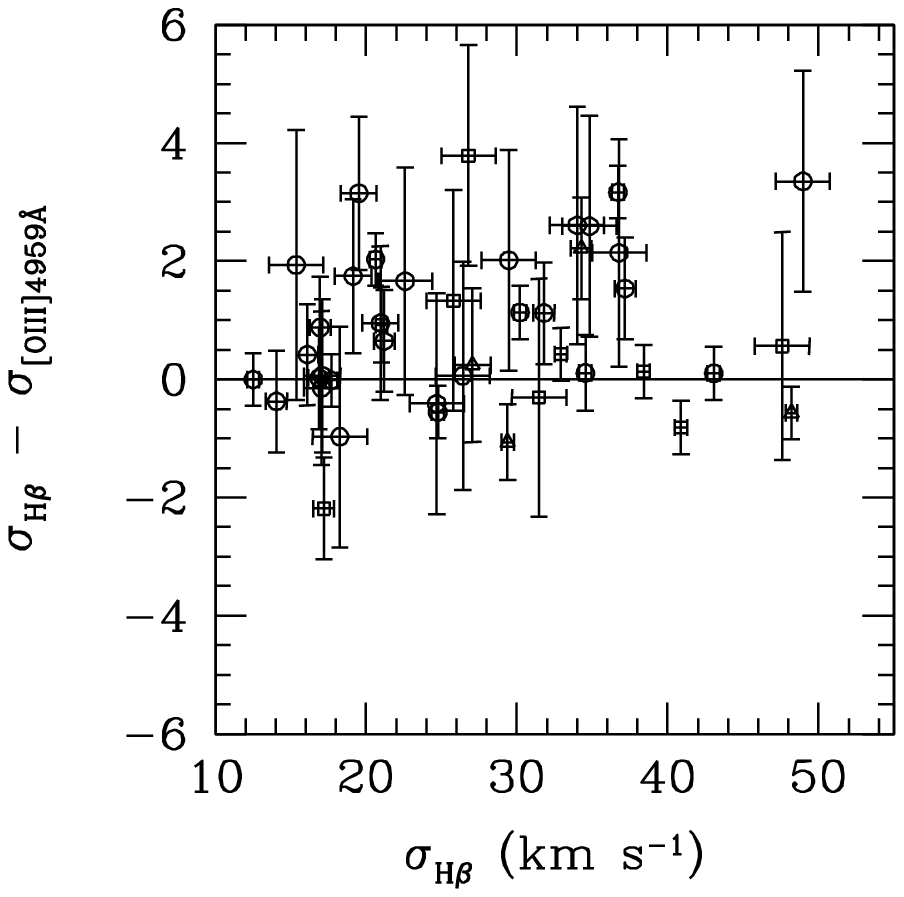}
\caption{Comparison between $\sigma$ derived from H$\beta$ and
[OIII]$\lambda$4959. (\textit{left panel}) All galaxies included.
(\textit{right panel}) Only galaxies presenting $\delta\sigma<2$ km
s$^{-1}$. A small systematic difference
$\sigma_{\scriptsize\mbox{H}}-\sigma_{\scriptsize\mbox{[OIII]}}=1-2$
km s$^{-1}$ is verified for galaxies with $\sigma\lesssim60$ km
s$^{-1}$. \label{fig13a}}
\end{figure*}

Another hint about the origin of velocity widths found in H{\sc
ii}Gs is provided by the systematic difference found between the
width of H{\sc ii} and [OIII] lines. This difference was firstly
found in GH{\sc ii}Rs by \cite{hip86}. 
Figure~\ref{fig13a} shows the differences found for our sample of
H{\sc ii}Gs.
In the left panel of Figure~\ref{fig13a} we plot all galaxies with
two lines observed, H$\beta$ and [OIII] 4959\AA, whereas in the right
panel we only plot the lines with the highest S/N with the smallest
$\sigma$ errors ($\delta\sigma<1.2$ km s$^{-1}$). We do not see any
systematic difference between these measurements above $\sim50$ km
s$^{-1}$. In fact, we note high discrepancies (5-20 km s$^{-1}$)
between $\sigma_{\scriptsize\mbox{H}}$ and
$\sigma_{\scriptsize\mbox{[OIII]}}$ with no trend. On the other
hand, we see that H{\sc ii}Gs showing $\sigma$ values in the regime
$15<\sigma<50$ km s$^{-1}$ present the intrinsic kinematic property
also found in
GH{\sc ii}Rs.
There is a small systematic difference between $\sigma$ derived from
both ions, $\langle\sigma_{\scriptsize\mbox{H}} -
\sigma_{\scriptsize\mbox{[OIII]}}\rangle = 1-2$ km s$^{-1}$. It
seems to be a special consequence of coupled kinematic and physical
conditions in these systems. A possible explanation would be the
existence of a well behaved ionization structure, in which more
excited ions are concentrated closer to the ionizing sources and
densest regions. However a gradient of velocity dispersion is not
observed at scales of tens to hundreds of parsecs in giant H{\sc ii}
regions. Other kinematic mechanisms that can be invoked to account
for the observed systematic difference can be associated with
turbulence \citep{hip86}.

We note that many H{\sc ii}Gs presented in the
Figure~\ref{fig13a} (right panel) are in fact those classified as
Gaussian line profiles (open circles). It is another indication that
these systems define a homogeneous sample in terms of their internal
kinematics. Although our sample presents few galaxies with both
measurements in high $\sigma$ regime, Figure~\ref{fig13a} (left
panel) suggests that H{\sc ii}Gs with $\sigma>$ 60 km s$^{-1}$ do
not seem to share the same kinematic properties as the ones with
$\sigma<$ 60 km s$^{-1}$.

\subsection{Age Effects: Short- and Long-Term Evolution}\label{sec5_2}

\begin{figure*}
\epsscale{1}\plotone{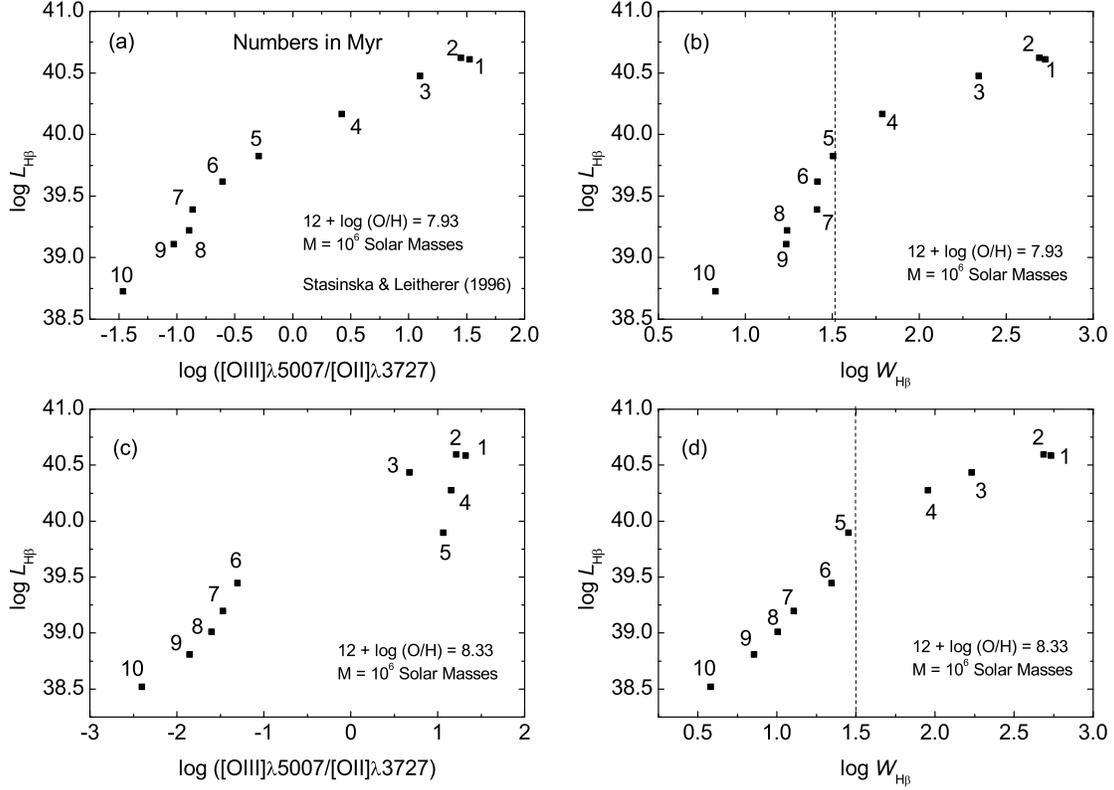} \caption{H$\beta$ luminosity as a
function of the [OIII]$\lambda5007$/[OII]$\lambda3727$ (a and c
panels) and H$\beta$ equivalent width (b and d) in logarithmic units
as predicted by single stellar population models by \cite{sta96}
with $M=10^{6}M_{\odot}$ and two metallicities, 12 + $\log$ (O/H) =
7.93 (a and b panels) and 8.33 (c and d panels). Numbers indicate
the age of the starburst in Myr. Vertical dotted lines in panels b
and d indicate the cutoff value
$W_{\scriptsize\mbox{H}\beta}\sim30${\AA} which coincides with age
$\sim$ 5 Myr of the starburst in both models. \label{fig14a}}
\end{figure*}

We have shown that the age effect (short-term evolution) is the
first order effect over the $L$-$\sigma$ relation and should not be
neglected explicitly in an empirical relation in order to derive a
distance indicator. Here we have tested
$W_{\scriptsize\mbox{H}\beta}$ and [OIII]/[OII] as fiducial
chronometers of the starburst. Both are virtually equally efficient
in order to explain the scatter in the $L$-$\sigma$ plane.
In order to reduce the scatter due to an age effect, one does not
need a parameter that measures the burst age, but that measures
efficiently the luminosity variation of the starburst as it ages.
Photoionized models have shown that the most robust parameter to
measure the luminosity evolution of the starburst is in fact
$W_{\scriptsize\mbox{H}\beta}$ \citep{sta96,mar08}. Its advantage is
due to its weak dependence on the metallicity. [OIII]/[OII] and
$W_{\scriptsize\mbox{[OIII]}5007}$ are more sensitive to
metallicity. Figure~\ref{fig14a} shows the predictions from
photoionized models by \cite{sta96} for H$\beta$ luminosity
evolution of the starburst as a function of
$W_{\scriptsize\mbox{H}\beta}$ (panels b and d) and [OIII]/[OII]
(panels a and c), for a given starburst mass of 10$^{6}$M$_{\odot}$
and two metallicities $12+\log$(O/H)=7.93 (panels a and b) and 8.33
(c and d). It is possible to see that
$L$($W_{\scriptsize\mbox{H}\beta}$) presents nearly constant curve
shape in both abundance sets, whereas $L$([OIII]/[OII]) does not
present a monotonic curve in $12+\log$(O/H)=8.33 set, especially in
the first 5 Myrs of the starburst. Note that
$L$($W_{\scriptsize\mbox{H}\beta}$) still presents a more stable
curve for $\log W_{\scriptsize\mbox{H}\beta}>1.5$ or $\sim$ 30\AA.
Models considering the continuous burst scenario have shown that
$W_{\scriptsize\mbox{H}\beta}$ is in fact a robust age indicator of
the starburst even for a significant presence of an underlying
stellar population \citep{mar08}.

Metallicity itself is also a tracer of the evolution, though it is
more closely linked to the long-term history of the star formation
of the galaxy. The metallicity given by the O/H ratio, as proposed
by MTM, may represent a second parameter in the
$L$-$\sigma$ relation, since successive chemical enrichment promoted
by generations of stars may affect globally the luminosity evolution
(short-term) of each starburst.

Recent efforts have been made by \cite{pli10} in order to
investigate the precision required on distance determinations of
high-$z$ H{\sc ii}Gs to achieve cosmological goals. Following
\cite{mel00}, these authors propose to apply the scaling relations,
like the ones derived here, to measure the dark energy equation of
state, $w$($z$), and the matter content of the Universe
$\Omega_{\scriptsize\mbox{m}}$, by using the Hubble diagram for
H{\sc ii}Gs as an alternative to Supernovae type Ia above $z\sim1$.
Considerations made here about the systematic effects over
the $L$-$\sigma$ relation may be valuable for its further
application.

\section{Summary}\label{sec_6}


We have presented a new large data set of line width measurements
for star-forming regions in over 100 H{\sc ii}Gs. This was
used to analyse in some detail, the line profile class as a
function of the galaxy morphology, line fitting methodologies,
principal component analysis of the observed physical parameters,
and the statistical methods of data selection, rejection
and linear regression to reach our final conclusions on the
calibration of the $L$-$\sigma$ relation for local H{\sc ii}Gs.

(i) The family of H{\sc ii}Gs as firstly catalogued mainly by
emission line detection does not represent a homogeneous sample in
terms of their internal kinematics. Gaussianity of their emission
line profiles is closely related to the physical origin of the
$L$-$\sigma$ relation, since galaxies showing the most Gaussian line
profiles seem to produce a steeper $L\propto\sigma^{4}$ relation
with the least scatter ($\delta\log
L_{\scriptsize\mbox{H}\alpha}=0.30$). These objects represent a more
homogeneous class in terms of kinematics and define the key property
of the H{\sc ii}G targets of the $L$-$\sigma$ relation.
These objects also seem to suffer less from aperture effects.

(ii) The $L$-$\sigma$ relation is simultaneously affected by the
short-
and the long-term evolution of the starburst.
Both effects can be accounted for, to predict Balmer
luminosity in a functional form in order to obtain a more accurate
distance indicator. Here we have quantified the contribution of
H$\beta$ equivalent width and the ionization parameter
[OIII]$\lambda\lambda5007,4959$/[OII]$\lambda3727$ as measures of
short-term evolution and O/H as measures the long-term evolution.

(iii) We calibrated the $L$-$\sigma$-O/H empirical relation,
previously presented by MTM as the distance indicator of the H{\sc
ii}Gs and provided an alternative set of new empirical relations,
especially the $L$-$\sigma$-$W_{\scriptsize\mbox{H}\beta}$ relation,
that can be used alternatively, with greater accuracy ($\delta\log
L_{\scriptsize\mbox{H}\alpha}=0.20$ or 0.5 mag) and
simplicity, to determine distances of H{\sc ii}Gs at high redshifts
with application to cosmology questions.

\acknowledgments VB acknowledges his  CAPES scholarship.  We are
grateful to the staff at ESO who have helped us with the many
observing runs at La Silla. We are very thankful to Roberto
Terlevich, Elena Terlevich, Guillermo Tenorio-Tagle and Casiana
Mu\~noz-Tu\~n\'on for innumerous discussions on topics related to
this work over the years. Finally, we are thankful to the anonymous
referee for his/her comments to improve the paper.

\bibliographystyle{apj}
\bibliography{mybib}

\end{document}